\documentclass[aps,pra,superscriptaddress,notitlepage,twocolumn]{revtex4-2}
\pdfoutput=1
\usepackage[english]{babel}
\usepackage[utf8]{inputenc}
\usepackage[dvipsnames]{xcolor}
\usepackage{graphicx}
\usepackage{url}
\definecolor{VibrantBlue}{HTML}{0077BB}
\definecolor{VibrantCyan}{HTML}{33BBEE}
\definecolor{VibrantTeal}{HTML}{009988}
\definecolor{VibrantOrange}{HTML}{EE7733}
\definecolor{VibrantRed}{HTML}{CC3311}
\definecolor{VibrantMagenta}{HTML}{EE3377}
\definecolor{VibrantGrey}{HTML}{BBBBBB}
\definecolor{PaleBlue}{HTML}{BBCCEE}
\definecolor{PaleCyan}{HTML}{CCEEFF}
\definecolor{PaleGreen}{HTML}{CCDDAA}
\definecolor{PaleYellow}{HTML}{EEEEBB}
\definecolor{PaleRed}{HTML}{FFCCCC}
\definecolor{PaleGrey}{HTML}{DDDDDD}
\definecolor{LightBlue}{HTML}{77AADD}
\definecolor{LightCyan}{HTML}{99DDFF}
\definecolor{LightMint}{HTML}{44BB99}
\definecolor{LightPear}{HTML}{BBCC33}
\definecolor{LightOlive}{HTML}{AAAA00}
\definecolor{LightYellow}{HTML}{EEDD88}
\definecolor{LightOrange}{HTML}{EE8866}
\definecolor{LightPink}{HTML}{FFAABB}
\definecolor{LightGrey}{HTML}{DDDDDD}
\usepackage[colorlinks=true, urlcolor=VibrantBlue, citecolor=VibrantBlue, linkcolor=VibrantMagenta,anchorcolor=VibrantMagenta]{hyperref}
\usepackage{bibentry}
\usepackage{enumerate}
\usepackage{dcolumn}
\usepackage{multirow}
\usepackage{physics}
\usepackage{bm}
\usepackage{bbm}
\usepackage{soul}
\usepackage[labelfont=bf]{caption}
\usepackage{subcaption}
\usepackage[makeroom]{cancel}
\usepackage{adjustbox}
\usepackage{mathtools}
\usepackage{amssymb}
\usepackage{amsmath}
\newcommand{\eq}[2]{\begin{equation}\begin{split} \label{eq:#1} #2 \end{split}\end{equation}}

\DeclareMathOperator{\true}{\mathsf{tr}}
\DeclareMathOperator{\fal}{\mathsf{fa}}
\begin{document}

\title{Variational inference with a quantum computer}

\author{Marcello Benedetti}
\email{marcello.benedetti@cambridgequantum.com}
\affiliation{Cambridge Quantum Computing Limited, SW1E 6DR London, United Kingdom}

\author{Brian Coyle}
\affiliation{Cambridge Quantum Computing Limited, SW1E 6DR London, United Kingdom}
\affiliation{School of Informatics, University of Edinburgh,  EH8 9AB Edinburgh, United Kingdom}

\author{Mattia Fiorentini}
\affiliation{Cambridge Quantum Computing Limited, SW1E 6DR London, United Kingdom}

\author{Michael Lubasch}
\affiliation{Cambridge Quantum Computing Limited, SW1E 6DR London, United Kingdom}

\author{Matthias Rosenkranz}
\email{matthias.rosenkranz@cambridgequantum.com}
\affiliation{Cambridge Quantum Computing Limited, SW1E 6DR London, United Kingdom}

\date{October 28, 2021}

\begin{abstract} 
Inference is the task of drawing conclusions about unobserved variables given observations of related variables. Applications range from identifying diseases from symptoms to classifying economic regimes from price movements. Unfortunately, performing exact inference is intractable in general. One alternative is variational inference, where a candidate probability distribution is optimized to approximate the posterior distribution over unobserved variables. For good approximations, a flexible and highly expressive candidate distribution is desirable. In this work, we use quantum Born machines as variational distributions over discrete variables. We apply the framework of operator variational inference to achieve this goal. In particular, we adopt two specific realizations: one with an adversarial objective and one based on the kernelized Stein discrepancy. We demonstrate the approach numerically using examples of Bayesian networks, and implement an experiment on an IBM quantum computer. Our techniques enable efficient variational inference with distributions beyond those that are efficiently representable on a classical computer.
\end{abstract}

\maketitle

\section{Introduction}

Probabilistic graphical models describe the dependencies of random variables in complex systems~\cite{koller2009probabilistic}. This framework enables two important tasks: learning and inference. Learning yields a model that approximates the observed data distribution. Inference uses the model to answer queries about unobserved variables given observations of other variables. In general, exact inference is intractable and so producing good \emph{approximate} solutions becomes a desirable goal. This article introduces approximate inference solutions using a hybrid quantum-classical framework.

Prominent examples of probabilistic graphical models are Bayesian and Markov networks. Applications across many domains employ inference on those models, including in health care and medicine~\cite{morrisRecognitionNetworksApproximate2001,richensImprovingAccuracyMedical2020}, biology, genetics, and forensics~\cite{maathuis2018handbook,wiegerinckBayesianNetworksExpert2010}, finance~\cite{denev2015probabilistic}, and fault diagnosis~\cite{caiBayesianNetworksFault2017}. These are applications where qualifying and quantifying the uncertainty of conclusions is crucial. The posterior distribution can be used to quantify this uncertainty. It can also be used in other downstream tasks such as determining the likeliest configuration of unobserved variables that best explains observed data.

Approximate inference methods broadly fall into two categories: Markov chain Monte Carlo (MCMC) and variational inference (VI). MCMC methods produce samples from the true posterior distribution in an asymptotic limit~\cite{nealProbabilisticInferenceUsing1993,brooksHandbookMarkovChain2011}. VI is a machine learning technique that casts inference as an optimization problem over a parameterized family of probability distributions~\cite{Blei_2017}. If quantum computers can deliver even a small improvement to these methods, the impact across science and engineering could be large.

Initial progress in combining MCMC with quantum computing was achieved by replacing standard MCMC with quantum annealing hardware in the training of some types of Bayesian and Markov networks~\cite{adachi2015application, Benedetti_2016, Benedetti_2017, korenkevych2016benchmarking, Benedetti_2018, Khoshaman_2018, wilson2019quantumassisted}. Despite promising empirical results, it proved difficult to show \emph{if} and \emph{when} a quantum advantage could be delivered. An arguably more promising path towards quantum advantage is through gate-based quantum computers. In this context, there exist algorithms for MCMC with proven asymptotic advantage~\cite{low2014quantum, Montanaro_2015, chowdhury2016quantum, Wittek_2017}, but they require error correction and other features that go beyond the capability of existing machines. Researchers have recently shifted their focus to near-term machines, proposing new quantum algorithms for sampling thermal distributions~\cite{wu2019variational, verdonQuantumHamiltonianBasedModels2019, chowdhury2020variational, wang2020variational, shingu2020boltzmann}.

The advantage of combining quantum computing with VI has not been explored to the same extent as MCMC. Previous work includes classical VI algorithms using ideas from quantum annealing~\cite{satoQuantumAnnealingVariational2009,ogorman2015bayesian,miyaharaQuantumExtensionVariational2018}. In this work, we turn our attention to performing VI on a quantum computer. We adopt a distinct approach that focuses on improving inference in classical probabilistic models by using quantum resources. We use \emph{Born machines}, which are quantum machine learning models that exhibit high expressivity. We show how to employ gradient-based methods and \textit{amortization} in the training phase. These choices are inspired by recent advances in classical VI~\cite{zhang2018advances}.

Finding a quantum advantage in machine learning in \emph{any} capacity is an exciting research goal, and promising theoretical works in this direction have recently been developed across several prominent subfields. Advantages in supervised learning have been proposed considering: information theoretic arguments~\cite{abbas_power_2020, huang_information-theoretic_2021, poland2020free, sharma2020reformulation}, probably approximately correct (PAC) learning~\cite{servedio_equivalences_2004, arunachalam_guest_2017, arunachalam_optimal_2018, liu_rigorous_2020}, and the representation of data in such models~\cite{havlicek_supervised_2019, schuld2019quantum, huang_power_2020}. More relevant for our purposes are results in \emph{unsupervised} learning, which have considered complexity and learning theory arguments for distributions~\cite{gao_quantum_2018, Coyle_2020, sweke_quantum_2020} and quantum nonlocality and contextuality~\cite{gao_enhancing_2021}. Furthermore, some advantages have been observed experimentally~\cite{riste_demonstration_2017, coyle_quantum_2020, alcazar_classical_2020, johri_nearest_2020}. For further reading, see Refs.~\cite{ciliberto_quantum_2018, benedetti_parameterized_2019, lamata_quantum_2020} for recent overviews of some advances in quantum machine learning.

In Sec.~\ref{s:variational_inference} we describe VI and its applications. In Sec.~\ref{s:born_as_implicit} we describe using Born machines to approximate posterior distributions. In Sec.~\ref{s:methods} we use the framework of operator VI to derive two suitable objective functions, and we employ classical techniques to deal with the problematic terms in the objectives. In Sec.~\ref{s:experiments} we demonstrate the methods on Bayesian networks. We conclude in Sec.~\ref{s:discussion} with a discussion of possible generalizations and future work.

\section{Variational inference and applications} 
\label{s:variational_inference}

It is important to clarify what type of inference we are referring to. Consider a probabilistic model $p$ over some set of random variables, $\mathcal{Y}$. The variables in $\mathcal{Y}$ can be continuous or discrete. Furthermore, assume that we are given \emph{evidence} for some variables in the model. This set of variables, denoted $\mathcal{X} \subseteq \mathcal{Y}$, is then \emph{observed} (fixed to the values of the evidence) and we use the vector notation $\bm{x}$ to denote a realization of these observed variables. We now want to \emph{infer} the posterior distribution of the unobserved variables, those in the set $\mathcal{Z} := \mathcal{Y}\backslash\mathcal{X}$. Denoting these by a vector $\bm{z}$, our target is that of computing the posterior distribution $p(\bm{z}|\bm{x})$, the conditional probability of $\bm{z}$ given $\bm{x}$. By definition, the conditional can be expressed in terms of the joint divided by the marginal: $p(\bm{z}|\bm{x})=p(\bm{x},\bm{z})/p(\bm{x})$. Also, recall that the joint can be written as $p(\bm{x}, \bm{z}) = p(\bm{x} |\bm{z}) p(\bm{z})$. Bayes' theorem combines the two identities and yields $p(\bm{z}|\bm{x}) = p(\bm{x} |\bm{z}) p(\bm{z}) / p(\bm{x})$.

\begin{figure}
    \centering
    \begin{subfigure}[t]{0.47\textwidth}
        \centering
        \includegraphics[]{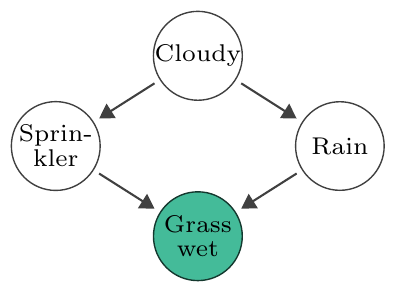}
        \caption{``Sprinkler'' network}
        \label{fig:sprinkler}
    \end{subfigure}
    \vskip5mm
    \begin{subfigure}[t]{0.47\textwidth}
        \centering
        \includegraphics[]{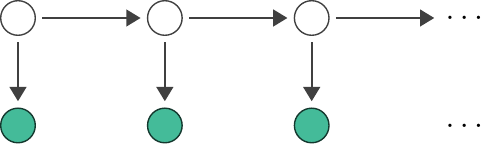}
        \caption{Regime switching in time series}
        \label{fig:hmm}
    \end{subfigure}
    \vskip5mm
    \begin{subfigure}[t]{0.47\textwidth}
        \centering
        \includegraphics[]{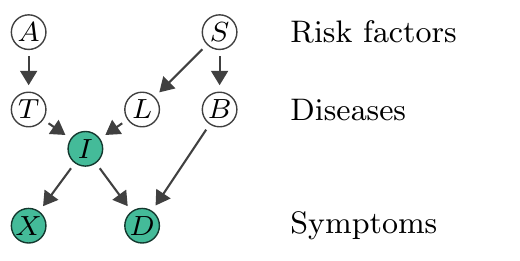}
        \caption{``Lung cancer'' network}
        \label{fig:lung_cancer}
     \end{subfigure}
    \vskip5mm
    \begin{subfigure}[t]{0.47\textwidth}
        \centering
        \includegraphics[]{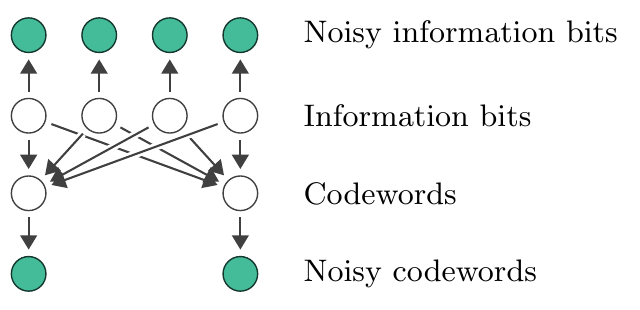}
        \caption{Error correction codes}
        \label{fig:turbo_codes}
    \end{subfigure}
    \caption{Some applications of inference on Bayesian networks. Filled (empty) circles indicate observed (unobserved) random variables. \subref{fig:sprinkler} ``Sprinkler'' network. \subref{fig:hmm} Regime switching in time series. \subref{fig:lung_cancer} ``Lung cancer'' network~\cite{lauritzen_local_1988}. \subref{fig:turbo_codes} Error correction codes.}
    \label{fig:examples}
\end{figure}

As described above, the inference problem is rather general. To set the scene, let us discuss some concrete examples in the context of Bayesian networks. These are commonly targeted in inference problems, and as such shall be our test bed for this work. A Bayesian network describes a set of random variables with a clear conditional probability structure. This structure is a directed acyclic graph where the conditional probabilities are modeled by tables, by explicit distributions, or even by neural networks. Figure~\ref{fig:sprinkler} is the textbook example of a Bayesian network for the distribution of binary variables: cloudy $(C)$, sprinkler $(S)$, rain $(R)$, and the grass being wet $(W)$. According to the graph this distribution factorizes as $P(C,S,R,W)=P(C)P(S|C)P(R|C)P(W|S,R)$. A possible inferential question is: what is the probability distribution of $C$, $S$, and $R$ given that $W=\true$? This can be estimated by ``inverting'' the probabilities using Bayes' theorem: $p(C, S, R | W=\true) = p(W=\true | C, S, R) p(C, S, R) / p(W)$. 

Figures~\ref{fig:hmm}--\ref{fig:turbo_codes} show a few additional applications of inference on Bayesian networks. The hidden Markov model in Fig.~\ref{fig:hmm} describes the joint probability distribution of a time series of asset returns and an unobserved ``market regime'' (e.g., a booming versus a recessive economic regime). 
A typical inference task is to detect regime switches by observing asset returns~\cite{kritzmanRegimeShiftsImplications2012}.  Figure~\ref{fig:lung_cancer} illustrates a modified version of the ``lung cancer'' Bayesian network that is an example from medical diagnosis (see, e.g., Ref.~\cite{richensImprovingAccuracyMedical2020} and the references therein). This network encodes expert knowledge about the relationship between risk factors, diseases, and symptoms. In health care, careful design of the network and algorithm are critical in order to reduce biases, e.g., in relation to health care access~\cite{obermeyerDissectingRacialBias2019}. Note that inference in medical diagnosis is often causal instead of associative~\cite{richensImprovingAccuracyMedical2020}. Bayesian networks can be interpreted causally and help answer causal queries using Pearl's do-calculus~\cite{pearlCausalityModelsReasoning2009}. Finally, Fig.~\ref{fig:turbo_codes} shows a Bayesian network representation of turbo codes, which are error correction schemes used in 3G and 4G mobile communications. The inference task for the receiver is to recover the original information bits from the information bits and codewords received over a noisy channel~\cite{mcelieceTurboDecodingInstance1998}.

Inference is a computationally hard task in all but the simplest probabilistic models. Roth~\cite{dan_roth_hardness_1996} extended results of Cooper~\cite{cooper1990computational} and showed that \emph{exact} inference in Bayesian networks with discrete variables is sharp-P-complete. Dagum and Luby~\cite{dagum1993approximating} showed that even \emph{approximate} inference is NP-hard. Therefore, unless some particular constraints are in place, these calculations are intractable. In many cases one is able to perform a ``forward pass'' and obtain unbiased samples from the joint $(\bm{x},\bm{z}) \sim p(\bm{x},\bm{z}) = p(\bm{x}|\bm{z}) p(\bm{z})$. However, obtaining unbiased samples from the posterior $\bm{z} \sim p(\bm{z}|\bm{x}) = p(\bm{x}, \bm{z}) / p(\bm{x})$ is intractable due to the unknown normalization constant. One can perform MCMC sampling by constructing an ergodic Markov chain whose stationary distribution is the desired posterior. MCMC methods have nice theoretical guarantees, but they may converge slowly in practice~\cite{brooksHandbookMarkovChain2011}. In contrast, VI is often faster in high dimensions but does not come with guarantees. The idea in VI is to optimize a variational distribution $q$ by minimizing its ``distance'' from the true posterior $p$ (see Ref.~\cite{Blei_2017} for an introduction to the topic and Ref.~\cite{zhang2018advances} for a review of the recent advances).

VI has experienced a resurgence in recent years due to substantial developments. First, generic methods have reduced the amount of analytical calculations required and have made VI much more user friendly (e.g., black-box VI~\cite{ranganath2013black}). Second, machine learning has enabled the use of highly expressive variational distributions implemented via neural networks~\cite{kingma2014autoencoding}, probabilistic programs~\cite{wingate2013automated}, nonparametric models~\cite{gershman2012nonparametric}, normalizing flows~\cite{rezende2016variational}, and others. In contrast, the original VI methods were mostly limited to analytically tractable distributions such as those that could be factorized. Third, amortization methods have reduced the costs by optimizing $q_{\bm{\theta}}(\bm{z}|\bm{x})$ where the vector of parameters $\bm{\theta}$ is ``shared'' among all possible observations $\bm{x}$ instead of optimizing individual parameters for each observation. This approach can also generalize across inferences.

Casting the inference problem as an optimization problem comes itself with challenges, in particular when the unobserved variables are discrete. \mbox{REINFORCE~\cite{williams1992simple}} is a generic method that requires calculation of the score $\partial_{\bm{\theta}} \log q_{\bm{\theta}}(\bm{z}|\bm{x})$ and may suffer from high variance. Gumbel-softmax reparameterizations~\cite{maddison2017concrete,jang2017categorical} use a continuous relaxation that does not follow the exact distribution. 

We now show that near-term quantum computers provide an alternative tool for VI. We use the quantum Born machine as a candidate for variational distributions. These models are highly expressive and they naturally represent discrete distributions as a result of quantum measurement. Furthermore, the Born machine can be trained by gradient-based optimization.

\section{Born machines as implicit variational distributions}
\label{s:born_as_implicit}

\begin{figure}
    \centering
    \includegraphics[]{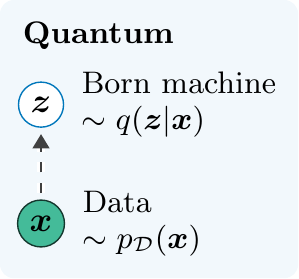}
    \includegraphics[]{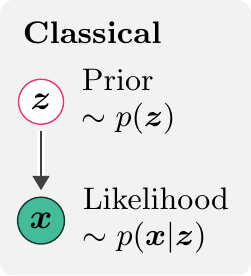}
    \caption{Probabilistic models used in our methods. The classical model comprises a prior over unobserved discrete variables and a likelihood over observed variables. The quantum model approximates the posterior distribution of the unobserved variables given observed data. All distributions can be sampled efficiently as long as one follows the arrows and uses a suitable computer.}
    \label{fig:models}
\end{figure}

By exploiting the inherent probabilistic nature of quantum mechanics, one can model the probability distribution of classical data using a pure quantum state. This model based on the Born rule in quantum mechanics is referred to as the Born machine~\cite{Cheng_2018}. Let us consider binary vectors $\bm{z} \in \{0,1\}^n$ where $n$ is the number of variables. The Born machine is a normalized quantum state $\ket{\psi(\bm{\theta})}$ parameterized by $\bm{\theta}$ that outputs $n$-bit strings with probabilities $q_{\bm{\theta}}(\bm{z}) = \abs{\braket{\bm{z}}{\psi(\bm{\theta})}}^2$. Here $\ket{\bm{z}}$ are computational basis states, thus sampling the above probability boils down to a simple measurement. Other forms of discrete variables can be dealt with by a suitable encoding. When using amortization, the variational distribution requires conditioning on observed variables. To extend the Born machine and include this feature, we let $\bm{x}$ play the role of additional parameters. This yields a pure state where the output probabilities are $q_{\bm{\theta}}(\bm{z}|\bm{x}) = \abs{\braket{\bm{z}}{\psi(\bm{\theta},\bm{x})}}^2$. Figure~\ref{fig:models} shows the relation between the classical model and the quantum model for approximate inference.

Born machines have been applied for benchmarking hybrid quantum-classical systems~\cite{Benedetti_2019,Hamilton_2019,leytonortega2019robust,Zhu_2019}, generative modeling~\cite{Han_2018,rudolph2020generation, cepaite_continuous_2020}, finance~\cite{alcazar_classical_2020,Zoufal_2019,coyle_quantum_2020}, anomaly detection~\cite{herr2020anomaly}, and have been proposed for demonstrating quantum advantage~\cite{Coyle_2020}\footnote{Some literature (e.g., Ref.~\cite{Zoufal_2019}) refer to Born machines as quantum generative adversarial networks (QGANs) when using an adversarial objective function. It is more appropriate to reserve the term QGAN to adversarial methods for quantum data \mbox{(e.g., Refs.~\cite{Dallaire_2018, benedetti2019adversarial, chakrabarti2019quantum}).}}. These models can be realized in a variety of ways, in both classical and quantum computers. When realized via certain classes of quantum circuits, they are classically intractable to simulate.\footnote{A quantum computer cannot efficiently implement arbitrary circuits, thus some Born machines are intractable even for a quantum computer.} For example, instantaneous quantum polytime (IQP) circuits are Born machines with $O(\text{poly}(n))$ parameters that yield classically intractable distributions in the average case under widely accepted complexity theoretic assumptions~\cite{Bremner_2016}.
Additional examples for such classically hard circuits are boson sampling~\cite{Aaronson2011} and random circuits~\cite{Boixo2018, Bouland2019}. Thus, quantum Born machines have expressive power larger than that of classical models, including neural networks~\cite{Du_2020} and partially matrix product states~\cite{glasserExpressivePowerTensornetwork2019}. It can be shown that the model remains classically intractable throughout training~\cite{Coyle_2020}. We return to this discussion in Appendix~\ref{a:quantum_advantage_inference}.

A useful way to classify probabilistic models is the following: \emph{prescribed} models provide an explicit parametric specification of the distribution, \emph{implicit} models define only the data generation process~\cite{mohamed2017learning}. Born machines can be effectively regarded as implicit models. It is easy to obtain an unbiased sample as long as we can execute the corresponding circuit on a quantum computer and measure the computational basis. However, it requires exponential resources to estimate the probability of a sample with multiplicative error.

Implicit models are challenging to train with standard methods. The major challenge is the design of the objective function precisely because likelihoods are ``prohibited''. Valid objectives involve only statistical quantities (such as expectation values) that can be efficiently estimated from samples~\footnote{Bias and variance of the estimates need to be analyzed and controlled, which is a challenge in itself.}. For generative modeling with Born machines, progress has been made towards practical objectives such as moment matching~\cite{Benedetti_2019}, maximum mean discrepancy~\cite{Liu_2018}, Stein and Sinkhorn divergences~\cite{Coyle_2020}, adversarial objectives~\cite{Situ_2020, Zeng_2019}, as well as incorporating Bayesian priors on the parameters~\cite{Du_2020}.

In the next section we make some progress towards practical objectives for VI with Born machines. First, we mention some related work. The approaches in Ref.~\cite{verdonQuantumHamiltonianBasedModels2019} use VI ideas to deal with thermal states and quantum data. In contrast, our inference methods apply to classical graphical models and classical data. The approaches of Refs.~\cite{low2014quantum, borujeni_quantum_2020,tucciQuantumBayesianNets1995} aim to encode Bayesian networks directly in a quantum state, and subsequently perform exact inference. In contrast, our inference methods are approximate, efficient, and apply also to graphical models that are not Bayesian networks.

\section{Operator variational inference}
\label{s:methods}

Operator variational inference (OPVI)~\cite{ranganath2018operator} is a rather general method that uses mathematical operators to design objectives for the approximate posterior. Suitable operators are those for which (i) the minima of the variational objective is attained at the true posterior, and (ii) it is possible to estimate the objective without computing the true posterior. In general, the amortized OPVI objective is
\eq{opvi}{
    \mathbb{E}_{\bm{x} \sim p_\mathcal{D}(\bm{x})} \sup_{f \in \mathcal{F}} h\left( \mathbb{E}_{\bm{z} \sim q(\bm{z}|\bm{x})} \Big[ (O^{p,q} f)(\bm{z}) \Big] \right) ,
}
where $f(\cdot) \in \mathbb{R}^d$ is a test function within the chosen family $\mathcal{F}$, $O^{p,q}$ is an operator that depends on $p(\bm{z}|\bm{x})$ and $q(\bm{z}|\bm{x})$, and $h(\cdot) \in [0, \infty]$ yields a non-negative objective. Note that we have an expectation over the data distribution $p_\mathcal{D}(\bm{x})$. This indicates the average over a dataset $\mathcal{D}= \{ \bm{x}^{(i)} \}_i$ of observations to condition on.

We present two methods that follow directly from two operator choices. These operator choices result in objectives based on the Kullback-Leibler (KL) divergence and the Stein discrepancy. The former utilizes an \emph{adversary} to make the computation tractable, whereas in the latter, the tractability arises from the use of a \emph{kernel} function.

The KL divergence is an example of an \emph{$f$ divergence}, while the Stein discrepancy is in the class of \emph{integral probability metrics}, two fundamentally different families of probability distance measures. OPVI can yield methods from these two different classes under a suitable choice of operator. As shown in Ref.~\cite{sriperumbudur2009integral}, these two families intersect nontrivially only at the total variation distance (TVD). It is for this reason that we choose the TVD as our benchmark in later numerical results. Integral probability metrics~\cite{Liu_2018, Coyle_2020} and $f$ divergences~\cite{Benedetti_2019, Zhu_2019} have both been used to train Born machines for the task of generative modeling, and we allow the more thorough incorporation of these methods in future work. Furthermore, these works demonstrate that the choice of training metric has a significant effect on the trainability of the model. It is for this reason that we adopt the OPVI framework in generality, rather than focusing on any specific instance of it. This allows the switching between different metrics depending on the need.

\subsection{The adversarial method}
\label{s:method1}

\begin{figure}
    \centering
    \includegraphics[]{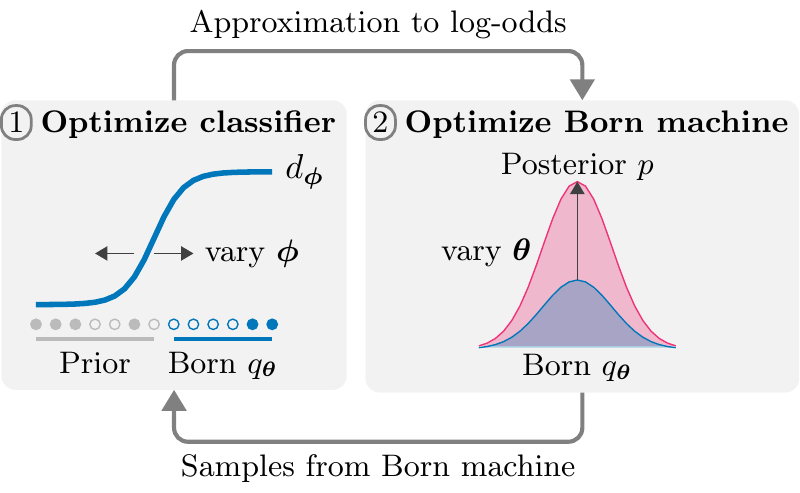}
    \caption{Adversarial variational inference with a Born machine. Step 1 optimizes a classifier $d_{\bm{\phi}}$ to output the probabilities that the observed samples come from the Born machine rather than from the prior. Step 2 optimizes the Born machine $q_{\bm{\theta}}$ to better match the true posterior. The updated Born machine is fed back into step 1 and the process repeats until convergence of the Born machine.}
    \label{fig:adversarial}
\end{figure}

One possible objective function for VI is the KL divergence of the true posterior relative to the approximate one. This is obtained from Eq.~\eqref{eq:opvi} by choosing $h$ to be the identity function, and by choosing the operator $(O^{p,q}f)(\bm{z}) = \log \frac{q(\bm{z}|\bm{x})}{p(\bm{z}|\bm{x})}$ for all $f$:
\eq{kl_vi}{
    \mathbb{E}_{\bm{x} \sim p_\mathcal{D}(\bm{x})} \underbrace{\mathbb{E}_{\bm{z} \sim q_{\bm{\theta}}(\bm{z}|\bm{x})} \left[ \log  \frac{ q_{\bm{\theta}}(\bm{z}|\bm{x}) }{ p(\bm{z}|\bm{x}) }  \right]}_{\text{Kullback-Leibler divergence}}.
}
Here $q_{\bm{\theta}}$ is the variational distribution parameterized \mbox{by $\bm{\theta}$}. To avoid singularities we assume that $p(\bm{z}|\bm{x}) > 0$ for all $\bm{x}$ and all $\bm{z}$. The objective's minimum is zero and is attained by the true posterior $q_{\bm{\theta}}(\bm{z}|\bm{x}) = p(\bm{z}|\bm{x})$ for all $\bm{x}$. 

There are many ways to rewrite this objective. Here we focus on the form known as \emph{prior contrastive}. Substituting Bayes' formula ${ p(\bm{z}|\bm{x}) = p(\bm{x}|\bm{z})p(\bm{z})/p(\bm{x}) }$ in the equation above we obtain
\eq{prior_contrastive}{
    \mathbb{E}_{\bm{x} \sim p_\mathcal{D}(\bm{x})} \mathbb{E}_{\bm{z} \sim q_{\bm{\theta}}(\bm{z}|\bm{x})} \left[ \log \frac{q_{\bm{\theta}}(\bm{z}|\bm{x})}{ p(\bm{z}) } - \log p(\bm{x}|\bm{z}) \right] + \text{const}. ,
}
where we ignore the term $\mathbb{E}_{\bm{x} \sim p_\mathcal{D}(\bm{x})} [\log p(\bm{x})]$ since it is constant with respect to $q$. This objective has been used in many generative models, including the celebrated variational autoencoder (VAE)~\cite{kingma2014autoencoding, rezende2014stochastic}.  While the original VAE relies on tractable variational posteriors, one could also use more powerful, implicit distributions as demonstrated in adversarial variational Bayes~\cite{mescheder2018adversarial} and in prior-contrastive adversarial VI~\cite{huszar2017variational}.

Let us use a Born machine to model the variational distribution $q_{\bm{\theta}}(\bm{z}|\bm{x}) = \abs{\braket{\bm{z}}{\psi(\bm{\theta},\bm{x})}}^2$. Since this model is implicit, the ratio $q_{\bm{\theta}}(\bm{z}|\bm{x}) / p(\bm{z})$ in Eq.~\eqref{eq:prior_contrastive} cannot be computed efficiently. Therefore, we introduce an adversarial method for approximately estimating the ratio above. The key insight here is that this ratio can be estimated from the output of a binary classifier~\cite{mohamed2017learning}. Suppose that we ascribe samples ${(\bm{z}, \bm{x}) \sim q_{\bm{\theta}}(\bm{z}|\bm{x})p_\mathcal{D}(\bm{x})}$ to the first class, and samples ${(\bm{z}, \bm{x}) \sim p(\bm{z})p_\mathcal{D}(\bm{x})}$ to the second class. A binary classifier $d_{\bm{\phi}}$ parameterized by $\bm{\phi}$ outputs the probability $d_{\bm{\phi}}(\bm{z},\bm{x})$ that the pair $(\bm{z}, \bm{x})$ belongs to one of two classes. Hence, ${1-d_{\bm{\phi}}(\bm{z},\bm{x})}$ indicates the probability that $(\bm{z}, \bm{x})$ belongs to the other class. There exist many possible choices of objective function for the classifier~\cite{mohamed2017learning}. In this work we consider the cross-entropy
\eq{objective_G}{
    \mathcal{G}_\text{KL}(\bm{\phi}; \bm{\theta}) &= \mathbb{E}_{\bm{x} \sim p_\mathcal{D}(\bm{x})} \mathbb{E}_{ \bm{z} \sim q_{\bm{\theta}}(\bm{z}|\bm{x}) } \big[ \log d_{\bm{\phi}}(\bm{z},\bm{x}) \big] \\
    &+ \mathbb{E}_{\bm{x} \sim p_\mathcal{D}(\bm{x})} \mathbb{E}_{ \bm{z} \sim p(\bm{z}) } \big[ \log (1- d_{\bm{\phi}}(\bm{z},\bm{x})) \big] .
}
The optimal classifier that maximizes this equation is~\cite{mescheder2018adversarial}
\eq{optimal_classifier}{
    d^*(\bm{z},\bm{x}) = \frac{q_{\bm{\theta}}(\bm{z}|\bm{x})} {q_{\bm{\theta}}(\bm{z}|\bm{x}) + p(\bm{z})} .
}
Since the probabilities in Eq.~\ref{eq:optimal_classifier} are unknown, the classifier must be trained on a dataset of samples. This does not pose a problem because samples from the Born machine $q_{\bm{\theta}}(\bm{z}|\bm{x})$ and prior $p(\bm{z})$ are easy to obtain by assumption. Once the classifier is trained, the logit transformation provides the log odds of a data point coming from the Born machine joint $q_{\bm\theta}(\bm z | \bm x)p_\mathcal{D}(\bm x)$ versus the prior joint $p(\bm z)p_\mathcal{D}(\bm x)$. The log odds are an approximation to the log ratio of the two distributions, i.e.,
\eq{logit}{
    \text{logit}( d_{\bm{\phi}}(\bm{z},\bm{x})) \equiv \log \frac{ d_{\bm{\phi}}(\bm{z},\bm{x}) }{ 1-d_{\bm{\phi}}(\bm{z},\bm{x}) } \approx \log  \frac{ q_{\bm{\theta}}(\bm{z}|\bm{x}) }{ p(\bm{z})} ,
}
which is exact if $d_{\bm{\phi}}$ is the optimal classifier in Eq.~\eqref{eq:optimal_classifier}. Now, we can avoid the computation of the problematic term in the KL divergence. Substituting this result into Eq.~\eqref{eq:prior_contrastive} and ignoring the constant term, the final objective for the Born machine is
\eq{objective_L}{
    \mathcal{L}_\text{KL}(\bm{\theta}; \bm{\phi}) = \mathbb{E}_{\bm{x} \sim p_\mathcal{D}(\bm{x})} \mathbb{E}_{\bm{z} \sim q_{\bm{\theta}}(\bm{z}|\bm{x})} \big[ \text{logit}(d_{\bm{\phi}}(\bm{z},\bm{x})) \\
    - \log p(\bm{x}|\bm{z}) \big].
}
The optimization can be performed in tandem as
\eq{minimax}{
    &\max_{\bm{\phi}} \mathcal{G}_\text{KL}(\bm{\phi}; \bm{\theta}) \\
    &\min_{\bm{\theta}} \mathcal{L}_\text{KL}(\bm{\theta}; \bm{\phi}) ,
}
using gradient ascent and descent, respectively. It can be shown~\cite{huszar2017variational,mescheder2018adversarial} that the gradient of $\log \tfrac{q_{\bm{\theta}}(\bm{z}|\bm{x})}{p(\bm{z})}$ with respect to $\bm{\theta}$ vanishes. Thus, under the assumption of an optimal classifier $d_{\bm{\phi}}$, the gradient of Eq.~\eqref{eq:objective_L}
is significantly simplified. This gradient is derived in Appendix~\ref{a:optimization}.

A more intuitive interpretation of the procedure just described is as follows. The log likelihood in Eq.~\eqref{eq:prior_contrastive} can be expanded as $\log p(\bm{x}|\bm{z}) = \log \tfrac{p(\bm{z}|\bm{x})}{p(\bm{z})} + \log p(\bm{x})$. Then Eq.~\eqref{eq:prior_contrastive} can be rewritten as $
\mathbb{E}_{\bm{x} \sim p_\mathcal{D}(\bm{x})} \mathbb{E}_{\bm{z} \sim q_{\bm{\theta}}(\bm{z}|\bm{x})} [ \log  \tfrac{q_{\bm{\theta}}(\bm{z}|\bm{x})} { p(\bm{z}) }  - \log  \tfrac{p(\bm{z}|\bm{x})}{ p(\bm{z}) }  ]
$, revealing the difference between two log odds. The first is given by the optimal classifier in Eq.~\eqref{eq:logit} for the approximate posterior and prior. The second is given by a hypothetical classifier for the true posterior and prior. The adversarial method is illustrated in Fig.~\ref{fig:adversarial}.

\subsection{The kernelized method}

\begin{figure}
    \centering
    \includegraphics[]{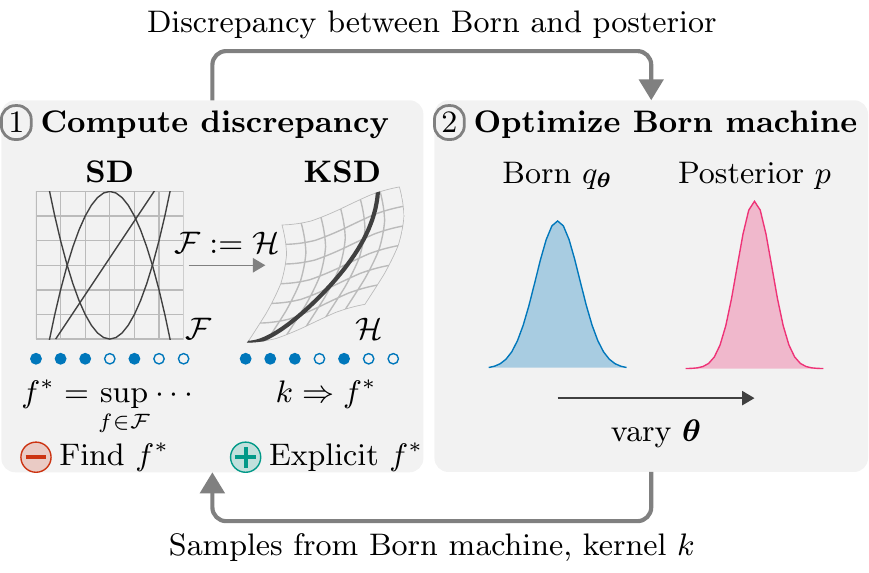}
    \caption{Kernelized Stein variational inference with a Born machine. Step 1 computes the Stein discrepancy (SD) between the Born machine and the true posterior using samples from the Born machine alone. By choosing a reproducing kernel Hilbert space $\mathcal{H}$ with kernel $k$, the kernelized Stein discrepancy (KSD) can be calculated in closed form. Step 2 optimizes the Born machine to reduce the discrepancy. The process repeats until convergence.}
    \label{fig:Stein}
\end{figure}

Another possible objective function for VI is the Stein discrepancy (SD) of the true posterior from the approximate one. This is obtained from Eq.~\eqref{eq:opvi}, assuming that the image of $f$ has the same dimension as $\bm{z}$, choosing $h$ to be the absolute value, and choosing $O^{p,q}$ to be a Stein operator. A Stein operator is independent from $q$ and is characterized by having zero expectation under the true posterior for all functions $f$ in the chosen family $\mathcal{F}$. 

For binary variables, a possible Stein operator is $(O^{p}f)(\bm{z}) = s_p(\bm{x},\bm{z})^T f(\bm{z}) - \tr(\Delta f(\bm{z}))$, where
\eq{stein_score}{
    (s_p(\bm{x},\bm{z}))_i = \frac{\Delta_{z_i} p(\bm{x}, \bm{z})}{p(\bm{x}, \bm{z})} = 1 - \frac{ p(\bm{x}, \neg_i \bm{z}) }{ p(\bm{x}, \bm{z}) } ,
}
is the difference score function. The partial difference operator is defined for any scalar function $g$ on binary vectors as
\eq{partial_difference_operator}{
    \Delta_{z_i} g(\bm{z}) = g(\bm{z}) - g(\neg_i  \bm{z}) ,
}
where $\neg_i \bm{z}$ flips the $i$th bit in binary vector $\bm{z}$. We have also defined $\Delta f(\bm{z})$ as the matrix with entries $(\Delta f(\bm{z}))_{ij} = \Delta_{z_i} f_j(\bm{z})$.
Under the assumption that $p(\bm{z}|\bm{x}) > 0$ for all $\bm{x}$ and all $\bm{z}$, we show in Appendix~\ref{a:stein_operator} that this is a valid Stein operator for binary variables. For more general discrete variables, we refer the interested reader to Ref.~\cite{yang2018goodness}. Substituting these definitions into Eq.~\eqref{eq:opvi} we obtain~\footnote{In contrast to Ref.~\cite{yang2018goodness} but in line with earlier work such as Ref.~\cite{gorhamMeasuringSampleQuality2018} we include the absolute value in the definition of the Stein discrepancy.}:
\eq{stein_vi}{
    \mathbb{E}_{\bm{x} \sim p_\mathcal{D}(\bm{x})} \underbrace{ \sup_{f \in \mathcal{F}} \left| \mathbb{E}_{\bm{z} \sim q_{\bm{\theta}}(\bm{z}|\bm{x})} \left[ s_p(\bm{x},\bm{z})^T f(\bm{z}) - \tr(\Delta f(\bm{z})) \right] \right| }_{\text{Stein discrepancy}}.
}
At this point one can parameterize the test function $f$ and obtain an adversarial objective similar in spirit to that presented in Sec.~\ref{s:method1}. Here, however, we take a different route. If we take $\mathcal{F}$ to be a reproducing kernel Hilbert space of vector-valued functions, $\mathcal{H}$, and restricting the Hilbert space norm of $f$ to be at most one ($||f||_{\mathcal{H}} \leq 1$), the supremum in Eq.~\eqref{eq:stein_vi} can be calculated in closed form using a kernel~\cite{yang2018goodness}. This is achieved using the ``reproducing'' properties of kernels, $f(\bm{z}) = \langle k(\bm{z}, \cdot), f(\cdot) \rangle_{\mathcal{H}}$, substituting into Eq.~\eqref{eq:stein_vi} and explicitly solving for the supremum (the explicit form of this supremal $f^*$ is not illuminating for our discussion; see~\cite{yang2018goodness} for details). Inserting this $f^*$ into the SD results in the kernelized Stein discrepancy (KSD),
\eq{kernelized_stein}{
    \mathbb{E}_{\bm{x} \sim p_\mathcal{D}(\bm{x})} \underbrace{ \sqrt{ \mathbb{E}_{\bm{z}, \bm{z}^\prime \sim q_{\bm{\theta}}(\bm{z}|\bm{x}) } [ \kappa_p (\bm{z}, \bm{z}^\prime | \bm{x}) ] } }_{\text{kernelized Stein discrepancy}},
}
where $\kappa_p$ is a so-called Stein kernel. For binary variables, this can be written as:
\begin{multline}
\label{eq:stein_kernel}
    \kappa_p(\bm{z}, \bm{z}^\prime | \bm{x}) = s_p(\bm{x},\bm{z})^T k(\bm{z},\bm{z}^\prime) s_p(\bm{x},\bm{z}^\prime) \\
    - s_p(\bm{x},\bm{z})^T \Delta_{\bm{z}^\prime} k(\bm{z},\bm{z}^\prime)
    - \Delta_{\bm{z}} k(\bm{z}, \bm{z}^\prime)^T s_p(\bm{x},\bm{z}^\prime) \\
    + \tr( \Delta_{\bm{z}, \bm{z}^\prime} k(\bm{z}, \bm{z}^\prime) ) .
\end{multline}
Here $\Delta_{\bm{z}} k(\bm{z},\cdot)$ is the vector of partial differences with components $\Delta_{z_i} k(\bm{z},\cdot)$, $\Delta_{\bm{z}, \bm{z}^\prime} k(\bm{z}, \bm{z}^\prime)$ is the matrix produced by applying $\Delta_{\bm{z}}$ to each component of the vector $\Delta_{\bm{z}^\prime} k(\cdot, \bm{z}^\prime)$, and $\tr(\cdot)$ is the usual trace operation on this matrix. The expression $\kappa_p(\bm{z}, \bm{z}^\prime | \bm{x})$ results in a scalar value.

Equation~\eqref{eq:kernelized_stein} shows that the discrepancy between the approximate and true posteriors can be calculated from samples of the approximate posterior alone. Note that the Stein kernel, $\kappa_p$, depends on \emph{another} kernel $k$. For $n$ unobserved Bernoulli variables, one possibility is the generic Hamming kernel ${ k(\bm{z}, \bm{z}^\prime) = \exp(-\tfrac{1}{n} \| \bm{z} - \bm{z}^\prime \|_1 ) }$. The KSD is a valid discrepancy measure if this ``internal'' kernel, $k$, is positive definite, which is the case for the Hamming kernel~\cite{yang2018goodness}.

In summary, constraining $\norm{f}_{\mathcal{H}} \leq 1$ in Eq.~\eqref{eq:stein_vi} and substituting the KSD we obtain
\eq{ksd_vi}{
    \mathcal{L}_\text{KSD}(\bm{\theta}) = \mathbb{E}_{\bm{x} \sim p_\mathcal{D}(\bm{x})} \sqrt{ \mathbb{E}_{\bm{z}, \bm{z}^\prime \sim q_{\bm{\theta}}(\bm{z}|\bm{x}) } [ \kappa_p (\bm{z}, \bm{z}^\prime | \bm{x}) ] } ,
}
and the problem consists of finding $\min_{\bm{\theta}} \mathcal{L}_\text{KSD}(\bm{\theta})$. The gradient of Eq.~\eqref{eq:ksd_vi} is derived in Appendix~\ref{a:optimization_stein}. The kernelized method is illustrated in Fig.~\ref{fig:Stein}.

The KSD was used in Ref.~\cite{Coyle_2020} for generative modeling with Born machines. In that context the distribution $p$ is unknown; thus, the authors derived methods to approximate the score $s_p$ from available data. VI is a suitable application of the KSD because in this context we do know the joint $p(\bm{z}, \bm{x})$. Moreover the score in Eq.~\eqref{eq:stein_score} can be computed efficiently even if we have the joint in unnormalized form.

\section{Experiments}
\label{s:experiments}

To demonstrate our approach, we employ three experiments as the subject of the following sections. First, we validate both methods using the canonical ``sprinkler'' Bayesian network. Second, we consider continuous observed variables in a hidden Markov model (HMM), and illustrate how multiple observations can be incorporated effectively via amortization. Finally, we consider a larger model, the ``lung cancer'' Bayesian network, and demonstrate our methods using an IBM quantum computer.

\begin{figure}
    \centering
    \includegraphics[]{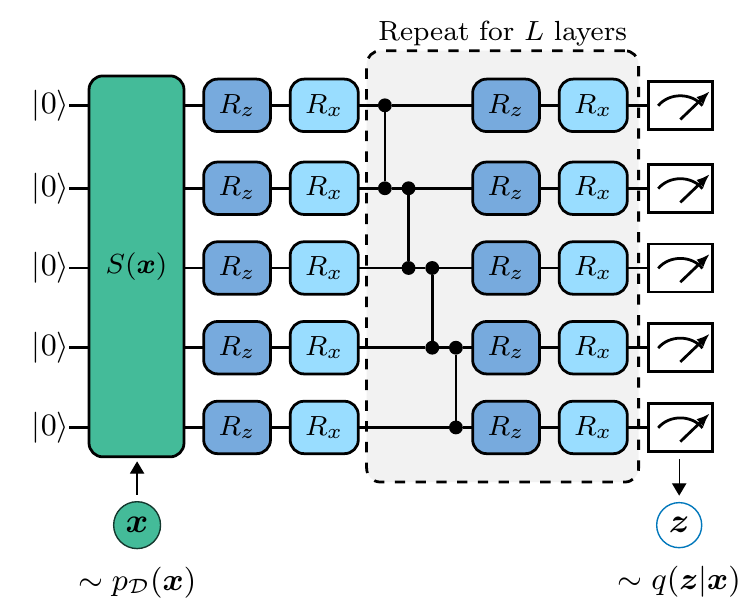}
    \caption{Example of hardware-efficient ansatz for the Born machine used in the experiments. All rotations $R_x$, $R_z$ are parameterized by individual angles. The layer denoted as $S(\bm{x})$ encodes the values of the observed variables $\bm{x}$. The particular choice of $S(\bm{x})$ depends on the application (see the main text). The corresponding classical random variables are indicated below the circuit.}
    \label{fig:ansatz}
\end{figure}

\subsection{Proof of principle with the ``sprinkler'' Bayesian network}
\label{s:experiment1_sprinkler}
  
As a first experiment, we classically simulate the methods on the ``sprinkler'' network in Fig.~\ref{fig:sprinkler}, one of the simplest possible examples. The purpose is to show that both methods are expected to work well even when using limited resources and without much fine tuning. First, we randomly generate the entries of each probability table from the uniform distribution $\mathcal{U}([0.01, 0.99])$ and produce a total of 30 instances of this network. For each instance, we condition on ``Grass wet'' being true, which means that our data distribution is $p_\mathcal{D}(W) = \delta(W = \true)$. We infer the posterior of the remaining three variables using the three-qubit version of the hardware-efficient ansatz shown in Fig.~\ref{fig:ansatz}. We use a layer of Hadamard gates as a state preparation, $S(\bm{x}) = H \otimes H \otimes H$, and initialize all parameters to approximately $0$. Such hardware-efficient \emph{ans\"{a}tze}, while simple, have been shown to be vulnerable to \emph{barren plateaus}~\cite{mcclean_barren_2018, cerezo_cost-function-dependent_2020, wang_noise-induced_2021, holmes_connecting_2021}, or regions of exponentially vanishing gradient magnitudes that make training untenable. Alternatively, one could consider \emph{ans\"{a}tze} that have been shown to be somewhat ``immune'' to this phenomenon~\cite{pesah_absence_2020, zhao_analyzing_2021}, but we leave such investigation to future work.

\begin{figure}
    \centering
    \begin{subfigure}[t]{0.49\textwidth}
        \includegraphics[width=.92\linewidth]{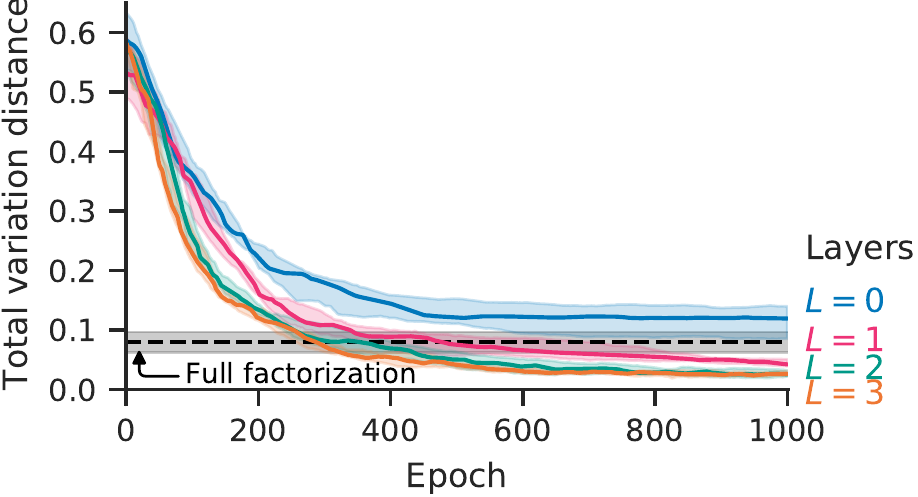}
        \caption{Adversarial objective}\label{fig:experiment1_KL}
    \end{subfigure}
    \begin{subfigure}[t]{0.49\textwidth}
        \centering
        \includegraphics[width=.92\linewidth]{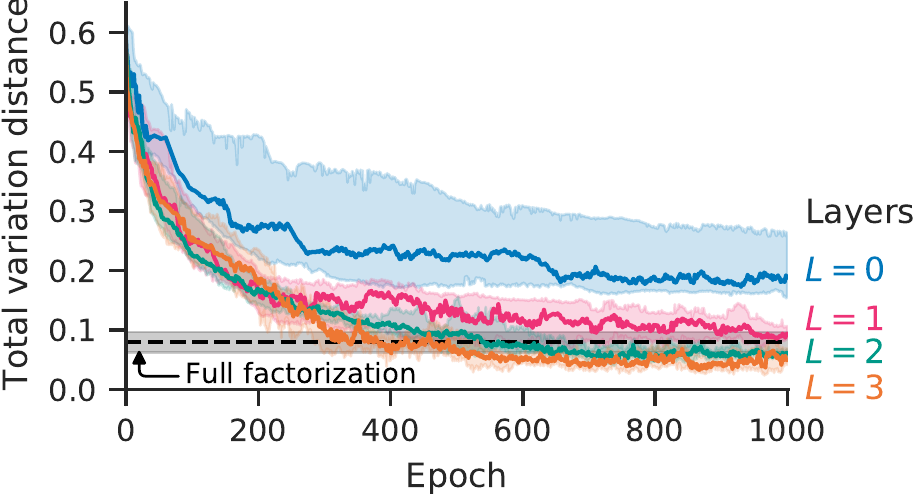}
        \caption{Kernelized Stein objective}\label{fig:experiment1_KSD}
    \end{subfigure}
    \caption{Total variation distance of the approximate and true posteriors. Median (lines) and $68\%$ confidence interval (shaded area) of $30$ instances for \subref{fig:experiment1_KL} KL and \subref{fig:experiment1_KSD} KSD objective functions for a different number of layers in the ansatz (cf. Fig.~\ref{fig:ansatz}). Each instance is a ``sprinkler'' Bayesian network as in Fig.~\ref{fig:sprinkler} where the conditional probabilities are chosen at random.}
    \label{fig:experiment1}
\end{figure}

For the KL objective, we utilize a multilayer perceptron (MLP) classifier made of three input units, six hidden rectified linear units, and one sigmoid output unit. The classifier is trained with a dataset of $100$ samples from the prior $p(C,R,S)$ and $100$ samples from the Born machine $q(C,R,S | W=\true)$. Here we use stochastic gradient descent with batches of size ten and a learning rate of $0.03$. For the KSD objective, we use the Hamming kernel as above. For both methods, the Born machine is trained using $100$ samples to estimate each expectation value for the gradients, and using vanilla gradient descent with a small learning rate of $0.003$. We compute the TVD of true and approximate posteriors at each epoch. We emphasize that TVD cannot be efficiently computed in general and can be shown only for small examples. Figure~\ref{fig:experiment1_KL} (Fig.~\ref{fig:experiment1_KSD}) shows the median TVD out of the $30$ instances for $1000$ epochs of training for the KL (KSD) objective. A shaded area indicates a $68\%$ confidence interval obtained by bootstrapping. Increasing the number of layers leads to better approximations to the posterior in all cases.

We compare the representation power of Born machines to that of fully factorized posteriors (e.g., those used in the na\"{i}ve mean-field approximation). We perform an exhaustive search for the best full factorization $q(\bm{z}) = \prod_{k=1}^3 q_k(z_k)$. This is computationally intractable, but can be done for our small examples. The horizontal gray lines in Figs.~\ref{fig:experiment1_KL} and ~\ref{fig:experiment1_KSD} show the median TVD across the $30$ instances for the full factorization. The $0$-layer Born machine generates unentangled states and could in principle match the horizontal line. This does not happen, which indicates the challenges of optimizing a fully factorized Born machine. On the other hand, Born machines with $L>0$ layers trained with the KL objective systematically outperform the full factorization. Born machines trained with the KSD objective yield suboptimal results in comparison to the KL objective.

Optimizing the KL objective is faster than the KSD objective. This is because training the MLP classifier in Eq.~\eqref{eq:objective_G} has cost linear in the dimension of the input, while calculating the Stein kernel in Eq.~\eqref{eq:stein_kernel} has quadratic cost. However, training the MLP classifier may still be nontrivial in some instances, as we show in Appendix~\ref{a:more_numerical_results}. Furthermore, at a large scale the performance of the classifier may suffer in quality, if kept efficient, due to the intractability of computing the KL divergence~\cite{kontonis_convergence_2020}. In contrast, the kernelized method is sample efficient at all scales. However, for the small problem sizes we use here, the adversarial method gives better results on average, so we choose it as the primary method in all subsequent experiments.

\subsection{Continuous observed variables and amortization with a hidden Markov model} 
\label{s:experiment2_HMM}

\begin{figure*}
    \centering
    \begin{subfigure}[t]{0.48\textwidth}
        \centering
        \includegraphics[]{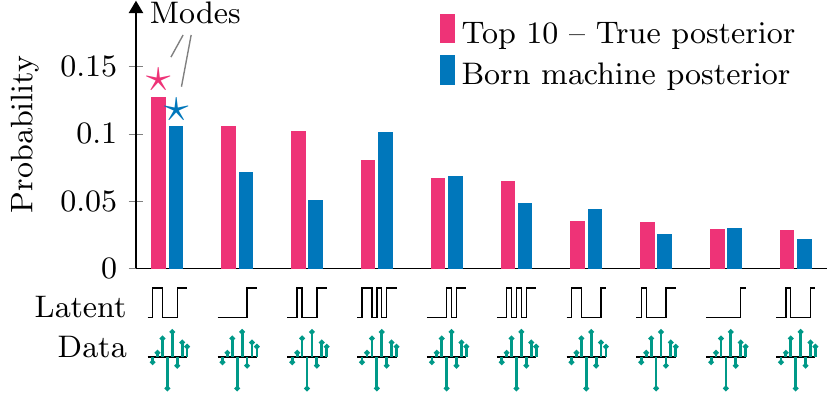}
    \caption{Posterior histogram given data $\bm{x}^{(1)}$}\label{fig:experiment2_histogram_s1}
    \end{subfigure}
    \hspace{0.02\textwidth}
    \begin{subfigure}[t]{0.48\textwidth}
        \centering
        \includegraphics[]{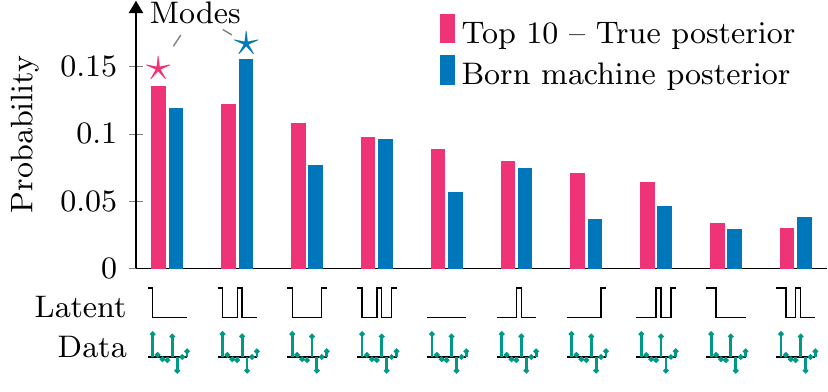}
        \caption{Posterior histogram given data $\bm{x}^{(2)}$}\label{fig:experiment2_histogram_s2}
    \end{subfigure}
    \vskip5mm
    \begin{subfigure}[t]{0.48\textwidth}
        \centering
        \includegraphics[]{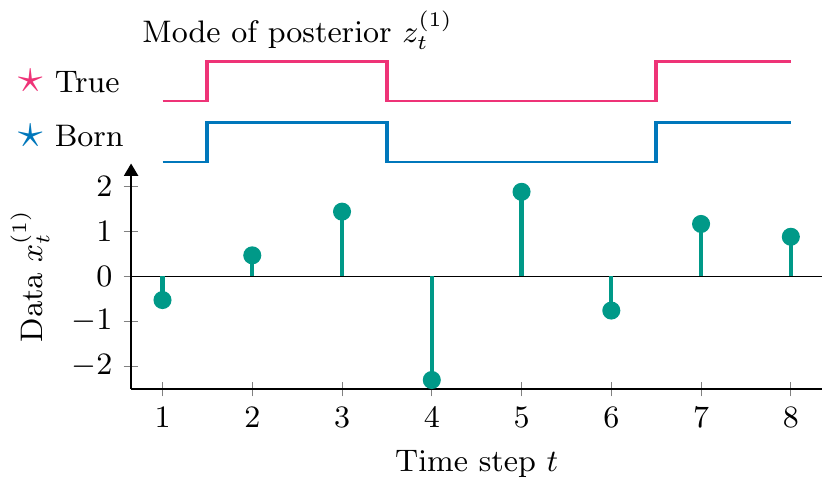}
    \caption{Time series and posterior modes for $\bm{x}^{(1)}$}\label{fig:experiment2_time_s1}
    \end{subfigure}
    \hspace{0.02\textwidth}
    \begin{subfigure}[t]{0.48\textwidth}
        \centering
        \includegraphics[]{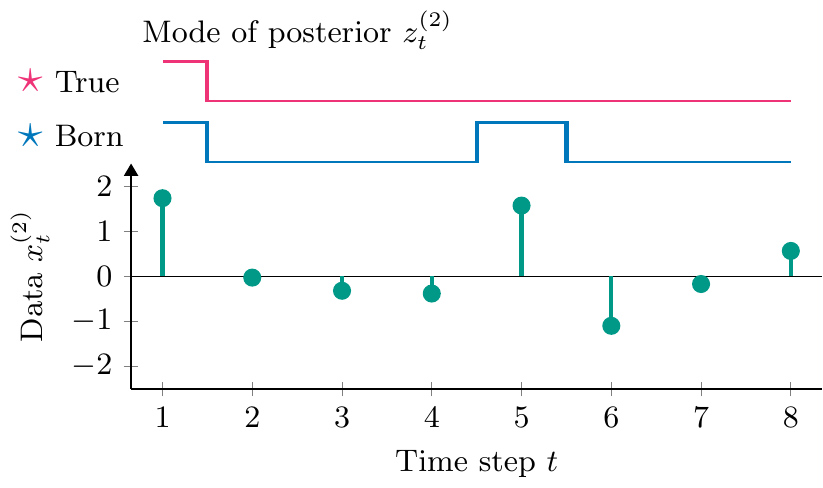}
        \caption{Time series and posterior modes for $\bm{x}^{(2)}$}\label{fig:experiment2_time_s2}
    \end{subfigure}
    \caption{Truncated, ordered histograms of the posteriors for two observed samples \subref{fig:experiment2_histogram_s1} $\bm{x}^{(1)}$ and \subref{fig:experiment2_histogram_s2} $\bm{x}^{(2)}$ of the hidden Markov model in Eqs.~\eqref{eq:hmm_z}--\eqref{eq:hmm_x}. The histograms are sorted by probability of the true posterior. The blue bars are the probabilities of the corresponding approximate posterior. The $x$ axis shows the latent state for each bar and the observed data point $\bm{x}$. The lower panels are the time series of the data \subref{fig:experiment2_time_s1} $\bm{x}^{(1)}$ and \subref{fig:experiment2_time_s2} $\bm{x}^{(2)}$, as well as the corresponding modes of the true posterior and Born machine posterior as indicated with stars in the upper panel.} 
    \label{fig:experiment2_histogram}
\end{figure*}

In our second experiment, we simulate the adversarial VI method on the HMM in Fig.~\ref{fig:hmm}. The purpose was to demonstrate two interesting features: continuous observed variables and amortization. We set up the HMM for $T=8$ time steps (white circles in Fig.~\ref{fig:hmm}), each represented by a Bernoulli latent variable with conditional dependency
\eq{hmm_z}{
    &z_1 \sim \mathcal{B}(\tfrac{1}{2}), \\ 
    &z_t \sim 
    \begin{cases}
        \mathcal{B}(\tfrac{1}{3}) & \text{if $z_{t-1} = 0$,} \\
        \mathcal{B}(\tfrac{2}{3}) & \text{if $z_{t-1} = 1$.}
    \end{cases}
}
These represent the unknown ``regime'' at time $t$. The regime affects how the observable data are generated. We use Gaussian observed variables (green circles in Fig.~\ref{fig:hmm}) whose mean and standard deviation depend on the latent variables as
\eq{hmm_x}{
    x_t \sim 
    \begin{cases}
        \mathcal{N}(0, 1) & \text{if $z_t = 0$} \\
        \mathcal{N}(1, \tfrac{1}{2}) & \text{if $z_t = 1$.}
    \end{cases} 
}
We sample two time series observations $\bm{x}^{(1)}, \bm{x}^{(2)}$ from the HMM and take this to be our data distribution. These time series are shown by green dots in Fig.~\ref{fig:experiment2_histogram}. Instead of fitting two approximate posteriors \emph{separately}, we use a single Born machine with amortization $\ket{\psi(\bm{\theta},\bm{x})}$. We used the ansatz in Fig.~\ref{fig:ansatz} for eight qubits with a state preparation layer $S(\bm{x}) = \otimes_{t=1}^8 R_x(x_t)$. In practice, this encoding step can be done under more careful considerations~\cite{larose2020robust, schuld_effect_2020, PerezSalinas2020datareuploading}. Parameters $\bm{\theta}$ are initialized to small values at random. We optimize the KL objective and a MLP classifier with $16$ inputs, 24 rectified linear units, and one sigmoid unit, and the system is trained for $3000$ epochs. The learning rates are set to be $0.006$ for the Born machine and $0.03$ for the MLP. The Born machine used $100$ samples to estimate each expectation value, the MLP used $100$ samples from each distribution and minibatches of size ten.

The histograms in Fig.~\ref{fig:experiment2_histogram} show the ten most probable configurations of latent variables for the true posterior along with the probabilities assigned by the Born machine. Conditioning on data point $\bm{x}^{(1)}$, the inferred most likely explanation is $\ket{01100011}$ (i.e., the mode of the Born machine). This coincides with the true posterior mode. For $\bm{x}^{(2)}$, the inferred mode is $\ket{10001000}$, which differs from the true posterior mode $\ket{10000000}$ by a single bit. The regime switching has therefore been modeled with high accuracy.

Rather than focusing on the mode, one can make use of the whole distribution to estimate some quantity of interest. This is done by taking samples from the Born machine and using them in a Monte Carlo estimate for such a quantity. For example, we could predict the expected value of the next latent variable $z_{T+1}$ given the available observations. For data point $\bm{x}^{(1)}$, this entails the estimation of
\eq{predict_nex_latent}{
    \mathbb{E}_{z_T \sim q_{\theta}( z_T | \bm{x}^{(1)} )} \mathbb{E}_{ z_{T+1} \sim p(z_{T+1} | z_{T} )} [ z_{T+1} ].
}
To conclude, we mention that inference in this HMM can be performed exactly and efficiently with classical algorithms. The most likely explanation for the observed data can be found with the Viterbi algorithm in time $O(T|S|^2)$, where $T$ is the length of the sequence and $|S|$ is the size of the unobserved variables (e.g., $|S|=2$ for Bernoulli variables). This is not the case for more general models. For example, a factorial HMM has multiple independent chains of unobserved variables~\cite{ghahramani1997factorial}. Exact inference costs $O(TM |S|^{M+1})$, where $M$ is number of chains, and is typically replaced by approximate inference~\cite{ghahramani1997factorial,zico2012approximate}. Our VI methods are generic and apply to factorial and other HMMs without changes.

\subsection{Demonstration on IBMQ with the \mbox{``lung cancer''} Bayesian network} \label{ssec:lung_cancer_ibmq}

As a final experiment, we test the performance on real quantum hardware using the slightly more complex ``lung cancer'' network~\cite{lauritzen_local_1988}, specifically using the five-qubit \texttt{ibmq\_rome} quantum processor. We access this device using \texttt{PyQuil}~\cite{smith2016practical} and \texttt{tket}~\cite{sivarajah_tvertketrangle_2020}.

The lung cancer network (also called the ``Asia'' network) is an example of a medical diagnosis Bayesian network, and is illustrated by Fig.~\ref{fig:lung_cancer}. This network was chosen since it is small enough to fit without adaptation onto the quantum device, but also comes closer to a ``real-world'' example. A more complicated version with extra variables can be found in Ref.~\cite{barberBayesianReasoningMachine2012}. We note that the version we use is slightly modified from that of Ref.~\cite{lauritzen_local_1988} for numerical reasons.

The network has eight variables: two ``symptoms'' for whether the patient presented with dyspnoea ($D$) (shortness of breath), or had a positive x ray ($X$); four possible ``diseases'' causing the symptoms bronchitis ($B$), tuberculosis ($T$), lung cancer ($L$), or an ``illness'' ($I$) (which could be either tuberculosis or lung cancer or something other than bronchitis); and two possible ``risk factors'' for whether the patient had traveled to Asia ($A$), or whether they had a history of smoking ($S$).

Based on the graph structure in Fig.~\ref{fig:lung_cancer}, the distribution over the variables, $p(A, T, S, L, I, X, B, D)$ can be factorized as
\eq{factorized_lung_probs}{
    p(A)p(T|A)p(X|I)p(I|T, L)p(D|B, I)p(B|S)p(L|S)p(S)
}
In Appendix~\ref{a:prob_table_lung_cancer_marginals} we show the explicit probability table we use (Table~\ref{tab:prob_table_lung_cancer_marginals}) for completeness. Modifying an illustrative example of a potential ``real-world'' use case in Ref.~\cite{lauritzen_local_1988}, a patient may present in a clinic with an ``illness'' ($I = \true$) but no shortness of breath ($D=\fal$) as a symptom. Furthermore, an x ray has revealed a negative result ($X = \fal$). However, we have no patient history and we do not know which underlying disease is actually present. As such, in the experiment, we condition on having observed the ``evidence'' variables, $\bm{x}$: $X, D$ and $I$. The remaining five are the latent variables, $\bm{z}$, so we will require five qubits.

\begin{figure*}[ht]
    \centering
    \includegraphics[]{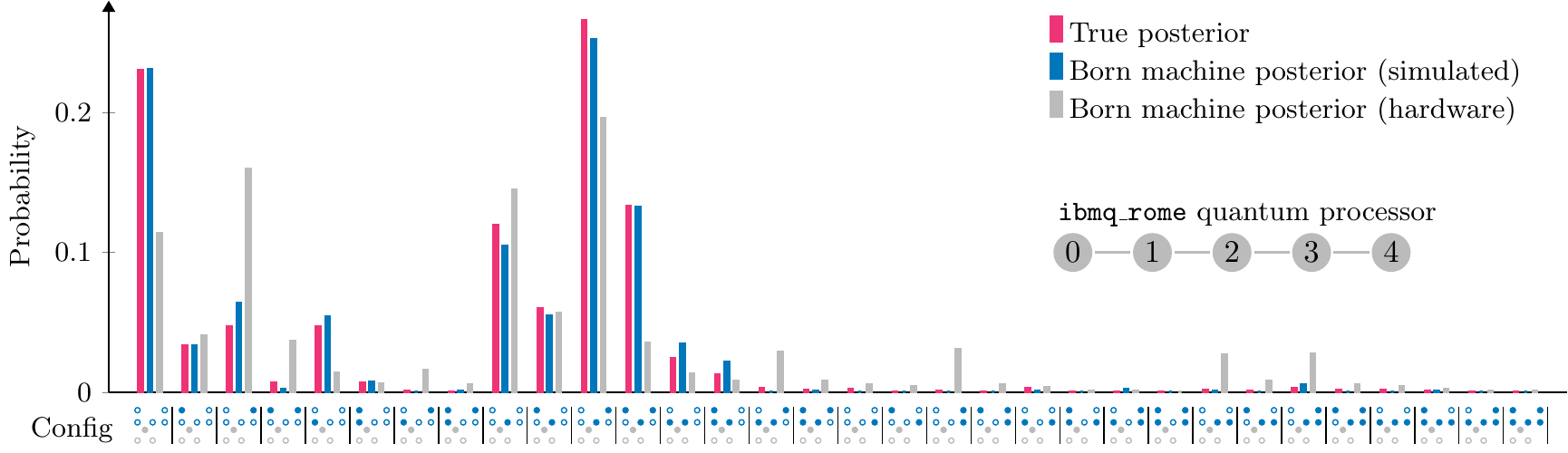}
    \caption{Histograms of true versus the learned posteriors with a simulated and hardware trained Born machine for the ``lung cancer'' network in Fig.~\ref{fig:lung_cancer}. Here we use the \texttt{ibmq\_rome} quantum processor (whose connectivity is shown in the inset). We condition on $X=\fal, D=\fal, I=\true$. The $x$ axis shows the configuration of observed (gray represents $X$, $D$, $I$) and unobserved variables (blue represents $A$, $S$, $T$, $L$, $B$) corresponding to each probability (filled circles $=\true$, empty circles $=\fal$). For hardware results, the histogram is generated using $1024$ samples.}\label{fig:experiment3_histogram_s1}
\end{figure*}

In Fig.~\ref{fig:experiment3_histogram_s1}, we show the results when using the five-qubit \texttt{ibmq\_rome} quantum processor, which has a linear topology. The topology is convenient since it introduces no major overheads when compiling from our ansatz in Fig.~\ref{fig:ansatz}. We plot the true posterior versus the one learned by the Born machine both simulated, and on the quantum processor using the best parameters found (after about $400$ epochs simulated and about $50$ epochs on the processor). To train in both cases, we use the same classifier as in Sec.~\ref{s:experiment2_HMM} with five inputs and ten hidden rectified linear units using the KL objective, which we observed to give the best results. We employ a two-layer ansatz in Fig.~\ref{fig:ansatz} ($L=2$), and $1024$ shots are taken from the Born machine in all cases. We observe that the simulated Born machine is able to learn the true posterior very well with these parameters, but the performance suffers on the hardware. That being said, the trained hardware model is still able to pick three out of four highest probability configurations of the network (shown on the x axis of Fig.~\ref{fig:experiment3_histogram_s1}). The hardware results could likely be improved with error mitigation techniques.

\section{Discussion}
\label{s:discussion}

We present two variational inference methods that can exploit highly expressive approximate posteriors given by quantum models. The first method is based on minimizing the Kullback-Leibler divergence to the true posterior and relies on a classifier that estimates probability ratios. The resulting adversarial training may be challenging due to the large number of hyperparameters and well-known stability issues~\cite{arjovsky2017principled,lucic2018gans}. On the other hand, it only requires the ability to (i) sample from the prior $p(\bm{z})$, (ii) sample from the Born machine $q_\theta(\bm{z}|\bm{x})$, and (iii) calculate the likelihood $p(\bm{x}|\bm{z})$. We can apply the method as is when the prior is implicit. If the likelihood is also implicit, we can reformulate Eq.~\eqref{eq:prior_contrastive} in \emph{joint contrastive} form~\cite{huszar2017variational} and apply the method with minor changes. This opens the possibility to train a model where both the generative and inference processes are described by Born machines (i.e., replacing the classical model in Fig.~\ref{fig:models} with a quantum one). This model can be thought of as an alternative way to implement the quantum-assisted Helmholtz machine~\cite{Benedetti_2018} and the quantum variational autoencoder~\cite{Khoshaman_2018}.

We also present a second VI method based on the kernelized Stein discrepancy. A limitation of this approach is that it requires explicit priors and likelihoods. On the other hand, it provides plenty of flexibility in the choice of kernel. We use a generic Hamming kernel to compute similarities between bit strings. For VI on graphical models, a graph-based kernel~\cite{kondor2002diffusion} that takes into account the structure could yield improved results. Moreover, one could attempt a demonstration of quantum advantage in VI using classically intractable kernels arising from quantum circuits~\cite{havlicek_supervised_2019, schuld2019quantum, huang_power_2020}.

Another interesting extension could be approximating the posterior with a Born machine with traced out additional qubits, i.e., locally purified states. This may allow trading a larger number of qubits for reduced circuit depth at constant expressivity~\cite{glasserExpressivePowerTensornetwork2019}.

We present a few examples with direct application to financial, medical, and other domains. Another area of potential application of our methods is natural language processing. Quantum circuits have been proposed as a way to encode and compose words to form meanings~\cite{coecke2020foundations,meichanetzidis2020grammaraware,lorenz2021qnlp}. Question answering with these models could be phrased as variational inference on quantum computers where our methods fit naturally.

\section*{Acknowledgments}

This research was funded by Cambridge Quantum Computing. We gratefully acknowledge the cloud computing resources received from the ``Microsoft for Startups'' program. We acknowledge the use of IBM Quantum Services for this work. We thank Bert Kappen, Konstantinos Meichanetzidis, Robin Lorenz, and David Amaro for helpful discussions.

\appendix

\section{Quantum advantage in inference}
\label{a:quantum_advantage_inference}

The intuitive reason for why approximate inference may be a suitable candidate for a quantum advantage is similar to that which has already been studied for the related problem of generative modeling using quantum computers~\cite{gao_quantum_2018, Du_2020, Coyle_2020, sweke_quantum_2020, gao_enhancing_2021}. Here, the primary argument for quantum advantage reduces to complexity theoretic arguments for the hardness of sampling from particular probability distributions by any efficient classical means. At one end of the spectrum, the result of Ref.~\cite{sweke_quantum_2020} is based on the classical hardness of the discrete logarithm problem and is not suitable for near-term quantum computers. On the other end, the authors of Refs.~\cite{Du_2020, Coyle_2020} leveraged arguments from ``quantum computational supremacy'' to show how Born machines are more expressive than their classical counterparts. Specifically, assuming the noncollapse of the polynomial hierarchy, the Born machine is able to represent IQP time computations~\cite{Bremner_2016}.

A similar argument can be made for approximate inference. The key distinction in this case over the previous arguments for generative modeling is that now we must construct a conditional distribution, the posterior $p(\bm{z}|\bm{x})$, which at least requires a Born machine to \emph{represent} it. For an IQP circuit, probability amplitudes are given by partition functions of complex Ising models~\cite{De_las_Cuevas_2011, Fujii_2017}. Thus, for a demonstration of quantum advantage in inference, we may seek posterior distributions of the form $p(\bm{z} | \bm{x}) \propto | \tr( e^{-i H(\bm{z}, \bm{x})}) |^2$. Here the Ising Hamiltonian $H(\bm{z}, \bm{x})$ is diagonal and its coefficients depend on both $\bm{z}$ and $\bm{x}$. Alternative ``quantum computational supremacy'' posteriors could arise from boson sampling~\cite{Aaronson2011} and random circuits~\cite{Boixo2018, Bouland2019}. In the latter case, experimental validations have been performed~\cite{arute_quantum_2019, wu_strong_2021} demonstrating the \emph{physical} existence of candidate distributions. The former experiment~\cite{arute_quantum_2019} used $53$ qubits in the \emph{Sycamore} quantum chip with $20$ ``cycles'', where a cycle is a layer of random single-qubit gates interleaved with alternating two-qubit gates. Notably, a cycle is not conceptually very different from a layer in our hardware-efficient ansatz in Fig.~\ref{fig:ansatz} in the main text. The average (isolated) single- (two-)qubit gate error is estimated to be $0.13\%$ ($0.36\%$) in this experiment, which is sufficient to generate a cross-entropy (a measure of distribution ``quality'' related to the KL divergence) benchmarking fidelity of $0.1\%$. The latter~\cite{wu_strong_2021} repeated this experiment by using $56$ qubits from the \emph{Zuchongzhi} processor, again with $20$ cycles. In this case, the single- (two-)qubit gates carried an average error of $0.14\%$ ($0.59\%$) to achieve a fidelity of approximately $0.07\%$. Clearly, quantum processors exist that can express classically intractable distributions, and so are also likely capable of representing intractable posterior distributions for suitably constructed examples. In all cases, the posterior shall be constructed nontrivially from the prior $p(\bm{z})$ and likelihood $p(\bm{x} | \bm{z})$ via Bayes' theorem. We leave this construction for future work. As a final comment, it is also important to note here that we are discussing only \emph{representational power} and not \emph{learnability}, which is a separate issue. It is still unknown whether a distribution family (\emph{a posterior} or otherwise) exists that is learnable by (near-term) quantum means, but not classically. See Refs.~\cite{Coyle_2020, sweke_quantum_2020} for discussions of this point.

\section{Gradients for the adversarial method}
\label{a:optimization}

In the adversarial VI method we optimize the objective functions given in Eqs.~\eqref{eq:objective_G} and \eqref{eq:objective_L} via gradient ascent and descent, respectively. Assuming that the optimal discriminator $d_{\bm{\phi}}=d^*$ has been found, the objective in Eq.~\eqref{eq:objective_L} becomes
\eq{kl_obj_opt_d}{
    \mathbb{E}_{\bm{x} \sim p_\mathcal{D}(\bm{x})} \mathbb{E}_{\bm{z} \sim q_{\bm{\theta}}(\bm{z}|\bm{x})}\big[ \text{logit}( d^*(\bm{x},\bm{z}) ) - \log p(\bm{x}|\bm{z}) \big] .
}
While $d^*$ implicitly depends on $\bm{\theta}$, the gradient  of its expectation vanishes. To see this, recall that $\text{logit}(d^*(\bm{z},\bm{x})) = \log q_{\bm{\theta}}(\bm{z}|\bm{x}) - \log p(\bm{z})$. Fixing $\bm{x}$ and taking the partial derivative for the $j$th parameter,
\eq{kl_grad_1}{
    \partial_{\theta_j} \sum_{\bm{z}} q_{\bm{\theta}}(\bm{z}|\bm{x}) \log q_{\bm{\theta}}(\bm{z}|\bm{x}) &=  
    \sum_{\bm{z}} \partial_{\theta_j} \Big( q_{\bm{\theta}}(\bm{z}|\bm{x}) \Big) \log ( q_{\bm{\theta}}(\bm{z}|\bm{x}) ) \\
    &+  \sum_{\bm{z}} q_{\bm{\theta}}(\bm{z}|\bm{x}) \partial_{\theta_j} \Big( \log q_{\bm{\theta}}(\bm{z}|\bm{x}) \Big).
}
The second term is zero since
\eq{kl_grad_2}{
    \sum_{\bm{z}} q_{\bm{\theta}}(\bm{z}|\bm{x}) \partial_{\theta_j} \Big( \log  q_{\bm{\theta}}(\bm{z}|\bm{x}) \Big) &= \sum_{\bm{z}} \partial_{\theta_j}  q_{\bm{\theta}}(\bm{z}|\bm{x}) \\  
    &= \partial_{\theta_j}  \sum_{\bm{z}} q_{\bm{\theta}}(\bm{z}|\bm{x}) \\
    &= \partial_{\theta_j} 1 = 0 .
    }
The approximate posterior in Eq.~\eqref{eq:kl_grad_1} is given by the expectations of observables $O_{\bm{z}}=\dyad{\bm{z}}{\bm{z}}$, i.e., $q_{\bm{\theta}}(\bm{z}|\bm{x}) = \langle O_{\bm{z}} \rangle_{\bm{\theta},\bm{x}}$. For a circuit parameterized by gates $U(\theta_j) = \exp(-i \tfrac{\theta_j}{2} H_j)$, where the Hermitian generator $H_j$ has eigenvalues $\pm 1$ (e.g., single-qubit rotations), the derivative of the first term in Eq.~\eqref{eq:kl_grad_1} can be evaluated with the parameter shift rule~\cite{mitaraiQuantumCircuitLearning2018,schuldEvaluatingAnalyticGradients2019, Banchi2021measuringanalytic}:
\eq{param_shift_rule}{
    \partial_{\theta_j} q_{\bm{\theta}}(\bm{z}|\bm{x}) = \frac{q_{\bm{\theta}_j^+}(\bm{z}|\bm{x}) - q_{\bm{\theta}_j^-}(\bm{z}|\bm{x})}{2},
}
with $\bm{\theta}_j^\pm = \bm{\theta} \pm \frac{\pi}{2} \bm{e}_j$ and $\bm{e}_j$ a unit vector in the $j$th direction. In summary, the partial derivative of the objective with respect to $\theta_j$ given the optimal discriminator is
\eq{kl_grad_3}{
    \frac{1}{2} \mathbb{E}_{\bm{x} \sim p_\mathcal{D}(\bm{x})} \Big[ &\mathbb{E}_{\bm{z} \sim q_{\bm{\theta}_j^+}(\bm{z}|\bm{x})}\big[ \text{logit}( d^*(\bm{x},\bm{z}) ) -
    \log p(\bm{x}|\bm{z}) \big] \\
    &- \mathbb{E}_{\bm{z} \sim q_{\bm{\theta}_j^-}(\bm{z}|\bm{x})}\big[ \text{logit}( d^*(\bm{x},\bm{z}) ) - \log p(\bm{x}|\bm{z}) \big] \Big].
}

\section{The Stein operator}
\label{a:stein_operator}

Here we prove that $(O^{p}f)(\bm{z}) = s_p(\bm{x},\bm{z})^T f(\bm{z}) - \tr(\Delta f(\bm{z}))$ is a Stein operator for any function $f\colon \{0,1\}^n \rightarrow \mathbb{R}^n$ and probability mass function $p(\bm{z}|\bm{x}) > 0$ on $\{0,1\}^n$. To do so, we need to show that its expectation under the true posterior vanishes. Taking the expectation with respect to the true posterior we have
\eq{so1}{
    \mathbb{E}_{\bm{z} \sim p(\bm{z}|\bm{x})} [ (O^{p}f)(\bm{z}) ] &= \sum_{\bm{z} \in \{0,1\}^n} p(\bm{z}|\bm{x}) \frac{\Delta p(\bm{x}, \bm{z})^T}{p(\bm{x}, \bm{z})} f(\bm{z}) \\
    &- \sum_{\bm{z} \in \{0,1\}^n} p(\bm{z}|\bm{x}) \tr(\Delta f(\bm{z}) )
}
The first summation is
\eq{so2}{
    &\sum_{\bm{z} \in \{0,1\}^n} \frac{1}{p(\bm{x})} \Delta p(\bm{x}, \bm{z})^T  f(\bm{z}) = \\
    &\qquad \sum_{\bm{z} \in \{0,1\}^n} \sum_{i=1}^n (p(\bm{z}|\bm{x}) - p(\neg_i \bm{z}|\bm{x})) f_i(\bm{z}),
}
while the second summation is
\eq{so3}{
    &\sum_{\bm{z} \in \{0,1\}^n} p(\bm{z}|\bm{x}) \tr(\Delta f(\bm{z})) = \\
    &\qquad \sum_{\bm{z} \in \{0,1\}^n} \sum_{i=1}^n p(\bm{z}|\bm{x}) ( f_i(\bm{z}) - f_i(\neg_i \bm{z}) ).
}
Subtracting \eqref{eq:so3} from \eqref{eq:so2} we have
\eq{so4}{
    \sum_{i=1}^n & \Big[ \sum_{\bm{z} \in \{0,1\}^n} p(\bm{z}|\bm{x}) f_i(\bm{z}) - \sum_{\bm{z} \in \{0,1\}^n} p(\neg_i \bm{z}|\bm{x}) f_i(\bm{z}) \\
    &- \sum_{\bm{z} \in \{0,1\}^n} p(\bm{z}|\bm{x}) f_i(\bm{z}) + \sum_{\bm{z} \in \{0,1\}^n} p(\bm{z}|\bm{x}) f_i(\neg_i \bm{z}) \Big] = 0 .
}
The first and third terms cancel. The second and fourth terms are identical and also cancel. This can be seen by substituting $\bm{y}_i = \neg_i \bm{z}$ into one of the summations and noting that $\bm{z} = \neg_i \bm{y}_i$. Thus, the operator $O^{p}$ satisfies Stein's identity $\mathbb{E}_{\bm{z} \sim p(\bm{z}|\bm{x})} [ (O^{p}f)(\bm{z}) ] = 0$ for any function $f$ as defined above, and is a valid Stein operator.

\section{Gradients for the kernelized method}
\label{a:optimization_stein}

We wish to optimize the kernelized Stein objective, Eq.~\eqref{eq:ksd_vi}, via gradient descent. This gradient was derived in Ref.\cite{Coyle_2020} and is given by
\eq{ksd_vi_grad_derivation}{
    &\partial_{\theta_j} \mathcal{L}_\text{KSD}(\bm{\theta}) = \partial_{\theta_j} \mathbb{E}_{\bm{x} \sim p_\mathcal{D}(\bm{x})}\sqrt{ \mathbb{E}_{\bm{z}, \bm{z}^\prime \sim q_{\bm{\theta}}(\bm{z}|\bm{x}) } [ \kappa_p (\bm{z}, \bm{z}^\prime | \bm{x}) ]}  \\
    & = \mathbb{E}_{\bm{x} \sim p_\mathcal{D}(\bm{x})} \partial_{\theta_j} \sqrt{ \mathbb{E}_{\bm{z}, \bm{z}^\prime \sim q_{\bm{\theta}}(\bm{z}|\bm{x}) } [ \kappa_p (\bm{z}, \bm{z}^\prime | \bm{x}) ]}  \\
    & = \frac{1}{2}\mathbb{E}_{\bm{x}\sim p_\mathcal{D}(\bm{x})} \Big[ \frac{1}{ \sqrt{ \mathbb{E}_{\bm{z}, \bm{z}^\prime \sim q_{\bm{\theta}}(\bm{z}|\bm{x})} \left[ \kappa_p (\bm{z}, \bm{z}^\prime | \bm{x})\right] }} \Big\{ \\
    & \qquad \qquad \qquad \qquad \partial_{\theta_j} \mathbb{E}_{\bm{z}, \bm{z}^\prime \sim q_{\bm{\theta}}(\bm{z}|\bm{x})} [ \kappa_p (\bm{z}, \bm{z}^\prime | \bm{x})] \Big\} \Big] \\
    &= \frac{1}{4} \mathbb{E}_{\bm{x} \sim p_\mathcal{D}(\bm{x})} \Big[ \frac{1}{ \sqrt{ \mathbb{E}_{\bm{z}, \bm{z}^\prime \sim q_{\bm{\theta}}(\bm{z}|\bm{x})} \left[ \kappa_p (\bm{z}, \bm{z}^\prime | \bm{x})\right] }} \Big\{ \\
    & \qquad \qquad \qquad \underset{{\substack{\bm{z}  \sim q_{\bm{\theta}}\\ \bm{z}^\prime \sim q_{\bm{\theta}^+_j}}}}{\mathbb{E}} [ \kappa_p (\bm{z}, \bm{z}^\prime | \bm{x})] - \underset{{\substack{\bm{z}  \sim q_{\bm{\theta}}\\ \bm{z}^\prime \sim q_{\bm{\theta}^-_j}}}}{\mathbb{E}}[ \kappa_p (\bm{z}, \bm{z}^\prime | \bm{x})] + \\
    &  \qquad \qquad \qquad \underset{{\substack{\bm{z}  \sim q_{\bm{\theta}^+_j}\\ \bm{z}^\prime \sim q_{\bm{\theta}}}}}{\mathbb{E}}[ \kappa_p (\bm{z}, \bm{z}^\prime | \bm{x})] -\underset{{\substack{\bm{z}  \sim q_{\bm{\theta}^-_j}\\ \bm{z}^\prime \sim q_{\bm{\theta}}}}}{\mathbb{E}}[ \kappa_p (\bm{z}, \bm{z}^\prime | \bm{x}) ]\Big\}\Big]
}
The final line (suppressing the dependence on the variables, $\bm{z}, \bm{x}$: $q_{\bm{\theta}} := q_{\bm{\theta}} (\bm{z}|\bm{x})$) in Eq.~\eqref{eq:ksd_vi_grad_derivation} follows from the parameter shift rule, Eq.~\eqref{eq:param_shift_rule}, for the gradient of the output probabilities of the Born machine circuit, under the same assumptions as in Appendix~\ref{a:optimization}. The relatively simple form of the gradient is due to the fact that the Stein kernel, $\kappa_p$, does not depend on the variational distribution, $q_{\bm{\theta}}(\bm{z}|\bm{x})$. This gradient can be estimated in a similar way to the KSD itself, by sampling from the original Born machine, $q_{\bm{\theta}}(\bm{z}|\bm{x})$, along with its parameter shifted versions, $q_{\bm{\theta}^{\pm}_j}(\bm{z}|\bm{x})$ for each parameter $j$.

\begin{figure*}[ht]
    \centering
    \begin{subfigure}[t]{0.48\textwidth}
        \centering
        \includegraphics[]{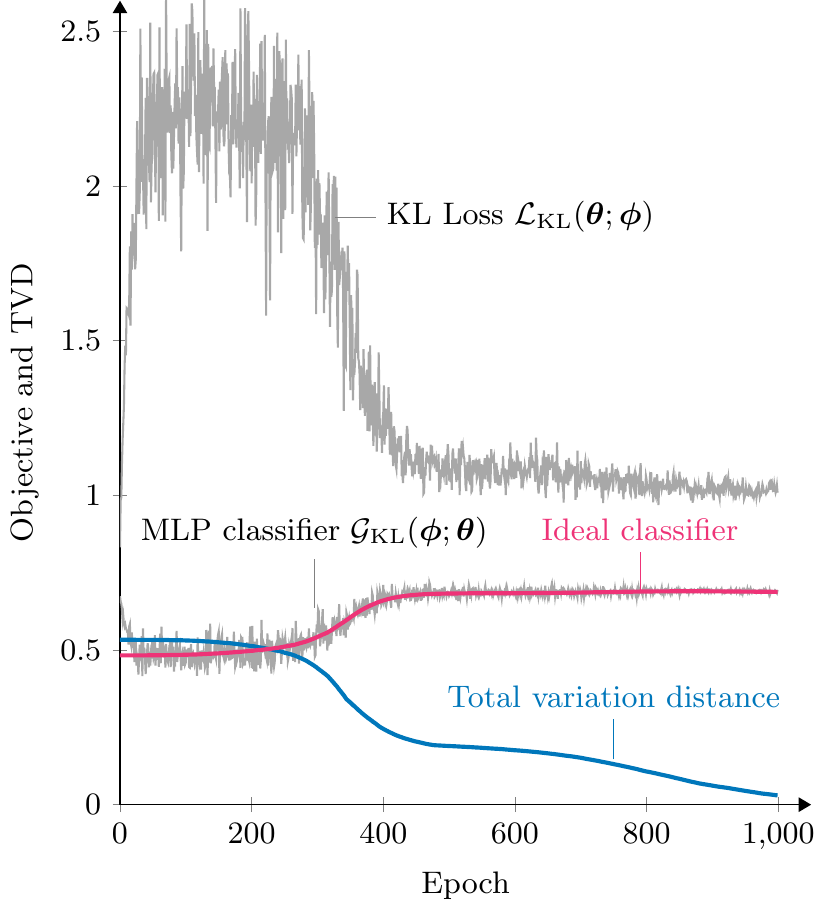}
    \caption{Successful example}\label{fig:good_example}
    \end{subfigure}
    \begin{subfigure}[t]{0.48\textwidth}
        \centering
        \includegraphics[]{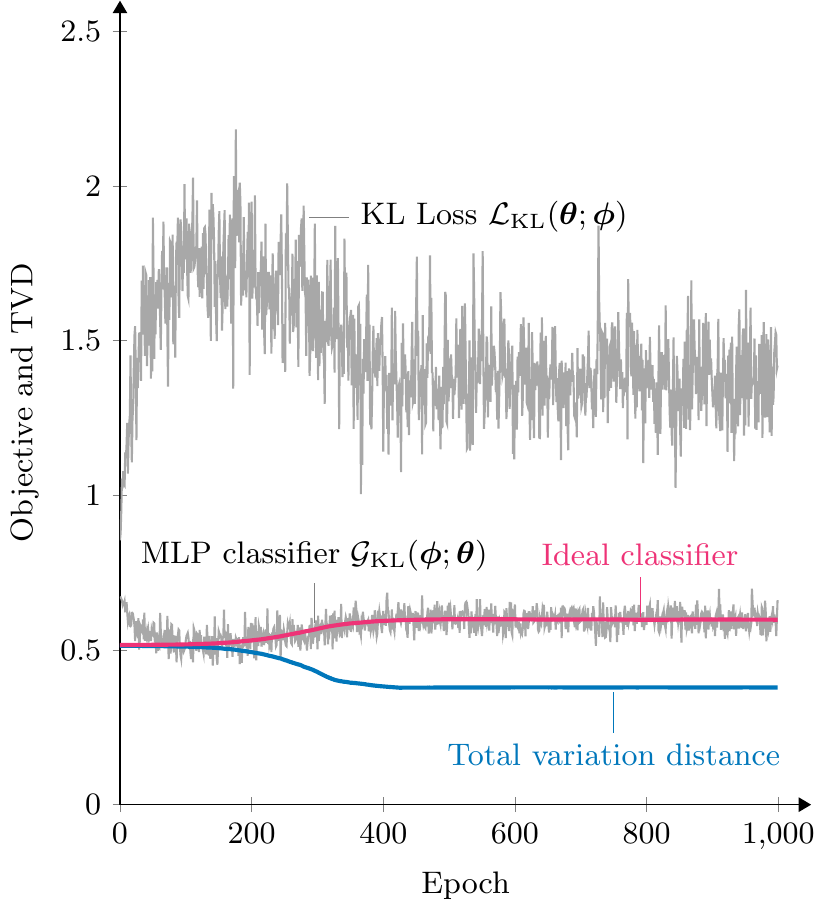}
        \caption{Unsuccessful example}\label{fig:bad_example}
    \end{subfigure}
    \caption{Examples of successful \subref{fig:good_example} and unsuccessful \subref{fig:bad_example} VI on the ``sprinkler'' Bayesian network in Sec.~\ref{s:experiment1_sprinkler} using the KL objective with the adversarial method.}
    \label{fig:experiment1_details}
\end{figure*}

\section{Learning curves for the adversarial method}
\label{a:more_numerical_results}

In this section we provide examples of successful and unsuccessful VI using the KL objective and adversarial methods. We inspect the random instances used in the ``sprinkler'' network experiment in Sec.~\ref{s:experiment1_sprinkler}. We cherry-pick two out of the 30 instances where a one-layer Born machine is employed. Figure \ref{fig:good_example} shows a successful experiment. Here the MLP classifier closely tracks the ideal classifier, providing a good signal for the Born machine. The Born machine is able to minimize its loss, which in turn leads to approximately $0$ total variation distance. Figure \ref{fig:bad_example} shows an unsuccessful experiment. Here, the MLP tracks the ideal classifier, but the Born machine is not able to find a ``direction'' that minimizes the loss (see epochs 400 to 1000). We verified that a more powerful two-layers Born machine performs much better for this instance (not shown).

In these figures gray lines correspond to quantities that can be estimated from samples, while blue and magenta lines represent exact quantities that are not practical to compute. It is challenging to assess the training looking at the gray lines. This discussion is to emphasize the need for new techniques to assess the models. These would provide an early stopping criterion for unsuccessful experiments as well as a validation criterion for successful ones. This is a common problem among adversarial approaches.

\begin{table*}[ht]
\begin{center}
\setlength{\tabcolsep}{12pt}
\renewcommand{\arraystretch}{1.25}
\scalebox{0.95}{
\begin{tabular}{|c|c|}
    \hline
    \textbf{Variable} & \textbf{Probabilities} \\ 
    \hline
    \multirow{2}{*}{Asia ($A$)} & $p(A = \true) = 0.01$  \\
    & $p(A = \fal) = 0.99$ \\
    \hline
    \multirow{4}{*}{Tuberculosis ($T$)} & $p(T = \true|A=\true) = 0.05$ \\
    &$p(T = \true|A=\fal) = 0.01$  \\
    & $p(T = \fal|A=\true) = 0.95$  \\
    &$p(T = \fal|A=\fal) = 0.99$  \\
    \hline
    \multirow{4}{*}{Lung cancer ($L$)}& $p(L = \true|S=\true) = 0.1\phantom{0}$ \\ 
    & $p(L = \true|S = \fal) = 0.01$  \\
    & $p(L = \fal|S=\true) = 0.9\phantom{0}$ \\
    & $p(L = \fal|S=\fal) = 0.99$\\
    \hline
    \multirow{4}{*}{Bronchitis ($B$)} & $p(B = \true|S=\true) = 0.6$ \\
    & $p(B = \true|S = \fal) = 0.3$ \\
    & $p(B = \fal|S=\true) = 0.4$ \\ 
    & $p(B = \fal|S=\fal) = 0.7$\\
    \hline
    \multirow{4}{*}{X-ray ($X$)} & $p(X = \true|I=\true) = 0.98$ \\
    & $p(X = \true|I=\fal) = 0.05$ \\
    & $p(X = \fal|I=\true) = 0.02$ \\
    & $p(X = \fal|I=\fal) = 0.95$  \\
    \hline
\end{tabular}
\quad
\begin{tabular}{|c|c|}
    \hline
    \textbf{Variable} & \textbf{Probabilities} \\ 
    \hline
    \multirow{2}{*}{Smoking (S)} & $p(S = \true) = 0.5$  \\
    & $p(S = \fal) = 0.5$ \\
    \hline
    \multirow{8}{*}{Illness (I)} & $p(I = \true|L=\true, T=\true) = 0.95\phantom{.00}$\\
    & $p(I = \true|L=\true, T=\fal) = 0.95\phantom{.00}$\\
    & $p(I = \true|L=\fal, T=\true) = 0.95\phantom{.00}$\\
    & $p(I = \true|L=\fal, T=\fal)= 0.05$\\
    & $p(I = \fal|L=\true, T=\true) = 0.05\phantom{.00}$\\
    & $p(I = \fal|L=\true, T=\fal) = 0.05\phantom{.00}$  \\
    & $p(I = \fal|L=\fal, T=\true) = 0.05\phantom{.00}$\\
    & $p(I = \fal|L=\fal, T=\fal)= 0.95$\\
    \hline
    \multirow{8}{*}{Dyspnoea (D)} 
    & $p(D = \true|B=\true, I=\true) = 0.9$\\
    & $p(D = \true|B=\true, I=\fal) = 0.8$ \\ 
    & $p(D = \true|B=\fal, I=\true) = 0.7$ \\
    & $p(D = \true|B=\fal, I=\fal)= 0.1$ \\
    & $p(D = \fal|B=\true, I=\true) = 0.1$\\
    & $p(D = \fal|B=\true, I=\fal) = 0.2$ \\
    & $p(D = \fal|B=\fal, I=\true) = 0.3$ \\
    & $p(D = \fal|B=\fal, I=\fal)= 0.9$ \\
    \hline
\end{tabular}
}
\end{center}
    \vspace{-2mm}
    \caption{Probability table corresponding to the network in Fig.~\ref{fig:lung_cancer}.}
    \label{tab:prob_table_lung_cancer_marginals}
\end{table*}

\section{Probability table for the ``lung cancer'' network}
\label{a:prob_table_lung_cancer_marginals}

We derived our VI methods under the assumption of nonzero posterior probabilities. In tabulated Bayesian networks, however, zero posterior probabilities may occur. The original ``lung cancer'' network~\cite{lauritzen_local_1988} contains a variable ``either'' (meaning that either lung cancer or tuberculosis are present). This variable is a deterministic \textsc{or} function of its parent variables and yields zero probabilities in some posterior distributions. In order to avoid numerical issues arising from zero probabilities, we replace ``either'' by ``illness'', which may be true even if both parent variables ``lung cancer'' and ``tuberculosis'' are false. In practice, this is done by adding a small $\epsilon > 0$ to the entries of the ``either'' table and then renormalizing. This simple techniques is known as additive smoothing, or Laplace smoothing. Table~\ref{tab:prob_table_lung_cancer_marginals} shows the modified probability table for the network in Fig.~\ref{fig:lung_cancer}.

An alternative approach to deal with a deterministic variable that implements the \textsc{or} function is to marginalize its parent variables. It can be verified that the entries of the resulting probability table are nonzero (unless the parent variables are also deterministic in which case one marginalizes these as well). One can perform variational inference on this modified network with less variables.


\begin{thebibliography}{132}%
\makeatletter
\providecommand \@ifxundefined [1]{%
 \@ifx{#1\undefined}
}%
\providecommand \@ifnum [1]{%
 \ifnum #1\expandafter \@firstoftwo
 \else \expandafter \@secondoftwo
 \fi
}%
\providecommand \@ifx [1]{%
 \ifx #1\expandafter \@firstoftwo
 \else \expandafter \@secondoftwo
 \fi
}%
\providecommand \natexlab [1]{#1}%
\providecommand \enquote  [1]{``#1''}%
\providecommand \bibnamefont  [1]{#1}%
\providecommand \bibfnamefont [1]{#1}%
\providecommand \citenamefont [1]{#1}%
\providecommand \href@noop [0]{\@secondoftwo}%
\providecommand \href [0]{\begingroup \@sanitize@url \@href}%
\providecommand \@href[1]{\@@startlink{#1}\@@href}%
\providecommand \@@href[1]{\endgroup#1\@@endlink}%
\providecommand \@sanitize@url [0]{\catcode `\\12\catcode `\$12\catcode
  `\&12\catcode `\#12\catcode `\^12\catcode `\_12\catcode `\%12\relax}%
\providecommand \@@startlink[1]{}%
\providecommand \@@endlink[0]{}%
\providecommand \url  [0]{\begingroup\@sanitize@url \@url }%
\providecommand \@url [1]{\endgroup\@href {#1}{\urlprefix }}%
\providecommand \urlprefix  [0]{URL }%
\providecommand \Eprint [0]{\href }%
\providecommand \doibase [0]{https://doi.org/}%
\providecommand \selectlanguage [0]{\@gobble}%
\providecommand \bibinfo  [0]{\@secondoftwo}%
\providecommand \bibfield  [0]{\@secondoftwo}%
\providecommand \translation [1]{[#1]}%
\providecommand \BibitemOpen [0]{}%
\providecommand \bibitemStop [0]{}%
\providecommand \bibitemNoStop [0]{.\EOS\space}%
\providecommand \EOS [0]{\spacefactor3000\relax}%
\providecommand \BibitemShut  [1]{\csname bibitem#1\endcsname}%
\let\auto@bib@innerbib\@empty
%</preamble>
\bibitem [{\citenamefont {Koller}\ and\ \citenamefont
  {Friedman}(2009)}]{koller2009probabilistic}%
  \BibitemOpen
  \bibfield  {author} {\bibinfo {author} {\bibfnamefont {D.}~\bibnamefont
  {Koller}}\ and\ \bibinfo {author} {\bibfnamefont {N.}~\bibnamefont
  {Friedman}},\ }\href@noop {} {\emph {\bibinfo {title} {Probabilistic
  Graphical Models: {{Principles}} and Techniques}}}\ (\bibinfo  {publisher}
  {{MIT Press}},\ \bibinfo {year} {2009})\BibitemShut {NoStop}%
\bibitem [{\citenamefont
  {Morris}(2001)}]{morrisRecognitionNetworksApproximate2001}%
  \BibitemOpen
  \bibfield  {author} {\bibinfo {author} {\bibfnamefont {Q.}~\bibnamefont
  {Morris}},\ }\bibfield  {title} {\bibinfo {title} {Recognition networks for
  approximate inference in {{BN20}} networks},\ }in\ \href@noop {} {\emph
  {\bibinfo {booktitle} {Proceedings of the {{Seventeenth Conference}} on
  {{Uncertainty}} in {{Artificial Intelligence}} ({{UAI2001}})}}},\ \bibinfo
  {editor} {edited by\ \bibinfo {editor} {\bibfnamefont {J.}~\bibnamefont
  {Breese}}\ and\ \bibinfo {editor} {\bibfnamefont {D.}~\bibnamefont
  {Koller}}}\ (\bibinfo  {publisher} {{Morgan Kaufmann Publishers Inc.}},\
  \bibinfo {address} {{Seattle, Washington, USA}},\ \bibinfo {year} {2001})\
  pp.\ \bibinfo {pages} {370--377},\ \Eprint {https://arxiv.org/abs/1301.2295}
  {arXiv:1301.2295} \BibitemShut {NoStop}%
\bibitem [{\citenamefont {Richens}\ \emph {et~al.}(2020)\citenamefont
  {Richens}, \citenamefont {Lee},\ and\ \citenamefont
  {Johri}}]{richensImprovingAccuracyMedical2020}%
  \BibitemOpen
  \bibfield  {author} {\bibinfo {author} {\bibfnamefont {J.~G.}\ \bibnamefont
  {Richens}}, \bibinfo {author} {\bibfnamefont {C.~M.}\ \bibnamefont {Lee}},\
  and\ \bibinfo {author} {\bibfnamefont {S.}~\bibnamefont {Johri}},\ }\bibfield
   {title} {\bibinfo {title} {Improving the accuracy of medical diagnosis with
  causal machine learning},\ }\href
  {https://doi.org/10.1038/s41467-020-17419-7} {\bibfield  {journal} {\bibinfo
  {journal} {Nature Communications}\ }\textbf {\bibinfo {volume} {11}},\
  \bibinfo {pages} {3923} (\bibinfo {year} {2020})}\BibitemShut {NoStop}%
\bibitem [{\citenamefont {Maathuis}\ \emph {et~al.}(2018)\citenamefont
  {Maathuis}, \citenamefont {Drton}, \citenamefont {Lauritzen},\ and\
  \citenamefont {Wainwright}}]{maathuis2018handbook}%
  \BibitemOpen
  \bibfield  {author} {\bibinfo {author} {\bibfnamefont {M.}~\bibnamefont
  {Maathuis}}, \bibinfo {author} {\bibfnamefont {M.}~\bibnamefont {Drton}},
  \bibinfo {author} {\bibfnamefont {S.}~\bibnamefont {Lauritzen}},\ and\
  \bibinfo {author} {\bibfnamefont {M.}~\bibnamefont {Wainwright}},\
  }\href@noop {} {\emph {\bibinfo {title} {Handbook of Graphical Models}}},\
  Chapman \& {{Hall}}/{{CRC}} Handbooks of Modern Statistical Methods\
  (\bibinfo  {publisher} {{CRC Press}},\ \bibinfo {address} {Boca Raton},\
  \bibinfo {year} {2018})\BibitemShut {NoStop}%
\bibitem [{\citenamefont {Wiegerinck}\ \emph {et~al.}(2010)\citenamefont
  {Wiegerinck}, \citenamefont {Kappen},\ and\ \citenamefont
  {Burgers}}]{wiegerinckBayesianNetworksExpert2010}%
  \BibitemOpen
  \bibfield  {author} {\bibinfo {author} {\bibfnamefont {W.}~\bibnamefont
  {Wiegerinck}}, \bibinfo {author} {\bibfnamefont {B.}~\bibnamefont {Kappen}},\
  and\ \bibinfo {author} {\bibfnamefont {W.}~\bibnamefont {Burgers}},\
  }\bibinfo {title} {Bayesian networks for expert systems: Theory and practical
  applications},\ in\ \href {https://doi.org/10.1007/978-3-642-11688-9_20}
  {\emph {\bibinfo {booktitle} {Interactive Collaborative Information
  Systems}}},\ \bibinfo {editor} {edited by\ \bibinfo {editor} {\bibfnamefont
  {R.}~\bibnamefont {Babu{\v{s}}ka}}\ and\ \bibinfo {editor} {\bibfnamefont
  {F.~C.~A.}\ \bibnamefont {Groen}}}\ (\bibinfo  {publisher} {Springer Berlin
  Heidelberg},\ \bibinfo {address} {Berlin, Heidelberg},\ \bibinfo {year}
  {2010})\ pp.\ \bibinfo {pages} {547--578}\BibitemShut {NoStop}%
\bibitem [{\citenamefont {Denev}(2015)}]{denev2015probabilistic}%
  \BibitemOpen
  \bibfield  {author} {\bibinfo {author} {\bibfnamefont {A.}~\bibnamefont
  {Denev}},\ }\href@noop {} {\emph {\bibinfo {title} {Probabilistic Graphical
  Models: {{A}} New Way of Thinking in Financial Modelling}}}\ (\bibinfo
  {publisher} {{Risk Books}},\ \bibinfo {year} {2015})\BibitemShut {NoStop}%
\bibitem [{\citenamefont {Cai}\ \emph {et~al.}(2017)\citenamefont {Cai},
  \citenamefont {Huang},\ and\ \citenamefont
  {Xie}}]{caiBayesianNetworksFault2017}%
  \BibitemOpen
  \bibfield  {author} {\bibinfo {author} {\bibfnamefont {B.}~\bibnamefont
  {Cai}}, \bibinfo {author} {\bibfnamefont {L.}~\bibnamefont {Huang}},\ and\
  \bibinfo {author} {\bibfnamefont {M.}~\bibnamefont {Xie}},\ }\bibfield
  {title} {\bibinfo {title} {Bayesian {{Networks}} in {{Fault Diagnosis}}},\
  }\href {https://doi.org/10.1109/TII.2017.2695583} {\bibfield  {journal}
  {\bibinfo  {journal} {IEEE Transactions on Industrial Informatics}\ }\textbf
  {\bibinfo {volume} {13}},\ \bibinfo {pages} {2227} (\bibinfo {year}
  {2017})}\BibitemShut {NoStop}%
\bibitem [{\citenamefont {Neal}(1993)}]{nealProbabilisticInferenceUsing1993}%
  \BibitemOpen
  \bibfield  {author} {\bibinfo {author} {\bibfnamefont {R.~M.}\ \bibnamefont
  {Neal}},\ }\href@noop {} {\emph {\bibinfo {title} {Probabilistic {{Inference
  Using {Markov Chain Monte Carlo} Methods}}}}},\ \bibinfo {type} {Technical
  {{Report}}}\ \bibinfo {number} {CRG-TR-93-1}\ (\bibinfo  {institution}
  {{Department of Computer Science, University of Toronto}},\ \bibinfo
  {address} {{Toronto}},\ \bibinfo {year} {1993})\BibitemShut {NoStop}%
\bibitem [{\citenamefont {Brooks}\ \emph {et~al.}(2011)\citenamefont {Brooks},
  \citenamefont {Gelman}, \citenamefont {Jones},\ and\ \citenamefont
  {Meng}}]{brooksHandbookMarkovChain2011}%
  \BibitemOpen
  \bibinfo {editor} {\bibfnamefont {S.}~\bibnamefont {Brooks}}, \bibinfo
  {editor} {\bibfnamefont {A.}~\bibnamefont {Gelman}}, \bibinfo {editor}
  {\bibfnamefont {G.~L.}\ \bibnamefont {Jones}},\ and\ \bibinfo {editor}
  {\bibfnamefont {X.-L.}\ \bibnamefont {Meng}},\ eds.,\ \href@noop {} {\emph
  {\bibinfo {title} {Handbook of {{Markov Chain Monte Carlo Methods}}}}}\
  (\bibinfo  {publisher} {{Chapman \& Hall/CRC}},\ \bibinfo {address} {{New
  York, NY, USA}},\ \bibinfo {year} {2011})\BibitemShut {NoStop}%
\bibitem [{\citenamefont {Blei}\ \emph {et~al.}(2017)\citenamefont {Blei},
  \citenamefont {Kucukelbir},\ and\ \citenamefont {McAuliffe}}]{Blei_2017}%
  \BibitemOpen
  \bibfield  {author} {\bibinfo {author} {\bibfnamefont {D.~M.}\ \bibnamefont
  {Blei}}, \bibinfo {author} {\bibfnamefont {A.}~\bibnamefont {Kucukelbir}},\
  and\ \bibinfo {author} {\bibfnamefont {J.~D.}\ \bibnamefont {McAuliffe}},\
  }\bibfield  {title} {\bibinfo {title} {{Variational Inference: A Review for
  Statisticians}},\ }\href {https://doi.org/10.1080/01621459.2017.1285773}
  {\bibfield  {journal} {\bibinfo  {journal} {Journal of the American
  Statistical Association}\ }\textbf {\bibinfo {volume} {112}},\ \bibinfo
  {pages} {859} (\bibinfo {year} {2017})}\BibitemShut {NoStop}%
\bibitem [{\citenamefont {Adachi}\ and\ \citenamefont
  {Henderson}(2015)}]{adachi2015application}%
  \BibitemOpen
  \bibfield  {author} {\bibinfo {author} {\bibfnamefont {S.~H.}\ \bibnamefont
  {Adachi}}\ and\ \bibinfo {author} {\bibfnamefont {M.~P.}\ \bibnamefont
  {Henderson}},\ }\href@noop {} {\bibinfo {title} {Application of quantum
  annealing to training of deep neural networks}} (\bibinfo {year} {2015}),\
  \Eprint {https://arxiv.org/abs/1510.06356} {arXiv:1510.06356 [quant-ph]}
  \BibitemShut {NoStop}%
\bibitem [{\citenamefont {Benedetti}\ \emph {et~al.}(2016)\citenamefont
  {Benedetti}, \citenamefont {Realpe-G\'omez}, \citenamefont {Biswas},\ and\
  \citenamefont {Perdomo-Ortiz}}]{Benedetti_2016}%
  \BibitemOpen
  \bibfield  {author} {\bibinfo {author} {\bibfnamefont {M.}~\bibnamefont
  {Benedetti}}, \bibinfo {author} {\bibfnamefont {J.}~\bibnamefont
  {Realpe-G\'omez}}, \bibinfo {author} {\bibfnamefont {R.}~\bibnamefont
  {Biswas}},\ and\ \bibinfo {author} {\bibfnamefont {A.}~\bibnamefont
  {Perdomo-Ortiz}},\ }\bibfield  {title} {\bibinfo {title} {Estimation of
  effective temperatures in quantum annealers for sampling applications: A case
  study with possible applications in deep learning},\ }\href
  {https://doi.org/10.1103/PhysRevA.94.022308} {\bibfield  {journal} {\bibinfo
  {journal} {Phys. Rev. A}\ }\textbf {\bibinfo {volume} {94}},\ \bibinfo
  {pages} {022308} (\bibinfo {year} {2016})}\BibitemShut {NoStop}%
\bibitem [{\citenamefont {Benedetti}\ \emph {et~al.}(2017)\citenamefont
  {Benedetti}, \citenamefont {Realpe-G\'omez}, \citenamefont {Biswas},\ and\
  \citenamefont {Perdomo-Ortiz}}]{Benedetti_2017}%
  \BibitemOpen
  \bibfield  {author} {\bibinfo {author} {\bibfnamefont {M.}~\bibnamefont
  {Benedetti}}, \bibinfo {author} {\bibfnamefont {J.}~\bibnamefont
  {Realpe-G\'omez}}, \bibinfo {author} {\bibfnamefont {R.}~\bibnamefont
  {Biswas}},\ and\ \bibinfo {author} {\bibfnamefont {A.}~\bibnamefont
  {Perdomo-Ortiz}},\ }\bibfield  {title} {\bibinfo {title} {Quantum-assisted
  learning of hardware-embedded probabilistic graphical models},\ }\href
  {https://doi.org/10.1103/PhysRevX.7.041052} {\bibfield  {journal} {\bibinfo
  {journal} {Phys. Rev. X}\ }\textbf {\bibinfo {volume} {7}},\ \bibinfo {pages}
  {041052} (\bibinfo {year} {2017})}\BibitemShut {NoStop}%
\bibitem [{\citenamefont {Korenkevych}\ \emph {et~al.}(2016)\citenamefont
  {Korenkevych}, \citenamefont {Xue}, \citenamefont {Bian}, \citenamefont
  {Chudak}, \citenamefont {Macready}, \citenamefont {Rolfe},\ and\
  \citenamefont {Andriyash}}]{korenkevych2016benchmarking}%
  \BibitemOpen
  \bibfield  {author} {\bibinfo {author} {\bibfnamefont {D.}~\bibnamefont
  {Korenkevych}}, \bibinfo {author} {\bibfnamefont {Y.}~\bibnamefont {Xue}},
  \bibinfo {author} {\bibfnamefont {Z.}~\bibnamefont {Bian}}, \bibinfo {author}
  {\bibfnamefont {F.}~\bibnamefont {Chudak}}, \bibinfo {author} {\bibfnamefont
  {W.~G.}\ \bibnamefont {Macready}}, \bibinfo {author} {\bibfnamefont
  {J.}~\bibnamefont {Rolfe}},\ and\ \bibinfo {author} {\bibfnamefont
  {E.}~\bibnamefont {Andriyash}},\ }\href@noop {} {\bibinfo {title}
  {Benchmarking quantum hardware for training of fully visible {Boltzmann}
  machines}} (\bibinfo {year} {2016}),\ \Eprint
  {https://arxiv.org/abs/1611.04528} {arXiv:1611.04528 [quant-ph]} \BibitemShut
  {NoStop}%
\bibitem [{\citenamefont {Benedetti}\ \emph {et~al.}(2018)\citenamefont
  {Benedetti}, \citenamefont {Realpe-G{\'{o}}mez},\ and\ \citenamefont
  {Perdomo-Ortiz}}]{Benedetti_2018}%
  \BibitemOpen
  \bibfield  {author} {\bibinfo {author} {\bibfnamefont {M.}~\bibnamefont
  {Benedetti}}, \bibinfo {author} {\bibfnamefont {J.}~\bibnamefont
  {Realpe-G{\'{o}}mez}},\ and\ \bibinfo {author} {\bibfnamefont
  {A.}~\bibnamefont {Perdomo-Ortiz}},\ }\bibfield  {title} {\bibinfo {title}
  {{Quantum-assisted Helmholtz machines: A quantum{\textendash}classical deep
  learning framework for industrial datasets in near-term devices}},\ }\href
  {https://doi.org/10.1088/2058-9565/aabd98} {\bibfield  {journal} {\bibinfo
  {journal} {Quantum Science and Technology}\ }\textbf {\bibinfo {volume}
  {3}},\ \bibinfo {pages} {034007} (\bibinfo {year} {2018})}\BibitemShut
  {NoStop}%
\bibitem [{\citenamefont {Khoshaman}\ \emph {et~al.}(2018)\citenamefont
  {Khoshaman}, \citenamefont {Vinci}, \citenamefont {Denis}, \citenamefont
  {Andriyash}, \citenamefont {Sadeghi},\ and\ \citenamefont
  {Amin}}]{Khoshaman_2018}%
  \BibitemOpen
  \bibfield  {author} {\bibinfo {author} {\bibfnamefont {A.}~\bibnamefont
  {Khoshaman}}, \bibinfo {author} {\bibfnamefont {W.}~\bibnamefont {Vinci}},
  \bibinfo {author} {\bibfnamefont {B.}~\bibnamefont {Denis}}, \bibinfo
  {author} {\bibfnamefont {E.}~\bibnamefont {Andriyash}}, \bibinfo {author}
  {\bibfnamefont {H.}~\bibnamefont {Sadeghi}},\ and\ \bibinfo {author}
  {\bibfnamefont {M.~H.}\ \bibnamefont {Amin}},\ }\bibfield  {title} {\bibinfo
  {title} {Quantum variational autoencoder},\ }\href
  {https://doi.org/10.1088/2058-9565/aada1f} {\bibfield  {journal} {\bibinfo
  {journal} {Quantum Science and Technology}\ }\textbf {\bibinfo {volume}
  {4}},\ \bibinfo {pages} {014001} (\bibinfo {year} {2018})}\BibitemShut
  {NoStop}%
\bibitem [{\citenamefont {Wilson}\ \emph {et~al.}(2021)\citenamefont {Wilson},
  \citenamefont {Vandal}, \citenamefont {Hogg},\ and\ \citenamefont
  {Rieffel}}]{wilson2019quantumassisted}%
  \BibitemOpen
  \bibfield  {author} {\bibinfo {author} {\bibfnamefont {M.}~\bibnamefont
  {Wilson}}, \bibinfo {author} {\bibfnamefont {T.}~\bibnamefont {Vandal}},
  \bibinfo {author} {\bibfnamefont {T.}~\bibnamefont {Hogg}},\ and\ \bibinfo
  {author} {\bibfnamefont {E.~G.}\ \bibnamefont {Rieffel}},\ }\bibfield
  {title} {\bibinfo {title} {Quantum-assisted associative adversarial network:
  Applying quantum annealing in deep learning},\ }\href
  {https://doi.org/10.1007/s42484-021-00047-9} {\bibfield  {journal} {\bibinfo
  {journal} {Quantum Machine Intelligence}\ }\textbf {\bibinfo {volume} {3}},\
  \bibinfo {pages} {1} (\bibinfo {year} {2021})}\BibitemShut {NoStop}%
\bibitem [{\citenamefont {Low}\ \emph {et~al.}(2014)\citenamefont {Low},
  \citenamefont {Yoder},\ and\ \citenamefont {Chuang}}]{low2014quantum}%
  \BibitemOpen
  \bibfield  {author} {\bibinfo {author} {\bibfnamefont {G.~H.}\ \bibnamefont
  {Low}}, \bibinfo {author} {\bibfnamefont {T.~J.}\ \bibnamefont {Yoder}},\
  and\ \bibinfo {author} {\bibfnamefont {I.~L.}\ \bibnamefont {Chuang}},\
  }\bibfield  {title} {\bibinfo {title} {Quantum inference on {Bayesian}
  networks},\ }\href {https://doi.org/10.1103/PhysRevA.89.062315} {\bibfield
  {journal} {\bibinfo  {journal} {Phys. Rev. A}\ }\textbf {\bibinfo {volume}
  {89}},\ \bibinfo {pages} {062315} (\bibinfo {year} {2014})}\BibitemShut
  {NoStop}%
\bibitem [{\citenamefont {Montanaro}(2015)}]{Montanaro_2015}%
  \BibitemOpen
  \bibfield  {author} {\bibinfo {author} {\bibfnamefont {A.}~\bibnamefont
  {Montanaro}},\ }\bibfield  {title} {\bibinfo {title} {{Quantum speedup of
  Monte Carlo methods}},\ }\href {https://doi.org/10.1098/rspa.2015.0301}
  {\bibfield  {journal} {\bibinfo  {journal} {Proceedings of the Royal Society
  A: Mathematical, Physical and Engineering Sciences}\ }\textbf {\bibinfo
  {volume} {471}},\ \bibinfo {pages} {20150301} (\bibinfo {year}
  {2015})}\BibitemShut {NoStop}%
\bibitem [{\citenamefont {Chowdhury}\ and\ \citenamefont
  {Somma}(2017)}]{chowdhury2016quantum}%
  \BibitemOpen
  \bibfield  {author} {\bibinfo {author} {\bibfnamefont {A.~N.}\ \bibnamefont
  {Chowdhury}}\ and\ \bibinfo {author} {\bibfnamefont {R.~D.}\ \bibnamefont
  {Somma}},\ }\bibfield  {title} {\bibinfo {title} {Quantum algorithms for
  {Gibbs} sampling and hitting-time estimation},\ }\href@noop {} {\bibfield
  {journal} {\bibinfo  {journal} {Quant. Inf. Comp.}\ }\textbf {\bibinfo
  {volume} {17}},\ \bibinfo {pages} {0041} (\bibinfo {year} {2017})},\ \Eprint
  {https://arxiv.org/abs/1603.02940} {arXiv:1603.02940 [quant-ph]} \BibitemShut
  {NoStop}%
\bibitem [{\citenamefont {Wittek}\ and\ \citenamefont
  {Gogolin}(2017)}]{Wittek_2017}%
  \BibitemOpen
  \bibfield  {author} {\bibinfo {author} {\bibfnamefont {P.}~\bibnamefont
  {Wittek}}\ and\ \bibinfo {author} {\bibfnamefont {C.}~\bibnamefont
  {Gogolin}},\ }\bibfield  {title} {\bibinfo {title} {{Quantum Enhanced
  Inference in Markov Logic Networks}},\ }\href
  {https://doi.org/10.1038/srep45672} {\bibfield  {journal} {\bibinfo
  {journal} {Scientific Reports}\ }\textbf {\bibinfo {volume} {7}} (\bibinfo
  {year} {2017})}\BibitemShut {NoStop}%
\bibitem [{\citenamefont {Wu}\ and\ \citenamefont
  {Hsieh}(2019)}]{wu2019variational}%
  \BibitemOpen
  \bibfield  {author} {\bibinfo {author} {\bibfnamefont {J.}~\bibnamefont
  {Wu}}\ and\ \bibinfo {author} {\bibfnamefont {T.~H.}\ \bibnamefont {Hsieh}},\
  }\bibfield  {title} {\bibinfo {title} {Variational thermal quantum simulation
  via thermofield double states},\ }\href
  {https://doi.org/10.1103/PhysRevLett.123.220502} {\bibfield  {journal}
  {\bibinfo  {journal} {Phys. Rev. Lett.}\ }\textbf {\bibinfo {volume} {123}},\
  \bibinfo {pages} {220502} (\bibinfo {year} {2019})}\BibitemShut {NoStop}%
\bibitem [{\citenamefont {Verdon}\ \emph {et~al.}(2019)\citenamefont {Verdon},
  \citenamefont {Marks}, \citenamefont {Nanda}, \citenamefont {Leichenauer},\
  and\ \citenamefont {Hidary}}]{verdonQuantumHamiltonianBasedModels2019}%
  \BibitemOpen
  \bibfield  {author} {\bibinfo {author} {\bibfnamefont {G.}~\bibnamefont
  {Verdon}}, \bibinfo {author} {\bibfnamefont {J.}~\bibnamefont {Marks}},
  \bibinfo {author} {\bibfnamefont {S.}~\bibnamefont {Nanda}}, \bibinfo
  {author} {\bibfnamefont {S.}~\bibnamefont {Leichenauer}},\ and\ \bibinfo
  {author} {\bibfnamefont {J.}~\bibnamefont {Hidary}},\ }\href@noop {}
  {\bibinfo {title} {Quantum {{Hamiltonian}}-{{Based Models}} and the
  {{Variational Quantum Thermalizer Algorithm}}}} (\bibinfo {year} {2019}),\
  \Eprint {https://arxiv.org/abs/1910.02071} {arXiv:1910.02071 [quant-ph]}
  \BibitemShut {NoStop}%
\bibitem [{\citenamefont {Chowdhury}\ \emph {et~al.}(2020)\citenamefont
  {Chowdhury}, \citenamefont {Low},\ and\ \citenamefont
  {Wiebe}}]{chowdhury2020variational}%
  \BibitemOpen
  \bibfield  {author} {\bibinfo {author} {\bibfnamefont {A.~N.}\ \bibnamefont
  {Chowdhury}}, \bibinfo {author} {\bibfnamefont {G.~H.}\ \bibnamefont {Low}},\
  and\ \bibinfo {author} {\bibfnamefont {N.}~\bibnamefont {Wiebe}},\
  }\href@noop {} {\bibinfo {title} {{A Variational Quantum Algorithm for
  Preparing Quantum Gibbs States}}} (\bibinfo {year} {2020}),\ \Eprint
  {https://arxiv.org/abs/2002.00055} {arXiv:2002.00055 [quant-ph]} \BibitemShut
  {NoStop}%
\bibitem [{\citenamefont {Wang}\ \emph {et~al.}(2020)\citenamefont {Wang},
  \citenamefont {Li},\ and\ \citenamefont {Wang}}]{wang2020variational}%
  \BibitemOpen
  \bibfield  {author} {\bibinfo {author} {\bibfnamefont {Y.}~\bibnamefont
  {Wang}}, \bibinfo {author} {\bibfnamefont {G.}~\bibnamefont {Li}},\ and\
  \bibinfo {author} {\bibfnamefont {X.}~\bibnamefont {Wang}},\ }\href@noop {}
  {\bibinfo {title} {{Variational quantum Gibbs state preparation with a
  truncated Taylor series}}} (\bibinfo {year} {2020}),\ \Eprint
  {https://arxiv.org/abs/2005.08797} {arXiv:2005.08797 [quant-ph]} \BibitemShut
  {NoStop}%
\bibitem [{\citenamefont {Shingu}\ \emph {et~al.}(2020)\citenamefont {Shingu},
  \citenamefont {Seki}, \citenamefont {Watabe}, \citenamefont {Endo},
  \citenamefont {Matsuzaki}, \citenamefont {Kawabata}, \citenamefont {Nikuni},\
  and\ \citenamefont {Hakoshima}}]{shingu2020boltzmann}%
  \BibitemOpen
  \bibfield  {author} {\bibinfo {author} {\bibfnamefont {Y.}~\bibnamefont
  {Shingu}}, \bibinfo {author} {\bibfnamefont {Y.}~\bibnamefont {Seki}},
  \bibinfo {author} {\bibfnamefont {S.}~\bibnamefont {Watabe}}, \bibinfo
  {author} {\bibfnamefont {S.}~\bibnamefont {Endo}}, \bibinfo {author}
  {\bibfnamefont {Y.}~\bibnamefont {Matsuzaki}}, \bibinfo {author}
  {\bibfnamefont {S.}~\bibnamefont {Kawabata}}, \bibinfo {author}
  {\bibfnamefont {T.}~\bibnamefont {Nikuni}},\ and\ \bibinfo {author}
  {\bibfnamefont {H.}~\bibnamefont {Hakoshima}},\ }\href@noop {} {\bibinfo
  {title} {Boltzmann machine learning with a variational quantum algorithm}}
  (\bibinfo {year} {2020}),\ \Eprint {https://arxiv.org/abs/2007.00876}
  {arXiv:2007.00876 [quant-ph]} \BibitemShut {NoStop}%
\bibitem [{\citenamefont {Sato}\ \emph {et~al.}(2009)\citenamefont {Sato},
  \citenamefont {Kurihara}, \citenamefont {Tanaka}, \citenamefont {Nakagawa},\
  and\ \citenamefont {Miyashita}}]{satoQuantumAnnealingVariational2009}%
  \BibitemOpen
  \bibfield  {author} {\bibinfo {author} {\bibfnamefont {I.}~\bibnamefont
  {Sato}}, \bibinfo {author} {\bibfnamefont {K.}~\bibnamefont {Kurihara}},
  \bibinfo {author} {\bibfnamefont {S.}~\bibnamefont {Tanaka}}, \bibinfo
  {author} {\bibfnamefont {H.}~\bibnamefont {Nakagawa}},\ and\ \bibinfo
  {author} {\bibfnamefont {S.}~\bibnamefont {Miyashita}},\ }\bibfield  {title}
  {\bibinfo {title} {Quantum annealing for variational {Bayes} inference},\
  }in\ \href@noop {} {\emph {\bibinfo {booktitle} {Proceedings of the
  Twenty-Fifth Conference on Uncertainty in Artificial Intelligence}}},\
  \bibinfo {series and number} {{{UAI}} '09}\ (\bibinfo  {publisher} {{AUAI
  Press}},\ \bibinfo {address} {{Arlington, Virginia, USA}},\ \bibinfo {year}
  {2009})\ pp.\ \bibinfo {pages} {479--486},\ \Eprint
  {https://arxiv.org/abs/1408.2037} {arXiv:1408.2037} \BibitemShut {NoStop}%
\bibitem [{\citenamefont {O’Gorman}\ \emph {et~al.}(2015)\citenamefont
  {O’Gorman}, \citenamefont {Babbush}, \citenamefont {Perdomo-Ortiz},
  \citenamefont {Aspuru-Guzik},\ and\ \citenamefont
  {Smelyanskiy}}]{ogorman2015bayesian}%
  \BibitemOpen
  \bibfield  {author} {\bibinfo {author} {\bibfnamefont {B.}~\bibnamefont
  {O’Gorman}}, \bibinfo {author} {\bibfnamefont {R.}~\bibnamefont {Babbush}},
  \bibinfo {author} {\bibfnamefont {A.}~\bibnamefont {Perdomo-Ortiz}}, \bibinfo
  {author} {\bibfnamefont {A.}~\bibnamefont {Aspuru-Guzik}},\ and\ \bibinfo
  {author} {\bibfnamefont {V.}~\bibnamefont {Smelyanskiy}},\ }\bibfield
  {title} {\bibinfo {title} {Bayesian network structure learning using quantum
  annealing},\ }\href {https://doi.org/10.1140/epjst/e2015-02349-9} {\bibfield
  {journal} {\bibinfo  {journal} {The European Physical Journal Special
  Topics}\ }\textbf {\bibinfo {volume} {224}},\ \bibinfo {pages} {163–188}
  (\bibinfo {year} {2015})}\BibitemShut {NoStop}%
\bibitem [{\citenamefont {Miyahara}\ and\ \citenamefont
  {Sughiyama}(2018)}]{miyaharaQuantumExtensionVariational2018}%
  \BibitemOpen
  \bibfield  {author} {\bibinfo {author} {\bibfnamefont {H.}~\bibnamefont
  {Miyahara}}\ and\ \bibinfo {author} {\bibfnamefont {Y.}~\bibnamefont
  {Sughiyama}},\ }\bibfield  {title} {\bibinfo {title} {Quantum extension of
  variational bayes inference},\ }\href
  {https://doi.org/10.1103/PhysRevA.98.022330} {\bibfield  {journal} {\bibinfo
  {journal} {Phys. Rev. A}\ }\textbf {\bibinfo {volume} {98}},\ \bibinfo
  {pages} {022330} (\bibinfo {year} {2018})}\BibitemShut {NoStop}%
\bibitem [{\citenamefont {Zhang}\ \emph {et~al.}(2018)\citenamefont {Zhang},
  \citenamefont {Butepage}, \citenamefont {Kjellstrom},\ and\ \citenamefont
  {Mandt}}]{zhang2018advances}%
  \BibitemOpen
  \bibfield  {author} {\bibinfo {author} {\bibfnamefont {C.}~\bibnamefont
  {Zhang}}, \bibinfo {author} {\bibfnamefont {J.}~\bibnamefont {Butepage}},
  \bibinfo {author} {\bibfnamefont {H.}~\bibnamefont {Kjellstrom}},\ and\
  \bibinfo {author} {\bibfnamefont {S.}~\bibnamefont {Mandt}},\ }\href@noop {}
  {\bibinfo {title} {Advances in variational inference}} (\bibinfo {year}
  {2018}),\ \Eprint {https://arxiv.org/abs/1711.05597} {arXiv:1711.05597
  [cs.LG]} \BibitemShut {NoStop}%
\bibitem [{\citenamefont {Abbas}\ \emph {et~al.}(2021)\citenamefont {Abbas},
  \citenamefont {Sutter}, \citenamefont {Zoufal}, \citenamefont {Lucchi},
  \citenamefont {Figalli},\ and\ \citenamefont {Woerner}}]{abbas_power_2020}%
  \BibitemOpen
  \bibfield  {author} {\bibinfo {author} {\bibfnamefont {A.}~\bibnamefont
  {Abbas}}, \bibinfo {author} {\bibfnamefont {D.}~\bibnamefont {Sutter}},
  \bibinfo {author} {\bibfnamefont {C.}~\bibnamefont {Zoufal}}, \bibinfo
  {author} {\bibfnamefont {A.}~\bibnamefont {Lucchi}}, \bibinfo {author}
  {\bibfnamefont {A.}~\bibnamefont {Figalli}},\ and\ \bibinfo {author}
  {\bibfnamefont {S.}~\bibnamefont {Woerner}},\ }\bibfield  {title} {\bibinfo
  {title} {The power of quantum neural networks},\ }\href
  {https://doi.org/10.1038/s43588-021-00084-1} {\bibfield  {journal} {\bibinfo
  {journal} {Nature Computational Science}\ }\textbf {\bibinfo {volume} {1}},\
  \bibinfo {pages} {403–409} (\bibinfo {year} {2021})}\BibitemShut {NoStop}%
\bibitem [{\citenamefont {Huang}\ \emph
  {et~al.}(2021{\natexlab{a}})\citenamefont {Huang}, \citenamefont {Kueng},\
  and\ \citenamefont {Preskill}}]{huang_information-theoretic_2021}%
  \BibitemOpen
  \bibfield  {author} {\bibinfo {author} {\bibfnamefont {H.-Y.}\ \bibnamefont
  {Huang}}, \bibinfo {author} {\bibfnamefont {R.}~\bibnamefont {Kueng}},\ and\
  \bibinfo {author} {\bibfnamefont {J.}~\bibnamefont {Preskill}},\ }\bibfield
  {title} {\bibinfo {title} {Information-theoretic bounds on quantum advantage
  in machine learning},\ }\href
  {https://doi.org/10.1103/PhysRevLett.126.190505} {\bibfield  {journal}
  {\bibinfo  {journal} {Phys. Rev. Lett.}\ }\textbf {\bibinfo {volume} {126}},\
  \bibinfo {pages} {190505} (\bibinfo {year} {2021}{\natexlab{a}})}\BibitemShut
  {NoStop}%
\bibitem [{\citenamefont {Poland}\ \emph {et~al.}(2020)\citenamefont {Poland},
  \citenamefont {Beer},\ and\ \citenamefont {Osborne}}]{poland2020free}%
  \BibitemOpen
  \bibfield  {author} {\bibinfo {author} {\bibfnamefont {K.}~\bibnamefont
  {Poland}}, \bibinfo {author} {\bibfnamefont {K.}~\bibnamefont {Beer}},\ and\
  \bibinfo {author} {\bibfnamefont {T.~J.}\ \bibnamefont {Osborne}},\
  }\href@noop {} {\bibinfo {title} {No free lunch for quantum machine
  learning}} (\bibinfo {year} {2020}),\ \Eprint
  {https://arxiv.org/abs/2003.14103} {arXiv:2003.14103 [quant-ph]} \BibitemShut
  {NoStop}%
\bibitem [{\citenamefont {Sharma}\ \emph {et~al.}(2020)\citenamefont {Sharma},
  \citenamefont {Cerezo}, \citenamefont {Holmes}, \citenamefont {Cincio},
  \citenamefont {Sornborger},\ and\ \citenamefont
  {Coles}}]{sharma2020reformulation}%
  \BibitemOpen
  \bibfield  {author} {\bibinfo {author} {\bibfnamefont {K.}~\bibnamefont
  {Sharma}}, \bibinfo {author} {\bibfnamefont {M.}~\bibnamefont {Cerezo}},
  \bibinfo {author} {\bibfnamefont {Z.}~\bibnamefont {Holmes}}, \bibinfo
  {author} {\bibfnamefont {L.}~\bibnamefont {Cincio}}, \bibinfo {author}
  {\bibfnamefont {A.}~\bibnamefont {Sornborger}},\ and\ \bibinfo {author}
  {\bibfnamefont {P.~J.}\ \bibnamefont {Coles}},\ }\href@noop {} {\bibinfo
  {title} {Reformulation of the no-free-lunch theorem for entangled data sets}}
  (\bibinfo {year} {2020}),\ \Eprint {https://arxiv.org/abs/2007.04900}
  {arXiv:2007.04900 [quant-ph]} \BibitemShut {NoStop}%
\bibitem [{\citenamefont {Servedio}\ and\ \citenamefont
  {Gortler}(2004)}]{servedio_equivalences_2004}%
  \BibitemOpen
  \bibfield  {author} {\bibinfo {author} {\bibfnamefont {R.~A.}\ \bibnamefont
  {Servedio}}\ and\ \bibinfo {author} {\bibfnamefont {S.~J.}\ \bibnamefont
  {Gortler}},\ }\bibfield  {title} {\bibinfo {title} {Equivalences and
  {Separations} {Between} {Quantum} and {Classical} {Learnability}},\ }\href
  {https://doi.org/10.1137/S0097539704412910} {\bibfield  {journal} {\bibinfo
  {journal} {SIAM Journal on Computing}\ }\textbf {\bibinfo {volume} {33}},\
  \bibinfo {pages} {1067} (\bibinfo {year} {2004})}\BibitemShut {NoStop}%
\bibitem [{\citenamefont {Arunachalam}\ and\ \citenamefont
  {de~Wolf}(2017)}]{arunachalam_guest_2017}%
  \BibitemOpen
  \bibfield  {author} {\bibinfo {author} {\bibfnamefont {S.}~\bibnamefont
  {Arunachalam}}\ and\ \bibinfo {author} {\bibfnamefont {R.}~\bibnamefont
  {de~Wolf}},\ }\bibfield  {title} {\bibinfo {title} {Guest {Column}: {A}
  {Survey} of {Quantum} {Learning} {Theory}},\ }\href
  {https://doi.org/10.1145/3106700.3106710} {\bibfield  {journal} {\bibinfo
  {journal} {ACM SIGACT News}\ }\textbf {\bibinfo {volume} {48}},\ \bibinfo
  {pages} {41} (\bibinfo {year} {2017})}\BibitemShut {NoStop}%
\bibitem [{\citenamefont {Arunachalam}\ and\ \citenamefont
  {Wolf}(2018)}]{arunachalam_optimal_2018}%
  \BibitemOpen
  \bibfield  {author} {\bibinfo {author} {\bibfnamefont {S.}~\bibnamefont
  {Arunachalam}}\ and\ \bibinfo {author} {\bibfnamefont {R.~d.}\ \bibnamefont
  {Wolf}},\ }\bibfield  {title} {\bibinfo {title} {Optimal {Quantum} {Sample}
  {Complexity} of {Learning} {Algorithms}},\ }\href
  {http://jmlr.org/papers/v19/18-195.html} {\bibfield  {journal} {\bibinfo
  {journal} {Journal of Machine Learning Research}\ }\textbf {\bibinfo {volume}
  {19}},\ \bibinfo {pages} {1} (\bibinfo {year} {2018})}\BibitemShut {NoStop}%
\bibitem [{\citenamefont {Liu}\ \emph {et~al.}(2021)\citenamefont {Liu},
  \citenamefont {Arunachalam},\ and\ \citenamefont
  {Temme}}]{liu_rigorous_2020}%
  \BibitemOpen
  \bibfield  {author} {\bibinfo {author} {\bibfnamefont {Y.}~\bibnamefont
  {Liu}}, \bibinfo {author} {\bibfnamefont {S.}~\bibnamefont {Arunachalam}},\
  and\ \bibinfo {author} {\bibfnamefont {K.}~\bibnamefont {Temme}},\ }\bibfield
   {title} {\bibinfo {title} {A rigorous and robust quantum speed-up in
  supervised machine learning},\ }\href
  {https://doi.org/10.1038/s41567-021-01287-z} {\bibfield  {journal} {\bibinfo
  {journal} {Nature Physics}\ }\textbf {\bibinfo {volume} {17}},\ \bibinfo
  {pages} {1013–1017} (\bibinfo {year} {2021})}\BibitemShut {NoStop}%
\bibitem [{\citenamefont {Havlíček}\ \emph {et~al.}(2019)\citenamefont
  {Havlíček}, \citenamefont {Córcoles}, \citenamefont {Temme}, \citenamefont
  {Harrow}, \citenamefont {Kandala}, \citenamefont {Chow},\ and\ \citenamefont
  {Gambetta}}]{havlicek_supervised_2019}%
  \BibitemOpen
  \bibfield  {author} {\bibinfo {author} {\bibfnamefont {V.}~\bibnamefont
  {Havlíček}}, \bibinfo {author} {\bibfnamefont {A.~D.}\ \bibnamefont
  {Córcoles}}, \bibinfo {author} {\bibfnamefont {K.}~\bibnamefont {Temme}},
  \bibinfo {author} {\bibfnamefont {A.~W.}\ \bibnamefont {Harrow}}, \bibinfo
  {author} {\bibfnamefont {A.}~\bibnamefont {Kandala}}, \bibinfo {author}
  {\bibfnamefont {J.~M.}\ \bibnamefont {Chow}},\ and\ \bibinfo {author}
  {\bibfnamefont {J.~M.}\ \bibnamefont {Gambetta}},\ }\bibfield  {title}
  {\bibinfo {title} {Supervised learning with quantum-enhanced feature
  spaces},\ }\href {https://doi.org/10.1038/s41586-019-0980-2} {\bibfield
  {journal} {\bibinfo  {journal} {Nature}\ }\textbf {\bibinfo {volume} {567}},\
  \bibinfo {pages} {209} (\bibinfo {year} {2019})}\BibitemShut {NoStop}%
\bibitem [{\citenamefont {Schuld}\ and\ \citenamefont
  {Killoran}(2019)}]{schuld2019quantum}%
  \BibitemOpen
  \bibfield  {author} {\bibinfo {author} {\bibfnamefont {M.}~\bibnamefont
  {Schuld}}\ and\ \bibinfo {author} {\bibfnamefont {N.}~\bibnamefont
  {Killoran}},\ }\bibfield  {title} {\bibinfo {title} {{Quantum Machine
  Learning in Feature Hilbert Spaces}},\ }\href
  {https://doi.org/10.1103/PhysRevLett.122.040504} {\bibfield  {journal}
  {\bibinfo  {journal} {Phys. Rev. Lett.}\ }\textbf {\bibinfo {volume} {122}},\
  \bibinfo {pages} {040504} (\bibinfo {year} {2019})}\BibitemShut {NoStop}%
\bibitem [{\citenamefont {Huang}\ \emph
  {et~al.}(2021{\natexlab{b}})\citenamefont {Huang}, \citenamefont {Broughton},
  \citenamefont {Mohseni}, \citenamefont {Babbush}, \citenamefont {Boixo},
  \citenamefont {Neven},\ and\ \citenamefont {McClean}}]{huang_power_2020}%
  \BibitemOpen
  \bibfield  {author} {\bibinfo {author} {\bibfnamefont {H.-Y.}\ \bibnamefont
  {Huang}}, \bibinfo {author} {\bibfnamefont {M.}~\bibnamefont {Broughton}},
  \bibinfo {author} {\bibfnamefont {M.}~\bibnamefont {Mohseni}}, \bibinfo
  {author} {\bibfnamefont {R.}~\bibnamefont {Babbush}}, \bibinfo {author}
  {\bibfnamefont {S.}~\bibnamefont {Boixo}}, \bibinfo {author} {\bibfnamefont
  {H.}~\bibnamefont {Neven}},\ and\ \bibinfo {author} {\bibfnamefont {J.~R.}\
  \bibnamefont {McClean}},\ }\bibfield  {title} {\bibinfo {title} {Power of
  data in quantum machine learning},\ }\href
  {https://doi.org/10.1038/s41467-021-22539-9} {\bibfield  {journal} {\bibinfo
  {journal} {Nature Communications}\ }\textbf {\bibinfo {volume} {12}},\
  \bibinfo {pages} {2631} (\bibinfo {year} {2021}{\natexlab{b}})}\BibitemShut
  {NoStop}%
\bibitem [{\citenamefont {Gao}\ \emph {et~al.}(2018)\citenamefont {Gao},
  \citenamefont {Zhang},\ and\ \citenamefont {Duan}}]{gao_quantum_2018}%
  \BibitemOpen
  \bibfield  {author} {\bibinfo {author} {\bibfnamefont {X.}~\bibnamefont
  {Gao}}, \bibinfo {author} {\bibfnamefont {Z.-Y.}\ \bibnamefont {Zhang}},\
  and\ \bibinfo {author} {\bibfnamefont {L.-M.}\ \bibnamefont {Duan}},\
  }\bibfield  {title} {\bibinfo {title} {A quantum machine learning algorithm
  based on generative models},\ }\href {https://doi.org/10.1126/sciadv.aat9004}
  {\bibfield  {journal} {\bibinfo  {journal} {Science Advances}\ }\textbf
  {\bibinfo {volume} {4}},\ \bibinfo {pages} {eaat9004} (\bibinfo {year}
  {2018})}\BibitemShut {NoStop}%
\bibitem [{\citenamefont {Coyle}\ \emph {et~al.}(2020)\citenamefont {Coyle},
  \citenamefont {Mills}, \citenamefont {Danos},\ and\ \citenamefont
  {Kashefi}}]{Coyle_2020}%
  \BibitemOpen
  \bibfield  {author} {\bibinfo {author} {\bibfnamefont {B.}~\bibnamefont
  {Coyle}}, \bibinfo {author} {\bibfnamefont {D.}~\bibnamefont {Mills}},
  \bibinfo {author} {\bibfnamefont {V.}~\bibnamefont {Danos}},\ and\ \bibinfo
  {author} {\bibfnamefont {E.}~\bibnamefont {Kashefi}},\ }\bibfield  {title}
  {\bibinfo {title} {The {Born} supremacy: quantum advantage and training of an
  {Ising Born} machine},\ }\href {https://doi.org/10.1038/s41534-020-00288-9}
  {\bibfield  {journal} {\bibinfo  {journal} {npj Quantum Information}\
  }\textbf {\bibinfo {volume} {6}} (\bibinfo {year} {2020})}\BibitemShut
  {NoStop}%
\bibitem [{\citenamefont {Sweke}\ \emph {et~al.}(2021)\citenamefont {Sweke},
  \citenamefont {Seifert}, \citenamefont {Hangleiter},\ and\ \citenamefont
  {Eisert}}]{sweke_quantum_2020}%
  \BibitemOpen
  \bibfield  {author} {\bibinfo {author} {\bibfnamefont {R.}~\bibnamefont
  {Sweke}}, \bibinfo {author} {\bibfnamefont {J.-P.}\ \bibnamefont {Seifert}},
  \bibinfo {author} {\bibfnamefont {D.}~\bibnamefont {Hangleiter}},\ and\
  \bibinfo {author} {\bibfnamefont {J.}~\bibnamefont {Eisert}},\ }\bibfield
  {title} {\bibinfo {title} {On the {Q}uantum versus {C}lassical {L}earnability
  of {D}iscrete {D}istributions},\ }\href
  {https://doi.org/10.22331/q-2021-03-23-417} {\bibfield  {journal} {\bibinfo
  {journal} {{Quantum}}\ }\textbf {\bibinfo {volume} {5}},\ \bibinfo {pages}
  {417} (\bibinfo {year} {2021})}\BibitemShut {NoStop}%
\bibitem [{\citenamefont {Gao}\ \emph {et~al.}(2021)\citenamefont {Gao},
  \citenamefont {Anschuetz}, \citenamefont {Wang}, \citenamefont {Cirac},\ and\
  \citenamefont {Lukin}}]{gao_enhancing_2021}%
  \BibitemOpen
  \bibfield  {author} {\bibinfo {author} {\bibfnamefont {X.}~\bibnamefont
  {Gao}}, \bibinfo {author} {\bibfnamefont {E.~R.}\ \bibnamefont {Anschuetz}},
  \bibinfo {author} {\bibfnamefont {S.-T.}\ \bibnamefont {Wang}}, \bibinfo
  {author} {\bibfnamefont {J.~I.}\ \bibnamefont {Cirac}},\ and\ \bibinfo
  {author} {\bibfnamefont {M.~D.}\ \bibnamefont {Lukin}},\ }\href@noop {}
  {\bibinfo {title} {Enhancing {Generative} {Models} via {Quantum}
  {Correlations}}} (\bibinfo {year} {2021}),\ \Eprint
  {https://arxiv.org/abs/2101.08354} {arXiv:2101.08354 [quant-ph]} \BibitemShut
  {NoStop}%
\bibitem [{\citenamefont {Ristè}\ \emph {et~al.}(2017)\citenamefont {Ristè},
  \citenamefont {da~Silva}, \citenamefont {Ryan}, \citenamefont {Cross},
  \citenamefont {Córcoles}, \citenamefont {Smolin}, \citenamefont {Gambetta},
  \citenamefont {Chow},\ and\ \citenamefont
  {Johnson}}]{riste_demonstration_2017}%
  \BibitemOpen
  \bibfield  {author} {\bibinfo {author} {\bibfnamefont {D.}~\bibnamefont
  {Ristè}}, \bibinfo {author} {\bibfnamefont {M.~P.}\ \bibnamefont
  {da~Silva}}, \bibinfo {author} {\bibfnamefont {C.~A.}\ \bibnamefont {Ryan}},
  \bibinfo {author} {\bibfnamefont {A.~W.}\ \bibnamefont {Cross}}, \bibinfo
  {author} {\bibfnamefont {A.~D.}\ \bibnamefont {Córcoles}}, \bibinfo {author}
  {\bibfnamefont {J.~A.}\ \bibnamefont {Smolin}}, \bibinfo {author}
  {\bibfnamefont {J.~M.}\ \bibnamefont {Gambetta}}, \bibinfo {author}
  {\bibfnamefont {J.~M.}\ \bibnamefont {Chow}},\ and\ \bibinfo {author}
  {\bibfnamefont {B.~R.}\ \bibnamefont {Johnson}},\ }\bibfield  {title}
  {\bibinfo {title} {Demonstration of quantum advantage in machine learning},\
  }\href {https://doi.org/10.1038/s41534-017-0017-3} {\bibfield  {journal}
  {\bibinfo  {journal} {npj Quantum Information}\ }\textbf {\bibinfo {volume}
  {3}},\ \bibinfo {pages} {1} (\bibinfo {year} {2017})}\BibitemShut {NoStop}%
\bibitem [{\citenamefont {Coyle}\ \emph {et~al.}(2021)\citenamefont {Coyle},
  \citenamefont {Henderson}, \citenamefont {Le}, \citenamefont {Kumar},
  \citenamefont {Paini},\ and\ \citenamefont {Kashefi}}]{coyle_quantum_2020}%
  \BibitemOpen
  \bibfield  {author} {\bibinfo {author} {\bibfnamefont {B.}~\bibnamefont
  {Coyle}}, \bibinfo {author} {\bibfnamefont {M.}~\bibnamefont {Henderson}},
  \bibinfo {author} {\bibfnamefont {J.~C.~J.}\ \bibnamefont {Le}}, \bibinfo
  {author} {\bibfnamefont {N.}~\bibnamefont {Kumar}}, \bibinfo {author}
  {\bibfnamefont {M.}~\bibnamefont {Paini}},\ and\ \bibinfo {author}
  {\bibfnamefont {E.}~\bibnamefont {Kashefi}},\ }\bibfield  {title} {\bibinfo
  {title} {Quantum versus classical generative modelling in finance},\ }\href
  {https://doi.org/10.1088/2058-9565/abd3db} {\bibfield  {journal} {\bibinfo
  {journal} {Quantum Science and Technology}\ }\textbf {\bibinfo {volume}
  {6}},\ \bibinfo {pages} {024013} (\bibinfo {year} {2021})}\BibitemShut
  {NoStop}%
\bibitem [{\citenamefont {Alcazar}\ \emph {et~al.}(2020)\citenamefont
  {Alcazar}, \citenamefont {Leyton-Ortega},\ and\ \citenamefont
  {Perdomo-Ortiz}}]{alcazar_classical_2020}%
  \BibitemOpen
  \bibfield  {author} {\bibinfo {author} {\bibfnamefont {J.}~\bibnamefont
  {Alcazar}}, \bibinfo {author} {\bibfnamefont {V.}~\bibnamefont
  {Leyton-Ortega}},\ and\ \bibinfo {author} {\bibfnamefont {A.}~\bibnamefont
  {Perdomo-Ortiz}},\ }\bibfield  {title} {\bibinfo {title} {Classical versus
  quantum models in machine learning: insights from a finance application},\
  }\href {https://doi.org/10.1088/2632-2153/ab9009} {\bibfield  {journal}
  {\bibinfo  {journal} {Machine Learning: Science and Technology}\ }\textbf
  {\bibinfo {volume} {1}},\ \bibinfo {pages} {035003} (\bibinfo {year}
  {2020})}\BibitemShut {NoStop}%
\bibitem [{\citenamefont {Johri}\ \emph {et~al.}(2020)\citenamefont {Johri},
  \citenamefont {Debnath}, \citenamefont {Mocherla}, \citenamefont {Singh},
  \citenamefont {Prakash}, \citenamefont {Kim},\ and\ \citenamefont
  {Kerenidis}}]{johri_nearest_2020}%
  \BibitemOpen
  \bibfield  {author} {\bibinfo {author} {\bibfnamefont {S.}~\bibnamefont
  {Johri}}, \bibinfo {author} {\bibfnamefont {S.}~\bibnamefont {Debnath}},
  \bibinfo {author} {\bibfnamefont {A.}~\bibnamefont {Mocherla}}, \bibinfo
  {author} {\bibfnamefont {A.}~\bibnamefont {Singh}}, \bibinfo {author}
  {\bibfnamefont {A.}~\bibnamefont {Prakash}}, \bibinfo {author} {\bibfnamefont
  {J.}~\bibnamefont {Kim}},\ and\ \bibinfo {author} {\bibfnamefont
  {I.}~\bibnamefont {Kerenidis}},\ }\href@noop {} {\bibinfo {title} {Nearest
  centroid classification on a trapped ion quantum computer}} (\bibinfo {year}
  {2020}),\ \Eprint {https://arxiv.org/abs/2012.04145} {arXiv:2012.04145
  [quant-ph]} \BibitemShut {NoStop}%
\bibitem [{\citenamefont {Ciliberto}\ \emph {et~al.}(2018)\citenamefont
  {Ciliberto}, \citenamefont {Herbster}, \citenamefont {Ialongo}, \citenamefont
  {Pontil}, \citenamefont {Rocchetto}, \citenamefont {Severini},\ and\
  \citenamefont {Wossnig}}]{ciliberto_quantum_2018}%
  \BibitemOpen
  \bibfield  {author} {\bibinfo {author} {\bibfnamefont {C.}~\bibnamefont
  {Ciliberto}}, \bibinfo {author} {\bibfnamefont {M.}~\bibnamefont {Herbster}},
  \bibinfo {author} {\bibfnamefont {A.~D.}\ \bibnamefont {Ialongo}}, \bibinfo
  {author} {\bibfnamefont {M.}~\bibnamefont {Pontil}}, \bibinfo {author}
  {\bibfnamefont {A.}~\bibnamefont {Rocchetto}}, \bibinfo {author}
  {\bibfnamefont {S.}~\bibnamefont {Severini}},\ and\ \bibinfo {author}
  {\bibfnamefont {L.}~\bibnamefont {Wossnig}},\ }\bibfield  {title} {\bibinfo
  {title} {Quantum machine learning: a classical perspective},\ }\href
  {https://doi.org/10.1098/rspa.2017.0551} {\bibfield  {journal} {\bibinfo
  {journal} {Proceedings of the Royal Society A: Mathematical, Physical and
  Engineering Sciences}\ }\textbf {\bibinfo {volume} {474}},\ \bibinfo {pages}
  {20170551} (\bibinfo {year} {2018})}\BibitemShut {NoStop}%
\bibitem [{\citenamefont {Benedetti}\ \emph
  {et~al.}(2019{\natexlab{a}})\citenamefont {Benedetti}, \citenamefont {Lloyd},
  \citenamefont {Sack},\ and\ \citenamefont
  {Fiorentini}}]{benedetti_parameterized_2019}%
  \BibitemOpen
  \bibfield  {author} {\bibinfo {author} {\bibfnamefont {M.}~\bibnamefont
  {Benedetti}}, \bibinfo {author} {\bibfnamefont {E.}~\bibnamefont {Lloyd}},
  \bibinfo {author} {\bibfnamefont {S.}~\bibnamefont {Sack}},\ and\ \bibinfo
  {author} {\bibfnamefont {M.}~\bibnamefont {Fiorentini}},\ }\bibfield  {title}
  {\bibinfo {title} {Parameterized quantum circuits as machine learning
  models},\ }\href {https://doi.org/10.1088/2058-9565/ab4eb5} {\bibfield
  {journal} {\bibinfo  {journal} {Quantum Science and Technology}\ }\textbf
  {\bibinfo {volume} {4}},\ \bibinfo {pages} {043001} (\bibinfo {year}
  {2019}{\natexlab{a}})}\BibitemShut {NoStop}%
\bibitem [{\citenamefont {Lamata}(2020)}]{lamata_quantum_2020}%
  \BibitemOpen
  \bibfield  {author} {\bibinfo {author} {\bibfnamefont {L.}~\bibnamefont
  {Lamata}},\ }\bibfield  {title} {\bibinfo {title} {Quantum machine learning
  and quantum biomimetics: {A} perspective},\ }\href
  {https://doi.org/10.1088/2632-2153/ab9803} {\bibfield  {journal} {\bibinfo
  {journal} {Machine Learning: Science and Technology}\ }\textbf {\bibinfo
  {volume} {1}},\ \bibinfo {pages} {033002} (\bibinfo {year}
  {2020})}\BibitemShut {NoStop}%
\bibitem [{\citenamefont {Lauritzen}\ and\ \citenamefont
  {Spiegelhalter}(1988)}]{lauritzen_local_1988}%
  \BibitemOpen
  \bibfield  {author} {\bibinfo {author} {\bibfnamefont {S.~L.}\ \bibnamefont
  {Lauritzen}}\ and\ \bibinfo {author} {\bibfnamefont {D.~J.}\ \bibnamefont
  {Spiegelhalter}},\ }\bibfield  {title} {\bibinfo {title} {Local
  {Computations} with {Probabilities} on {Graphical} {Structures} and {Their}
  {Application} to {Expert} {Systems}},\ }\href
  {https://www.jstor.org/stable/2345762} {\bibfield  {journal} {\bibinfo
  {journal} {Journal of the Royal Statistical Society. Series B
  (Methodological)}\ }\textbf {\bibinfo {volume} {50}},\ \bibinfo {pages} {157}
  (\bibinfo {year} {1988})}\BibitemShut {NoStop}%
\bibitem [{\citenamefont {Kritzman}\ \emph {et~al.}(2012)\citenamefont
  {Kritzman}, \citenamefont {Page},\ and\ \citenamefont
  {Turkington}}]{kritzmanRegimeShiftsImplications2012}%
  \BibitemOpen
  \bibfield  {author} {\bibinfo {author} {\bibfnamefont {M.}~\bibnamefont
  {Kritzman}}, \bibinfo {author} {\bibfnamefont {S.}~\bibnamefont {Page}},\
  and\ \bibinfo {author} {\bibfnamefont {D.}~\bibnamefont {Turkington}},\
  }\bibfield  {title} {\bibinfo {title} {Regime {{Shifts}}: {{Implications}}
  for {{Dynamic Strategies}} (corrected)},\ }\href
  {https://doi.org/10.2469/faj.v68.n3.3} {\bibfield  {journal} {\bibinfo
  {journal} {Financial Analysts Journal}\ }\textbf {\bibinfo {volume} {68}},\
  \bibinfo {pages} {22} (\bibinfo {year} {2012})}\BibitemShut {NoStop}%
\bibitem [{\citenamefont {Obermeyer}\ \emph {et~al.}(2019)\citenamefont
  {Obermeyer}, \citenamefont {Powers}, \citenamefont {Vogeli},\ and\
  \citenamefont {Mullainathan}}]{obermeyerDissectingRacialBias2019}%
  \BibitemOpen
  \bibfield  {author} {\bibinfo {author} {\bibfnamefont {Z.}~\bibnamefont
  {Obermeyer}}, \bibinfo {author} {\bibfnamefont {B.}~\bibnamefont {Powers}},
  \bibinfo {author} {\bibfnamefont {C.}~\bibnamefont {Vogeli}},\ and\ \bibinfo
  {author} {\bibfnamefont {S.}~\bibnamefont {Mullainathan}},\ }\bibfield
  {title} {\bibinfo {title} {Dissecting racial bias in an algorithm used to
  manage the health of populations},\ }\href
  {https://doi.org/10.1126/science.aax2342} {\bibfield  {journal} {\bibinfo
  {journal} {Science}\ }\textbf {\bibinfo {volume} {366}},\ \bibinfo {pages}
  {447} (\bibinfo {year} {2019})}\BibitemShut {NoStop}%
\bibitem [{\citenamefont {Pearl}(2009)}]{pearlCausalityModelsReasoning2009}%
  \BibitemOpen
  \bibfield  {author} {\bibinfo {author} {\bibfnamefont {J.}~\bibnamefont
  {Pearl}},\ }\href@noop {} {\emph {\bibinfo {title} {Causality: {{Models}},
  {{Reasoning}} and {{Inference}}}}},\ \bibinfo {edition} {2nd}\ ed.\ (\bibinfo
   {publisher} {{Cambridge University Press}},\ \bibinfo {address} {{New York,
  NY, USA}},\ \bibinfo {year} {2009})\BibitemShut {NoStop}%
\bibitem [{\citenamefont {McEliece}\ \emph {et~al.}(1998)\citenamefont
  {McEliece}, \citenamefont {MacKay},\ and\ \citenamefont {{Jung-Fu
  Cheng}}}]{mcelieceTurboDecodingInstance1998}%
  \BibitemOpen
  \bibfield  {author} {\bibinfo {author} {\bibfnamefont {R.}~\bibnamefont
  {McEliece}}, \bibinfo {author} {\bibfnamefont {D.}~\bibnamefont {MacKay}},\
  and\ \bibinfo {author} {\bibnamefont {{Jung-Fu Cheng}}},\ }\bibfield  {title}
  {\bibinfo {title} {Turbo decoding as an instance of {{Pearl}}'s ``belief
  propagation'' algorithm},\ }\href {https://doi.org/10.1109/49.661103}
  {\bibfield  {journal} {\bibinfo  {journal} {IEEE Journal on Selected Areas in
  Communications}\ }\textbf {\bibinfo {volume} {16}},\ \bibinfo {pages} {140}
  (\bibinfo {year} {1998})}\BibitemShut {NoStop}%
\bibitem [{\citenamefont {{Dan Roth}}(1996)}]{dan_roth_hardness_1996}%
  \BibitemOpen
  \bibfield  {author} {\bibinfo {author} {\bibnamefont {{Dan Roth}}},\
  }\bibfield  {title} {\bibinfo {title} {On the hardness of approximate
  reasoning},\ }\href {https://doi.org/10.1016/0004-3702(94)00092-1} {\bibfield
   {journal} {\bibinfo  {journal} {Artificial Intelligence}\ }\textbf {\bibinfo
  {volume} {82}},\ \bibinfo {pages} {273} (\bibinfo {year} {1996})}\BibitemShut
  {NoStop}%
\bibitem [{\citenamefont {Cooper}(1990)}]{cooper1990computational}%
  \BibitemOpen
  \bibfield  {author} {\bibinfo {author} {\bibfnamefont {G.~F.}\ \bibnamefont
  {Cooper}},\ }\bibfield  {title} {\bibinfo {title} {The computational
  complexity of probabilistic inference using bayesian belief networks},\
  }\href {https://doi.org/https://doi.org/10.1016/0004-3702(90)90060-D}
  {\bibfield  {journal} {\bibinfo  {journal} {Artificial Intelligence}\
  }\textbf {\bibinfo {volume} {42}},\ \bibinfo {pages} {393 } (\bibinfo {year}
  {1990})}\BibitemShut {NoStop}%
\bibitem [{\citenamefont {Dagum}\ and\ \citenamefont
  {Luby}(1993)}]{dagum1993approximating}%
  \BibitemOpen
  \bibfield  {author} {\bibinfo {author} {\bibfnamefont {P.}~\bibnamefont
  {Dagum}}\ and\ \bibinfo {author} {\bibfnamefont {M.}~\bibnamefont {Luby}},\
  }\bibfield  {title} {\bibinfo {title} {Approximating probabilistic inference
  in {Bayesian} belief networks is {NP-hard}},\ }\href
  {https://doi.org/https://doi.org/10.1016/0004-3702(93)90036-B} {\bibfield
  {journal} {\bibinfo  {journal} {Artificial Intelligence}\ }\textbf {\bibinfo
  {volume} {60}},\ \bibinfo {pages} {141 } (\bibinfo {year}
  {1993})}\BibitemShut {NoStop}%
\bibitem [{\citenamefont {Ranganath}\ \emph {et~al.}(2013)\citenamefont
  {Ranganath}, \citenamefont {Gerrish},\ and\ \citenamefont
  {Blei}}]{ranganath2013black}%
  \BibitemOpen
  \bibfield  {author} {\bibinfo {author} {\bibfnamefont {R.}~\bibnamefont
  {Ranganath}}, \bibinfo {author} {\bibfnamefont {S.}~\bibnamefont {Gerrish}},\
  and\ \bibinfo {author} {\bibfnamefont {D.~M.}\ \bibnamefont {Blei}},\
  }\href@noop {} {\bibinfo {title} {Black box variational inference}} (\bibinfo
  {year} {2013}),\ \Eprint {https://arxiv.org/abs/1401.0118} {arXiv:1401.0118
  [stat.ML]} \BibitemShut {NoStop}%
\bibitem [{\citenamefont {Kingma}\ and\ \citenamefont
  {Welling}(2014)}]{kingma2014autoencoding}%
  \BibitemOpen
  \bibfield  {author} {\bibinfo {author} {\bibfnamefont {D.~P.}\ \bibnamefont
  {Kingma}}\ and\ \bibinfo {author} {\bibfnamefont {M.}~\bibnamefont
  {Welling}},\ }\href@noop {} {\bibinfo {title} {Auto-encoding variational
  {Bayes}}} (\bibinfo {year} {2014}),\ \Eprint
  {https://arxiv.org/abs/1312.6114} {arXiv:1312.6114 [stat.ML]} \BibitemShut
  {NoStop}%
\bibitem [{\citenamefont {Wingate}\ and\ \citenamefont
  {Weber}(2013)}]{wingate2013automated}%
  \BibitemOpen
  \bibfield  {author} {\bibinfo {author} {\bibfnamefont {D.}~\bibnamefont
  {Wingate}}\ and\ \bibinfo {author} {\bibfnamefont {T.}~\bibnamefont
  {Weber}},\ }\href@noop {} {\bibinfo {title} {Automated variational inference
  in probabilistic programming}} (\bibinfo {year} {2013}),\ \Eprint
  {https://arxiv.org/abs/1301.1299} {arXiv:1301.1299 [stat.ML]} \BibitemShut
  {NoStop}%
\bibitem [{\citenamefont {Gershman}\ \emph {et~al.}(2012)\citenamefont
  {Gershman}, \citenamefont {Hoffman},\ and\ \citenamefont
  {Blei}}]{gershman2012nonparametric}%
  \BibitemOpen
  \bibfield  {author} {\bibinfo {author} {\bibfnamefont {S.}~\bibnamefont
  {Gershman}}, \bibinfo {author} {\bibfnamefont {M.}~\bibnamefont {Hoffman}},\
  and\ \bibinfo {author} {\bibfnamefont {D.}~\bibnamefont {Blei}},\ }\href@noop
  {} {\bibinfo {title} {Nonparametric variational inference}} (\bibinfo {year}
  {2012}),\ \Eprint {https://arxiv.org/abs/1206.4665} {arXiv:1206.4665 [cs.LG]}
  \BibitemShut {NoStop}%
\bibitem [{\citenamefont {Rezende}\ and\ \citenamefont
  {Mohamed}(2015)}]{rezende2016variational}%
  \BibitemOpen
  \bibfield  {author} {\bibinfo {author} {\bibfnamefont {D.~J.}\ \bibnamefont
  {Rezende}}\ and\ \bibinfo {author} {\bibfnamefont {S.}~\bibnamefont
  {Mohamed}},\ }\bibfield  {title} {\bibinfo {title} {Variational {{Inference}}
  with {{Normalizing Flows}}},\ }in\ \href@noop {} {\emph {\bibinfo {booktitle}
  {Proceedings of the 32nd {{International Conference}} on {{Machine
  Learning}}}}},\ Vol.~\bibinfo {volume} {37}\ (\bibinfo  {publisher}
  {{PMLR}},\ \bibinfo {year} {2015})\ pp.\ \bibinfo {pages} {1530--1538},\
  \Eprint {https://arxiv.org/abs/1505.05770} {arXiv:1505.05770 [stat.ML]}
  \BibitemShut {NoStop}%
\bibitem [{\citenamefont {Williams}(1992)}]{williams1992simple}%
  \BibitemOpen
  \bibfield  {author} {\bibinfo {author} {\bibfnamefont {R.~J.}\ \bibnamefont
  {Williams}},\ }\bibfield  {title} {\bibinfo {title} {Simple statistical
  gradient-following algorithms for connectionist reinforcement learning},\
  }\href {https://doi.org/10.1007/BF00992696} {\bibfield  {journal} {\bibinfo
  {journal} {Machine learning}\ }\textbf {\bibinfo {volume} {8}},\ \bibinfo
  {pages} {229} (\bibinfo {year} {1992})}\BibitemShut {NoStop}%
\bibitem [{\citenamefont {Maddison}\ \emph {et~al.}(2017)\citenamefont
  {Maddison}, \citenamefont {Mnih},\ and\ \citenamefont
  {Teh}}]{maddison2017concrete}%
  \BibitemOpen
  \bibfield  {author} {\bibinfo {author} {\bibfnamefont {C.~J.}\ \bibnamefont
  {Maddison}}, \bibinfo {author} {\bibfnamefont {A.}~\bibnamefont {Mnih}},\
  and\ \bibinfo {author} {\bibfnamefont {Y.~W.}\ \bibnamefont {Teh}},\
  }\href@noop {} {\bibinfo {title} {The concrete distribution: A continuous
  relaxation of discrete random variables}} (\bibinfo {year} {2017}),\ \Eprint
  {https://arxiv.org/abs/1611.00712} {arXiv:1611.00712 [cs.LG]} \BibitemShut
  {NoStop}%
\bibitem [{\citenamefont {Jang}\ \emph {et~al.}(2017)\citenamefont {Jang},
  \citenamefont {Gu},\ and\ \citenamefont {Poole}}]{jang2017categorical}%
  \BibitemOpen
  \bibfield  {author} {\bibinfo {author} {\bibfnamefont {E.}~\bibnamefont
  {Jang}}, \bibinfo {author} {\bibfnamefont {S.}~\bibnamefont {Gu}},\ and\
  \bibinfo {author} {\bibfnamefont {B.}~\bibnamefont {Poole}},\ }\href@noop {}
  {\bibinfo {title} {Categorical reparameterization with gumbel-softmax}}
  (\bibinfo {year} {2017}),\ \Eprint {https://arxiv.org/abs/1611.01144}
  {arXiv:1611.01144 [stat.ML]} \BibitemShut {NoStop}%
\bibitem [{\citenamefont {Cheng}\ \emph {et~al.}(2018)\citenamefont {Cheng},
  \citenamefont {Chen},\ and\ \citenamefont {Wang}}]{Cheng_2018}%
  \BibitemOpen
  \bibfield  {author} {\bibinfo {author} {\bibfnamefont {S.}~\bibnamefont
  {Cheng}}, \bibinfo {author} {\bibfnamefont {J.}~\bibnamefont {Chen}},\ and\
  \bibinfo {author} {\bibfnamefont {L.}~\bibnamefont {Wang}},\ }\bibfield
  {title} {\bibinfo {title} {Information perspective to probabilistic modeling:
  {Boltzmann} machines versus {Born} machines},\ }\href
  {https://doi.org/10.3390/e20080583} {\bibfield  {journal} {\bibinfo
  {journal} {Entropy}\ }\textbf {\bibinfo {volume} {20}},\ \bibinfo {pages}
  {583} (\bibinfo {year} {2018})}\BibitemShut {NoStop}%
\bibitem [{\citenamefont {Benedetti}\ \emph
  {et~al.}(2019{\natexlab{b}})\citenamefont {Benedetti}, \citenamefont
  {Garcia-Pintos}, \citenamefont {Perdomo}, \citenamefont {Leyton-Ortega},
  \citenamefont {Nam},\ and\ \citenamefont {Perdomo-Ortiz}}]{Benedetti_2019}%
  \BibitemOpen
  \bibfield  {author} {\bibinfo {author} {\bibfnamefont {M.}~\bibnamefont
  {Benedetti}}, \bibinfo {author} {\bibfnamefont {D.}~\bibnamefont
  {Garcia-Pintos}}, \bibinfo {author} {\bibfnamefont {O.}~\bibnamefont
  {Perdomo}}, \bibinfo {author} {\bibfnamefont {V.}~\bibnamefont
  {Leyton-Ortega}}, \bibinfo {author} {\bibfnamefont {Y.}~\bibnamefont {Nam}},\
  and\ \bibinfo {author} {\bibfnamefont {A.}~\bibnamefont {Perdomo-Ortiz}},\
  }\bibfield  {title} {\bibinfo {title} {A generative modeling approach for
  benchmarking and training shallow quantum circuits},\ }\href
  {https://doi.org/10.1038/s41534-019-0157-8} {\bibfield  {journal} {\bibinfo
  {journal} {npj Quantum Information}\ }\textbf {\bibinfo {volume} {5}}
  (\bibinfo {year} {2019}{\natexlab{b}})}\BibitemShut {NoStop}%
\bibitem [{\citenamefont {Hamilton}\ \emph {et~al.}(2019)\citenamefont
  {Hamilton}, \citenamefont {Dumitrescu},\ and\ \citenamefont
  {Pooser}}]{Hamilton_2019}%
  \BibitemOpen
  \bibfield  {author} {\bibinfo {author} {\bibfnamefont {K.~E.}\ \bibnamefont
  {Hamilton}}, \bibinfo {author} {\bibfnamefont {E.~F.}\ \bibnamefont
  {Dumitrescu}},\ and\ \bibinfo {author} {\bibfnamefont {R.~C.}\ \bibnamefont
  {Pooser}},\ }\bibfield  {title} {\bibinfo {title} {Generative model
  benchmarks for superconducting qubits},\ }\href
  {https://doi.org/10.1103/PhysRevA.99.062323} {\bibfield  {journal} {\bibinfo
  {journal} {Phys. Rev. A}\ }\textbf {\bibinfo {volume} {99}},\ \bibinfo
  {pages} {062323} (\bibinfo {year} {2019})}\BibitemShut {NoStop}%
\bibitem [{\citenamefont {{Leyton-Ortega}}\ \emph {et~al.}(2021)\citenamefont
  {{Leyton-Ortega}}, \citenamefont {{Perdomo-Ortiz}},\ and\ \citenamefont
  {Perdomo}}]{leytonortega2019robust}%
  \BibitemOpen
  \bibfield  {author} {\bibinfo {author} {\bibfnamefont {V.}~\bibnamefont
  {{Leyton-Ortega}}}, \bibinfo {author} {\bibfnamefont {A.}~\bibnamefont
  {{Perdomo-Ortiz}}},\ and\ \bibinfo {author} {\bibfnamefont {O.}~\bibnamefont
  {Perdomo}},\ }\bibfield  {title} {\bibinfo {title} {Robust implementation of
  generative modeling with parametrized quantum circuits},\ }\href
  {https://doi.org/10.1007/s42484-021-00040-2} {\bibfield  {journal} {\bibinfo
  {journal} {Quantum Machine Intelligence}\ }\textbf {\bibinfo {volume} {3}},\
  \bibinfo {pages} {17} (\bibinfo {year} {2021})}\BibitemShut {NoStop}%
\bibitem [{\citenamefont {Zhu}\ \emph {et~al.}(2019)\citenamefont {Zhu},
  \citenamefont {Linke}, \citenamefont {Benedetti}, \citenamefont {Landsman},
  \citenamefont {Nguyen}, \citenamefont {Alderete}, \citenamefont
  {Perdomo-Ortiz}, \citenamefont {Korda}, \citenamefont {Garfoot},
  \citenamefont {Brecque}, \citenamefont {Egan}, \citenamefont {Perdomo},\ and\
  \citenamefont {Monroe}}]{Zhu_2019}%
  \BibitemOpen
  \bibfield  {author} {\bibinfo {author} {\bibfnamefont {D.}~\bibnamefont
  {Zhu}}, \bibinfo {author} {\bibfnamefont {N.~M.}\ \bibnamefont {Linke}},
  \bibinfo {author} {\bibfnamefont {M.}~\bibnamefont {Benedetti}}, \bibinfo
  {author} {\bibfnamefont {K.~A.}\ \bibnamefont {Landsman}}, \bibinfo {author}
  {\bibfnamefont {N.~H.}\ \bibnamefont {Nguyen}}, \bibinfo {author}
  {\bibfnamefont {C.~H.}\ \bibnamefont {Alderete}}, \bibinfo {author}
  {\bibfnamefont {A.}~\bibnamefont {Perdomo-Ortiz}}, \bibinfo {author}
  {\bibfnamefont {N.}~\bibnamefont {Korda}}, \bibinfo {author} {\bibfnamefont
  {A.}~\bibnamefont {Garfoot}}, \bibinfo {author} {\bibfnamefont
  {C.}~\bibnamefont {Brecque}}, \bibinfo {author} {\bibfnamefont
  {L.}~\bibnamefont {Egan}}, \bibinfo {author} {\bibfnamefont {O.}~\bibnamefont
  {Perdomo}},\ and\ \bibinfo {author} {\bibfnamefont {C.}~\bibnamefont
  {Monroe}},\ }\bibfield  {title} {\bibinfo {title} {Training of quantum
  circuits on a hybrid quantum computer},\ }\href
  {https://doi.org/10.1126/sciadv.aaw9918} {\bibfield  {journal} {\bibinfo
  {journal} {Science Advances}\ }\textbf {\bibinfo {volume} {5}} (\bibinfo
  {year} {2019})}\BibitemShut {NoStop}%
\bibitem [{\citenamefont {Han}\ \emph {et~al.}(2018)\citenamefont {Han},
  \citenamefont {Wang}, \citenamefont {Fan}, \citenamefont {Wang},\ and\
  \citenamefont {Zhang}}]{Han_2018}%
  \BibitemOpen
  \bibfield  {author} {\bibinfo {author} {\bibfnamefont {Z.-Y.}\ \bibnamefont
  {Han}}, \bibinfo {author} {\bibfnamefont {J.}~\bibnamefont {Wang}}, \bibinfo
  {author} {\bibfnamefont {H.}~\bibnamefont {Fan}}, \bibinfo {author}
  {\bibfnamefont {L.}~\bibnamefont {Wang}},\ and\ \bibinfo {author}
  {\bibfnamefont {P.}~\bibnamefont {Zhang}},\ }\bibfield  {title} {\bibinfo
  {title} {Unsupervised generative modeling using matrix product states},\
  }\href {https://doi.org/10.1103/PhysRevX.8.031012} {\bibfield  {journal}
  {\bibinfo  {journal} {Phys. Rev. X}\ }\textbf {\bibinfo {volume} {8}},\
  \bibinfo {pages} {031012} (\bibinfo {year} {2018})}\BibitemShut {NoStop}%
\bibitem [{\citenamefont {Rudolph}\ \emph {et~al.}(2020)\citenamefont
  {Rudolph}, \citenamefont {Toussaint}, \citenamefont {Katabarwa},
  \citenamefont {Johri}, \citenamefont {Peropadre},\ and\ \citenamefont
  {Perdomo-Ortiz}}]{rudolph2020generation}%
  \BibitemOpen
  \bibfield  {author} {\bibinfo {author} {\bibfnamefont {M.~S.}\ \bibnamefont
  {Rudolph}}, \bibinfo {author} {\bibfnamefont {N.~B.}\ \bibnamefont
  {Toussaint}}, \bibinfo {author} {\bibfnamefont {A.}~\bibnamefont
  {Katabarwa}}, \bibinfo {author} {\bibfnamefont {S.}~\bibnamefont {Johri}},
  \bibinfo {author} {\bibfnamefont {B.}~\bibnamefont {Peropadre}},\ and\
  \bibinfo {author} {\bibfnamefont {A.}~\bibnamefont {Perdomo-Ortiz}},\
  }\href@noop {} {\bibinfo {title} {Generation of high-resolution handwritten
  digits with an ion-trap quantum computer}} (\bibinfo {year} {2020}),\ \Eprint
  {https://arxiv.org/abs/2012.03924} {arXiv:2012.03924 [quant-ph]} \BibitemShut
  {NoStop}%
\bibitem [{\citenamefont {Čepaitė}\ \emph {et~al.}(2020)\citenamefont
  {Čepaitė}, \citenamefont {Coyle},\ and\ \citenamefont
  {Kashefi}}]{cepaite_continuous_2020}%
  \BibitemOpen
  \bibfield  {author} {\bibinfo {author} {\bibfnamefont {I.}~\bibnamefont
  {Čepaitė}}, \bibinfo {author} {\bibfnamefont {B.}~\bibnamefont {Coyle}},\
  and\ \bibinfo {author} {\bibfnamefont {E.}~\bibnamefont {Kashefi}},\
  }\href@noop {} {\bibinfo {title} {A continuous variable {Born} machine}}
  (\bibinfo {year} {2020}),\ \Eprint {https://arxiv.org/abs/2011.00904}
  {arXiv:2011.00904 [quant-ph]} \BibitemShut {NoStop}%
\bibitem [{\citenamefont {Zoufal}\ \emph {et~al.}(2019)\citenamefont {Zoufal},
  \citenamefont {Lucchi},\ and\ \citenamefont {Woerner}}]{Zoufal_2019}%
  \BibitemOpen
  \bibfield  {author} {\bibinfo {author} {\bibfnamefont {C.}~\bibnamefont
  {Zoufal}}, \bibinfo {author} {\bibfnamefont {A.}~\bibnamefont {Lucchi}},\
  and\ \bibinfo {author} {\bibfnamefont {S.}~\bibnamefont {Woerner}},\
  }\bibfield  {title} {\bibinfo {title} {Quantum generative adversarial
  networks for learning and loading random distributions},\ }\href
  {https://doi.org/10.1038/s41534-019-0223-2} {\bibfield  {journal} {\bibinfo
  {journal} {npj Quantum Information}\ }\textbf {\bibinfo {volume} {5}}
  (\bibinfo {year} {2019})}\BibitemShut {NoStop}%
\bibitem [{\citenamefont {Herr}\ \emph {et~al.}(2021)\citenamefont {Herr},
  \citenamefont {Obert},\ and\ \citenamefont {Rosenkranz}}]{herr2020anomaly}%
  \BibitemOpen
  \bibfield  {author} {\bibinfo {author} {\bibfnamefont {D.}~\bibnamefont
  {Herr}}, \bibinfo {author} {\bibfnamefont {B.}~\bibnamefont {Obert}},\ and\
  \bibinfo {author} {\bibfnamefont {M.}~\bibnamefont {Rosenkranz}},\ }\bibfield
   {title} {\bibinfo {title} {Anomaly detection with variational quantum
  generative adversarial networks},\ }\href
  {https://doi.org/10.1088/2058-9565/ac0d4d} {\bibfield  {journal} {\bibinfo
  {journal} {Quantum Science and Technology}\ }\textbf {\bibinfo {volume}
  {6}},\ \bibinfo {pages} {045004} (\bibinfo {year} {2021})}\BibitemShut
  {NoStop}%
\bibitem [{Note1()}]{Note1}%
  \BibitemOpen
  \bibinfo {note} {Some literature (e.g., Ref.~\cite {Zoufal_2019}) refer to
  Born machines as quantum generative adversarial networks (QGANs) when using
  an adversarial objective function. It is more appropriate to reserve the term
  QGAN to adversarial methods for quantum data \protect \mbox {(e.g.,
  Refs.~\cite {Dallaire_2018, benedetti2019adversarial,
  chakrabarti2019quantum}).}}\BibitemShut {Stop}%
\bibitem [{Note2()}]{Note2}%
  \BibitemOpen
  \bibinfo {note} {A quantum computer cannot efficiently implement arbitrary
  circuits, thus some Born machines are intractable even for a quantum
  computer.}\BibitemShut {Stop}%
\bibitem [{\citenamefont {Bremner}\ \emph {et~al.}(2016)\citenamefont
  {Bremner}, \citenamefont {Montanaro},\ and\ \citenamefont
  {Shepherd}}]{Bremner_2016}%
  \BibitemOpen
  \bibfield  {author} {\bibinfo {author} {\bibfnamefont {M.~J.}\ \bibnamefont
  {Bremner}}, \bibinfo {author} {\bibfnamefont {A.}~\bibnamefont {Montanaro}},\
  and\ \bibinfo {author} {\bibfnamefont {D.~J.}\ \bibnamefont {Shepherd}},\
  }\bibfield  {title} {\bibinfo {title} {Average-case complexity versus
  approximate simulation of commuting quantum computations},\ }\href
  {https://doi.org/10.1103/PhysRevLett.117.080501} {\bibfield  {journal}
  {\bibinfo  {journal} {Phys. Rev. Lett.}\ }\textbf {\bibinfo {volume} {117}},\
  \bibinfo {pages} {080501} (\bibinfo {year} {2016})}\BibitemShut {NoStop}%
\bibitem [{\citenamefont {Aaronson}\ and\ \citenamefont
  {Arkhipov}(2011)}]{Aaronson2011}%
  \BibitemOpen
  \bibfield  {author} {\bibinfo {author} {\bibfnamefont {S.}~\bibnamefont
  {Aaronson}}\ and\ \bibinfo {author} {\bibfnamefont {A.}~\bibnamefont
  {Arkhipov}},\ }\bibfield  {title} {\bibinfo {title} {{The Computational
  Complexity of Linear Optics}},\ }in\ \href
  {https://doi.org/10.1145/1993636.1993682} {\emph {\bibinfo {booktitle}
  {Proceedings of the Forty-Third Annual ACM Symposium on Theory of
  Computing}}},\ \bibinfo {series and number} {STOC '11}\ (\bibinfo
  {publisher} {Association for Computing Machinery},\ \bibinfo {address} {New
  York, NY, USA},\ \bibinfo {year} {2011})\ p.\ \bibinfo {pages}
  {333–342}\BibitemShut {NoStop}%
\bibitem [{\citenamefont {Boixo}\ \emph {et~al.}(2018)\citenamefont {Boixo},
  \citenamefont {Isakov}, \citenamefont {Smelyanskiy}, \citenamefont {Babbush},
  \citenamefont {Ding}, \citenamefont {Jiang}, \citenamefont {Bremner},
  \citenamefont {Martinis},\ and\ \citenamefont {Neven}}]{Boixo2018}%
  \BibitemOpen
  \bibfield  {author} {\bibinfo {author} {\bibfnamefont {S.}~\bibnamefont
  {Boixo}}, \bibinfo {author} {\bibfnamefont {S.~V.}\ \bibnamefont {Isakov}},
  \bibinfo {author} {\bibfnamefont {V.~N.}\ \bibnamefont {Smelyanskiy}},
  \bibinfo {author} {\bibfnamefont {R.}~\bibnamefont {Babbush}}, \bibinfo
  {author} {\bibfnamefont {N.}~\bibnamefont {Ding}}, \bibinfo {author}
  {\bibfnamefont {Z.}~\bibnamefont {Jiang}}, \bibinfo {author} {\bibfnamefont
  {M.~J.}\ \bibnamefont {Bremner}}, \bibinfo {author} {\bibfnamefont {J.~M.}\
  \bibnamefont {Martinis}},\ and\ \bibinfo {author} {\bibfnamefont
  {H.}~\bibnamefont {Neven}},\ }\bibfield  {title} {\bibinfo {title}
  {Characterizing quantum supremacy in near-term devices},\ }\href
  {https://doi.org/10.1038/s41567-018-0124-x} {\bibfield  {journal} {\bibinfo
  {journal} {Nature Physics}\ }\textbf {\bibinfo {volume} {14}},\ \bibinfo
  {pages} {595} (\bibinfo {year} {2018})}\BibitemShut {NoStop}%
\bibitem [{\citenamefont {Bouland}\ \emph {et~al.}(2019)\citenamefont
  {Bouland}, \citenamefont {Fefferman}, \citenamefont {Nirkhe},\ and\
  \citenamefont {Vazirani}}]{Bouland2019}%
  \BibitemOpen
  \bibfield  {author} {\bibinfo {author} {\bibfnamefont {A.}~\bibnamefont
  {Bouland}}, \bibinfo {author} {\bibfnamefont {B.}~\bibnamefont {Fefferman}},
  \bibinfo {author} {\bibfnamefont {C.}~\bibnamefont {Nirkhe}},\ and\ \bibinfo
  {author} {\bibfnamefont {U.}~\bibnamefont {Vazirani}},\ }\bibfield  {title}
  {\bibinfo {title} {On the complexity and verification of quantum random
  circuit sampling},\ }\href {https://doi.org/10.1038/s41567-018-0318-2}
  {\bibfield  {journal} {\bibinfo  {journal} {Nature Physics}\ }\textbf
  {\bibinfo {volume} {15}},\ \bibinfo {pages} {159} (\bibinfo {year}
  {2019})}\BibitemShut {NoStop}%
\bibitem [{\citenamefont {Du}\ \emph {et~al.}(2020)\citenamefont {Du},
  \citenamefont {Hsieh}, \citenamefont {Liu},\ and\ \citenamefont
  {Tao}}]{Du_2020}%
  \BibitemOpen
  \bibfield  {author} {\bibinfo {author} {\bibfnamefont {Y.}~\bibnamefont
  {Du}}, \bibinfo {author} {\bibfnamefont {M.-H.}\ \bibnamefont {Hsieh}},
  \bibinfo {author} {\bibfnamefont {T.}~\bibnamefont {Liu}},\ and\ \bibinfo
  {author} {\bibfnamefont {D.}~\bibnamefont {Tao}},\ }\bibfield  {title}
  {\bibinfo {title} {Expressive power of parametrized quantum circuits},\
  }\href {https://doi.org/10.1103/physrevresearch.2.033125} {\bibfield
  {journal} {\bibinfo  {journal} {Physical Review Research}\ }\textbf {\bibinfo
  {volume} {2}} (\bibinfo {year} {2020})}\BibitemShut {NoStop}%
\bibitem [{\citenamefont {Glasser}\ \emph {et~al.}(2019)\citenamefont
  {Glasser}, \citenamefont {Sweke}, \citenamefont {Pancotti}, \citenamefont
  {Eisert},\ and\ \citenamefont
  {Cirac}}]{glasserExpressivePowerTensornetwork2019}%
  \BibitemOpen
  \bibfield  {author} {\bibinfo {author} {\bibfnamefont {I.}~\bibnamefont
  {Glasser}}, \bibinfo {author} {\bibfnamefont {R.}~\bibnamefont {Sweke}},
  \bibinfo {author} {\bibfnamefont {N.}~\bibnamefont {Pancotti}}, \bibinfo
  {author} {\bibfnamefont {J.}~\bibnamefont {Eisert}},\ and\ \bibinfo {author}
  {\bibfnamefont {I.}~\bibnamefont {Cirac}},\ }\bibfield  {title} {\bibinfo
  {title} {Expressive power of tensor-network factorizations for probabilistic
  modeling},\ }in\ \href@noop {} {\emph {\bibinfo {booktitle} {Advances in
  {{Neural Information Processing Systems}}}}},\ Vol.~\bibinfo {volume} {32},\
  \bibinfo {editor} {edited by\ \bibinfo {editor} {\bibfnamefont
  {H.}~\bibnamefont {Wallach}}, \bibinfo {editor} {\bibfnamefont
  {H.}~\bibnamefont {Larochelle}}, \bibinfo {editor} {\bibfnamefont
  {A.}~\bibnamefont {Beygelzimer}}, \bibinfo {editor} {\bibfnamefont
  {F.}~\bibnamefont {d\textquotesingle {Alch{\'e}-Buc}}}, \bibinfo {editor}
  {\bibfnamefont {E.}~\bibnamefont {Fox}},\ and\ \bibinfo {editor}
  {\bibfnamefont {R.}~\bibnamefont {Garnett}}}\ (\bibinfo  {publisher} {{Curran
  Associates, Inc.}},\ \bibinfo {year} {2019})\ pp.\ \bibinfo {pages}
  {1496--1508},\ \Eprint {https://arxiv.org/abs/1907.03741} {arXiv:1907.03741
  [cs.LG]} \BibitemShut {NoStop}%
\bibitem [{\citenamefont {Mohamed}\ and\ \citenamefont
  {Lakshminarayanan}(2017)}]{mohamed2017learning}%
  \BibitemOpen
  \bibfield  {author} {\bibinfo {author} {\bibfnamefont {S.}~\bibnamefont
  {Mohamed}}\ and\ \bibinfo {author} {\bibfnamefont {B.}~\bibnamefont
  {Lakshminarayanan}},\ }\href@noop {} {\bibinfo {title} {Learning in implicit
  generative models}} (\bibinfo {year} {2017}),\ \Eprint
  {https://arxiv.org/abs/1610.03483} {arXiv:1610.03483 [stat.ML]} \BibitemShut
  {NoStop}%
\bibitem [{\citenamefont {las Cuevas}\ \emph {et~al.}(2011)\citenamefont {las
  Cuevas}, \citenamefont {Dür}, \citenamefont {den Nest},\ and\ \citenamefont
  {Martin-Delgado}}]{De_las_Cuevas_2011}%
  \BibitemOpen
  \bibfield  {author} {\bibinfo {author} {\bibfnamefont {G.~D.}\ \bibnamefont
  {las Cuevas}}, \bibinfo {author} {\bibfnamefont {W.}~\bibnamefont {Dür}},
  \bibinfo {author} {\bibfnamefont {M.~V.}\ \bibnamefont {den Nest}},\ and\
  \bibinfo {author} {\bibfnamefont {M.~A.}\ \bibnamefont {Martin-Delgado}},\
  }\bibfield  {title} {\bibinfo {title} {Quantum algorithms for classical
  lattice models},\ }\href {https://doi.org/10.1088/1367-2630/13/9/093021}
  {\bibfield  {journal} {\bibinfo  {journal} {New Journal of Physics}\ }\textbf
  {\bibinfo {volume} {13}},\ \bibinfo {pages} {093021} (\bibinfo {year}
  {2011})}\BibitemShut {NoStop}%
\bibitem [{\citenamefont {Fujii}\ and\ \citenamefont
  {Morimae}(2017)}]{Fujii_2017}%
  \BibitemOpen
  \bibfield  {author} {\bibinfo {author} {\bibfnamefont {K.}~\bibnamefont
  {Fujii}}\ and\ \bibinfo {author} {\bibfnamefont {T.}~\bibnamefont
  {Morimae}},\ }\bibfield  {title} {\bibinfo {title} {Commuting quantum
  circuits and complexity of {Ising} partition functions},\ }\href
  {https://doi.org/10.1088/1367-2630/aa5fdb} {\bibfield  {journal} {\bibinfo
  {journal} {New Journal of Physics}\ }\textbf {\bibinfo {volume} {19}},\
  \bibinfo {pages} {033003} (\bibinfo {year} {2017})}\BibitemShut {NoStop}%
\bibitem [{Note3()}]{Note3}%
  \BibitemOpen
  \bibinfo {note} {Bias and variance of the estimates need to be analyzed and
  controlled, which is a challenge in itself.}\BibitemShut {Stop}%
\bibitem [{\citenamefont {Liu}\ and\ \citenamefont {Wang}(2018)}]{Liu_2018}%
  \BibitemOpen
  \bibfield  {author} {\bibinfo {author} {\bibfnamefont {J.-G.}\ \bibnamefont
  {Liu}}\ and\ \bibinfo {author} {\bibfnamefont {L.}~\bibnamefont {Wang}},\
  }\bibfield  {title} {\bibinfo {title} {Differentiable learning of quantum
  circuit {Born} machines},\ }\href
  {https://doi.org/10.1103/PhysRevA.98.062324} {\bibfield  {journal} {\bibinfo
  {journal} {Phys. Rev. A}\ }\textbf {\bibinfo {volume} {98}},\ \bibinfo
  {pages} {062324} (\bibinfo {year} {2018})}\BibitemShut {NoStop}%
\bibitem [{\citenamefont {Situ}\ \emph {et~al.}(2020)\citenamefont {Situ},
  \citenamefont {He}, \citenamefont {Wang}, \citenamefont {Li},\ and\
  \citenamefont {Zheng}}]{Situ_2020}%
  \BibitemOpen
  \bibfield  {author} {\bibinfo {author} {\bibfnamefont {H.}~\bibnamefont
  {Situ}}, \bibinfo {author} {\bibfnamefont {Z.}~\bibnamefont {He}}, \bibinfo
  {author} {\bibfnamefont {Y.}~\bibnamefont {Wang}}, \bibinfo {author}
  {\bibfnamefont {L.}~\bibnamefont {Li}},\ and\ \bibinfo {author}
  {\bibfnamefont {S.}~\bibnamefont {Zheng}},\ }\bibfield  {title} {\bibinfo
  {title} {Quantum generative adversarial network for generating discrete
  distribution},\ }\href {https://doi.org/10.1016/j.ins.2020.05.127} {\bibfield
   {journal} {\bibinfo  {journal} {Information Sciences}\ }\textbf {\bibinfo
  {volume} {538}},\ \bibinfo {pages} {193–208} (\bibinfo {year}
  {2020})}\BibitemShut {NoStop}%
\bibitem [{\citenamefont {Zeng}\ \emph {et~al.}(2019)\citenamefont {Zeng},
  \citenamefont {Wu}, \citenamefont {Liu}, \citenamefont {Wang},\ and\
  \citenamefont {Hu}}]{Zeng_2019}%
  \BibitemOpen
  \bibfield  {author} {\bibinfo {author} {\bibfnamefont {J.}~\bibnamefont
  {Zeng}}, \bibinfo {author} {\bibfnamefont {Y.}~\bibnamefont {Wu}}, \bibinfo
  {author} {\bibfnamefont {J.-G.}\ \bibnamefont {Liu}}, \bibinfo {author}
  {\bibfnamefont {L.}~\bibnamefont {Wang}},\ and\ \bibinfo {author}
  {\bibfnamefont {J.}~\bibnamefont {Hu}},\ }\bibfield  {title} {\bibinfo
  {title} {Learning and inference on generative adversarial quantum circuits},\
  }\href {https://doi.org/10.1103/PhysRevA.99.052306} {\bibfield  {journal}
  {\bibinfo  {journal} {Phys. Rev. A}\ }\textbf {\bibinfo {volume} {99}},\
  \bibinfo {pages} {052306} (\bibinfo {year} {2019})}\BibitemShut {NoStop}%
\bibitem [{\citenamefont {Borujeni}\ \emph {et~al.}(2020)\citenamefont
  {Borujeni}, \citenamefont {Nannapaneni}, \citenamefont {Nguyen},
  \citenamefont {Behrman},\ and\ \citenamefont
  {Steck}}]{borujeni_quantum_2020}%
  \BibitemOpen
  \bibfield  {author} {\bibinfo {author} {\bibfnamefont {S.~E.}\ \bibnamefont
  {Borujeni}}, \bibinfo {author} {\bibfnamefont {S.}~\bibnamefont
  {Nannapaneni}}, \bibinfo {author} {\bibfnamefont {N.~H.}\ \bibnamefont
  {Nguyen}}, \bibinfo {author} {\bibfnamefont {E.~C.}\ \bibnamefont
  {Behrman}},\ and\ \bibinfo {author} {\bibfnamefont {J.~E.}\ \bibnamefont
  {Steck}},\ }\href@noop {} {\bibinfo {title} {Quantum circuit representation
  of {Bayesian} networks}} (\bibinfo {year} {2020}),\ \Eprint
  {https://arxiv.org/abs/2004.14803} {arXiv:2004.14803 [quant-ph]} \BibitemShut
  {NoStop}%
\bibitem [{\citenamefont {Tucci}(1995)}]{tucciQuantumBayesianNets1995}%
  \BibitemOpen
  \bibfield  {author} {\bibinfo {author} {\bibfnamefont {R.~R.}\ \bibnamefont
  {Tucci}},\ }\bibfield  {title} {\bibinfo {title} {Quantum {{Bayesian
  Nets}}},\ }\href {https://doi.org/10.1142/S0217979295000148} {\bibfield
  {journal} {\bibinfo  {journal} {International Journal of Modern Physics B}\
  }\textbf {\bibinfo {volume} {09}},\ \bibinfo {pages} {295} (\bibinfo {year}
  {1995})}\BibitemShut {NoStop}%
\bibitem [{\citenamefont {Ranganath}\ \emph {et~al.}(2018)\citenamefont
  {Ranganath}, \citenamefont {Altosaar}, \citenamefont {Tran},\ and\
  \citenamefont {Blei}}]{ranganath2018operator}%
  \BibitemOpen
  \bibfield  {author} {\bibinfo {author} {\bibfnamefont {R.}~\bibnamefont
  {Ranganath}}, \bibinfo {author} {\bibfnamefont {J.}~\bibnamefont {Altosaar}},
  \bibinfo {author} {\bibfnamefont {D.}~\bibnamefont {Tran}},\ and\ \bibinfo
  {author} {\bibfnamefont {D.~M.}\ \bibnamefont {Blei}},\ }\href@noop {}
  {\bibinfo {title} {Operator variational inference}} (\bibinfo {year}
  {2018}),\ \Eprint {https://arxiv.org/abs/1610.09033} {arXiv:1610.09033
  [stat.ML]} \BibitemShut {NoStop}%
\bibitem [{\citenamefont {Sriperumbudur}\ \emph {et~al.}(2009)\citenamefont
  {Sriperumbudur}, \citenamefont {Fukumizu}, \citenamefont {Gretton},
  \citenamefont {Schölkopf},\ and\ \citenamefont
  {Lanckriet}}]{sriperumbudur2009integral}%
  \BibitemOpen
  \bibfield  {author} {\bibinfo {author} {\bibfnamefont {B.~K.}\ \bibnamefont
  {Sriperumbudur}}, \bibinfo {author} {\bibfnamefont {K.}~\bibnamefont
  {Fukumizu}}, \bibinfo {author} {\bibfnamefont {A.}~\bibnamefont {Gretton}},
  \bibinfo {author} {\bibfnamefont {B.}~\bibnamefont {Schölkopf}},\ and\
  \bibinfo {author} {\bibfnamefont {G.~R.~G.}\ \bibnamefont {Lanckriet}},\
  }\href@noop {} {\bibinfo {title} {On integral probability metrics,
  $\phi$-divergences and binary classification}} (\bibinfo {year} {2009}),\
  \Eprint {https://arxiv.org/abs/0901.2698} {arXiv:0901.2698 [cs.IT]}
  \BibitemShut {NoStop}%
\bibitem [{\citenamefont {Rezende}\ \emph {et~al.}(2014)\citenamefont
  {Rezende}, \citenamefont {Mohamed},\ and\ \citenamefont
  {Wierstra}}]{rezende2014stochastic}%
  \BibitemOpen
  \bibfield  {author} {\bibinfo {author} {\bibfnamefont {D.~J.}\ \bibnamefont
  {Rezende}}, \bibinfo {author} {\bibfnamefont {S.}~\bibnamefont {Mohamed}},\
  and\ \bibinfo {author} {\bibfnamefont {D.}~\bibnamefont {Wierstra}},\
  }\href@noop {} {\bibinfo {title} {Stochastic backpropagation and approximate
  inference in deep generative models}} (\bibinfo {year} {2014}),\ \Eprint
  {https://arxiv.org/abs/1401.4082} {arXiv:1401.4082 [stat.ML]} \BibitemShut
  {NoStop}%
\bibitem [{\citenamefont {Mescheder}\ \emph {et~al.}(2018)\citenamefont
  {Mescheder}, \citenamefont {Nowozin},\ and\ \citenamefont
  {Geiger}}]{mescheder2018adversarial}%
  \BibitemOpen
  \bibfield  {author} {\bibinfo {author} {\bibfnamefont {L.}~\bibnamefont
  {Mescheder}}, \bibinfo {author} {\bibfnamefont {S.}~\bibnamefont {Nowozin}},\
  and\ \bibinfo {author} {\bibfnamefont {A.}~\bibnamefont {Geiger}},\
  }\href@noop {} {\bibinfo {title} {Adversarial variational {Bayes}: Unifying
  variational autoencoders and generative adversarial networks}} (\bibinfo
  {year} {2018}),\ \Eprint {https://arxiv.org/abs/1701.04722} {arXiv:1701.04722
  [cs.LG]} \BibitemShut {NoStop}%
\bibitem [{\citenamefont {Husz{á}r}(2017)}]{huszar2017variational}%
  \BibitemOpen
  \bibfield  {author} {\bibinfo {author} {\bibfnamefont {F.}~\bibnamefont
  {Husz{á}r}},\ }\href@noop {} {\bibinfo {title} {Variational inference using
  implicit distributions}} (\bibinfo {year} {2017}),\ \Eprint
  {https://arxiv.org/abs/1702.08235} {arXiv:1702.08235 [stat.ML]} \BibitemShut
  {NoStop}%
\bibitem [{\citenamefont {Yang}\ \emph {et~al.}(2018)\citenamefont {Yang},
  \citenamefont {Liu}, \citenamefont {Rao},\ and\ \citenamefont
  {Neville}}]{yang2018goodness}%
  \BibitemOpen
  \bibfield  {author} {\bibinfo {author} {\bibfnamefont {J.}~\bibnamefont
  {Yang}}, \bibinfo {author} {\bibfnamefont {Q.}~\bibnamefont {Liu}}, \bibinfo
  {author} {\bibfnamefont {V.}~\bibnamefont {Rao}},\ and\ \bibinfo {author}
  {\bibfnamefont {J.}~\bibnamefont {Neville}},\ }\bibfield  {title} {\bibinfo
  {title} {Goodness-of-fit testing for discrete distributions via stein
  discrepancy},\ }in\ \href {http://proceedings.mlr.press/v80/yang18c.html}
  {\emph {\bibinfo {booktitle} {Proceedings of the 35th International
  Conference on Machine Learning}}},\ \bibinfo {series} {Proceedings of Machine
  Learning Research}, Vol.~\bibinfo {volume} {80},\ \bibinfo {editor} {edited
  by\ \bibinfo {editor} {\bibfnamefont {J.}~\bibnamefont {Dy}}\ and\ \bibinfo
  {editor} {\bibfnamefont {A.}~\bibnamefont {Krause}}}\ (\bibinfo  {publisher}
  {PMLR},\ \bibinfo {year} {2018})\ pp.\ \bibinfo {pages}
  {5561--5570}\BibitemShut {NoStop}%
\bibitem [{Note4()}]{Note4}%
  \BibitemOpen
  \bibinfo {note} {In contrast to Ref.~\cite {yang2018goodness} but in line
  with earlier work such as Ref.~\cite {gorhamMeasuringSampleQuality2018} we
  include the absolute value in the definition of the Stein
  discrepancy.}\BibitemShut {Stop}%
\bibitem [{\citenamefont {McClean}\ \emph {et~al.}(2018)\citenamefont
  {McClean}, \citenamefont {Boixo}, \citenamefont {Smelyanskiy}, \citenamefont
  {Babbush},\ and\ \citenamefont {Neven}}]{mcclean_barren_2018}%
  \BibitemOpen
  \bibfield  {author} {\bibinfo {author} {\bibfnamefont {J.~R.}\ \bibnamefont
  {McClean}}, \bibinfo {author} {\bibfnamefont {S.}~\bibnamefont {Boixo}},
  \bibinfo {author} {\bibfnamefont {V.~N.}\ \bibnamefont {Smelyanskiy}},
  \bibinfo {author} {\bibfnamefont {R.}~\bibnamefont {Babbush}},\ and\ \bibinfo
  {author} {\bibfnamefont {H.}~\bibnamefont {Neven}},\ }\bibfield  {title}
  {\bibinfo {title} {Barren plateaus in quantum neural network training
  landscapes},\ }\href {https://doi.org/10.1038/s41467-018-07090-4} {\bibfield
  {journal} {\bibinfo  {journal} {Nature Communications}\ }\textbf {\bibinfo
  {volume} {9}},\ \bibinfo {pages} {4812} (\bibinfo {year} {2018})}\BibitemShut
  {NoStop}%
\bibitem [{\citenamefont {Cerezo}\ \emph {et~al.}(2021)\citenamefont {Cerezo},
  \citenamefont {Sone}, \citenamefont {Volkoff}, \citenamefont {Cincio},\ and\
  \citenamefont {Coles}}]{cerezo_cost-function-dependent_2020}%
  \BibitemOpen
  \bibfield  {author} {\bibinfo {author} {\bibfnamefont {M.}~\bibnamefont
  {Cerezo}}, \bibinfo {author} {\bibfnamefont {A.}~\bibnamefont {Sone}},
  \bibinfo {author} {\bibfnamefont {T.}~\bibnamefont {Volkoff}}, \bibinfo
  {author} {\bibfnamefont {L.}~\bibnamefont {Cincio}},\ and\ \bibinfo {author}
  {\bibfnamefont {P.~J.}\ \bibnamefont {Coles}},\ }\bibfield  {title} {\bibinfo
  {title} {Cost function dependent barren plateaus in shallow parametrized
  quantum circuits},\ }\href {https://doi.org/10.1038/s41467-021-21728-w}
  {\bibfield  {journal} {\bibinfo  {journal} {Nature Communications}\ }\textbf
  {\bibinfo {volume} {12}} (\bibinfo {year} {2021})}\BibitemShut {NoStop}%
\bibitem [{\citenamefont {Wang}\ \emph {et~al.}(2021)\citenamefont {Wang},
  \citenamefont {Fontana}, \citenamefont {Cerezo}, \citenamefont {Sharma},
  \citenamefont {Sone}, \citenamefont {Cincio},\ and\ \citenamefont
  {Coles}}]{wang_noise-induced_2021}%
  \BibitemOpen
  \bibfield  {author} {\bibinfo {author} {\bibfnamefont {S.}~\bibnamefont
  {Wang}}, \bibinfo {author} {\bibfnamefont {E.}~\bibnamefont {Fontana}},
  \bibinfo {author} {\bibfnamefont {M.}~\bibnamefont {Cerezo}}, \bibinfo
  {author} {\bibfnamefont {K.}~\bibnamefont {Sharma}}, \bibinfo {author}
  {\bibfnamefont {A.}~\bibnamefont {Sone}}, \bibinfo {author} {\bibfnamefont
  {L.}~\bibnamefont {Cincio}},\ and\ \bibinfo {author} {\bibfnamefont {P.~J.}\
  \bibnamefont {Coles}},\ }\href@noop {} {\bibinfo {title} {Noise-induced
  barren plateaus in variational quantum algorithms}} (\bibinfo {year}
  {2021}),\ \Eprint {https://arxiv.org/abs/2007.14384} {arXiv:2007.14384
  [quant-ph]} \BibitemShut {NoStop}%
\bibitem [{\citenamefont {Holmes}\ \emph {et~al.}(2021)\citenamefont {Holmes},
  \citenamefont {Sharma}, \citenamefont {Cerezo},\ and\ \citenamefont
  {Coles}}]{holmes_connecting_2021}%
  \BibitemOpen
  \bibfield  {author} {\bibinfo {author} {\bibfnamefont {Z.}~\bibnamefont
  {Holmes}}, \bibinfo {author} {\bibfnamefont {K.}~\bibnamefont {Sharma}},
  \bibinfo {author} {\bibfnamefont {M.}~\bibnamefont {Cerezo}},\ and\ \bibinfo
  {author} {\bibfnamefont {P.~J.}\ \bibnamefont {Coles}},\ }\href@noop {}
  {\bibinfo {title} {Connecting ansatz expressibility to gradient magnitudes
  and barren plateaus}} (\bibinfo {year} {2021}),\ \Eprint
  {https://arxiv.org/abs/2101.02138} {arXiv:2101.02138 [quant-ph]} \BibitemShut
  {NoStop}%
\bibitem [{\citenamefont {Pesah}\ \emph {et~al.}(2020)\citenamefont {Pesah},
  \citenamefont {Cerezo}, \citenamefont {Wang}, \citenamefont {Volkoff},
  \citenamefont {Sornborger},\ and\ \citenamefont
  {Coles}}]{pesah_absence_2020}%
  \BibitemOpen
  \bibfield  {author} {\bibinfo {author} {\bibfnamefont {A.}~\bibnamefont
  {Pesah}}, \bibinfo {author} {\bibfnamefont {M.}~\bibnamefont {Cerezo}},
  \bibinfo {author} {\bibfnamefont {S.}~\bibnamefont {Wang}}, \bibinfo {author}
  {\bibfnamefont {T.}~\bibnamefont {Volkoff}}, \bibinfo {author} {\bibfnamefont
  {A.~T.}\ \bibnamefont {Sornborger}},\ and\ \bibinfo {author} {\bibfnamefont
  {P.~J.}\ \bibnamefont {Coles}},\ }\href@noop {} {\bibinfo {title} {Absence of
  barren plateaus in quantum convolutional neural networks}} (\bibinfo {year}
  {2020}),\ \Eprint {https://arxiv.org/abs/2011.02966} {arXiv:2011.02966
  [quant-ph]} \BibitemShut {NoStop}%
\bibitem [{\citenamefont {Zhao}\ and\ \citenamefont
  {Gao}(2021)}]{zhao_analyzing_2021}%
  \BibitemOpen
  \bibfield  {author} {\bibinfo {author} {\bibfnamefont {C.}~\bibnamefont
  {Zhao}}\ and\ \bibinfo {author} {\bibfnamefont {X.-S.}\ \bibnamefont {Gao}},\
  }\bibfield  {title} {\bibinfo {title} {Analyzing the barren plateau
  phenomenon in training quantum neural network with the {ZX}-calculus},\
  }\href {https://doi.org/10.22331/q-2021-06-04-466} {\bibfield  {journal}
  {\bibinfo  {journal} {Quantum}\ }\textbf {\bibinfo {volume} {5}},\ \bibinfo
  {pages} {466} (\bibinfo {year} {2021})}\BibitemShut {NoStop}%
\bibitem [{\citenamefont {Kontonis}\ \emph {et~al.}(2020)\citenamefont
  {Kontonis}, \citenamefont {Liu},\ and\ \citenamefont
  {Tzamos}}]{kontonis_convergence_2020}%
  \BibitemOpen
  \bibfield  {author} {\bibinfo {author} {\bibfnamefont {V.}~\bibnamefont
  {Kontonis}}, \bibinfo {author} {\bibfnamefont {S.}~\bibnamefont {Liu}},\ and\
  \bibinfo {author} {\bibfnamefont {C.}~\bibnamefont {Tzamos}},\ }\href@noop {}
  {\bibinfo {title} {Convergence and sample complexity of {SGD} in {GANs}}}
  (\bibinfo {year} {2020}),\ \Eprint {https://arxiv.org/abs/2012.00732}
  {arXiv:2012.00732 [cs.LG]} \BibitemShut {NoStop}%
\bibitem [{\citenamefont {LaRose}\ and\ \citenamefont
  {Coyle}(2020)}]{larose2020robust}%
  \BibitemOpen
  \bibfield  {author} {\bibinfo {author} {\bibfnamefont {R.}~\bibnamefont
  {LaRose}}\ and\ \bibinfo {author} {\bibfnamefont {B.}~\bibnamefont {Coyle}},\
  }\bibfield  {title} {\bibinfo {title} {Robust data encodings for quantum
  classifiers},\ }\href {https://doi.org/10.1103/PhysRevA.102.032420}
  {\bibfield  {journal} {\bibinfo  {journal} {Phys. Rev. A}\ }\textbf {\bibinfo
  {volume} {102}},\ \bibinfo {pages} {032420} (\bibinfo {year}
  {2020})}\BibitemShut {NoStop}%
\bibitem [{\citenamefont {Schuld}\ \emph {et~al.}(2021)\citenamefont {Schuld},
  \citenamefont {Sweke},\ and\ \citenamefont {Meyer}}]{schuld_effect_2020}%
  \BibitemOpen
  \bibfield  {author} {\bibinfo {author} {\bibfnamefont {M.}~\bibnamefont
  {Schuld}}, \bibinfo {author} {\bibfnamefont {R.}~\bibnamefont {Sweke}},\ and\
  \bibinfo {author} {\bibfnamefont {J.~J.}\ \bibnamefont {Meyer}},\ }\bibfield
  {title} {\bibinfo {title} {Effect of data encoding on the expressive power of
  variational quantum-machine-learning models},\ }\href
  {https://doi.org/10.1103/PhysRevA.103.032430} {\bibfield  {journal} {\bibinfo
   {journal} {Phys. Rev. A}\ }\textbf {\bibinfo {volume} {103}},\ \bibinfo
  {pages} {032430} (\bibinfo {year} {2021})}\BibitemShut {NoStop}%
\bibitem [{\citenamefont {P{\'{e}}rez-Salinas}\ \emph
  {et~al.}(2020)\citenamefont {P{\'{e}}rez-Salinas}, \citenamefont
  {Cervera-Lierta}, \citenamefont {Gil-Fuster},\ and\ \citenamefont
  {Latorre}}]{PerezSalinas2020datareuploading}%
  \BibitemOpen
  \bibfield  {author} {\bibinfo {author} {\bibfnamefont {A.}~\bibnamefont
  {P{\'{e}}rez-Salinas}}, \bibinfo {author} {\bibfnamefont {A.}~\bibnamefont
  {Cervera-Lierta}}, \bibinfo {author} {\bibfnamefont {E.}~\bibnamefont
  {Gil-Fuster}},\ and\ \bibinfo {author} {\bibfnamefont {J.~I.}\ \bibnamefont
  {Latorre}},\ }\bibfield  {title} {\bibinfo {title} {Data re-uploading for a
  universal quantum classifier},\ }\href
  {https://doi.org/10.22331/q-2020-02-06-226} {\bibfield  {journal} {\bibinfo
  {journal} {{Quantum}}\ }\textbf {\bibinfo {volume} {4}},\ \bibinfo {pages}
  {226} (\bibinfo {year} {2020})}\BibitemShut {NoStop}%
\bibitem [{\citenamefont {Ghahramani}\ and\ \citenamefont
  {Jordan}(1997)}]{ghahramani1997factorial}%
  \BibitemOpen
  \bibfield  {author} {\bibinfo {author} {\bibfnamefont {Z.}~\bibnamefont
  {Ghahramani}}\ and\ \bibinfo {author} {\bibfnamefont {M.~I.}\ \bibnamefont
  {Jordan}},\ }\bibfield  {title} {\bibinfo {title} {{Factorial hidden Markov
  models}},\ }\href {https://doi.org/10.1023/A:1007425814087} {\bibfield
  {journal} {\bibinfo  {journal} {Machine learning}\ }\textbf {\bibinfo
  {volume} {29}},\ \bibinfo {pages} {245} (\bibinfo {year} {1997})}\BibitemShut
  {NoStop}%
\bibitem [{\citenamefont {Kolter}\ and\ \citenamefont
  {Jaakkola}(2012)}]{zico2012approximate}%
  \BibitemOpen
  \bibfield  {author} {\bibinfo {author} {\bibfnamefont {J.~Z.}\ \bibnamefont
  {Kolter}}\ and\ \bibinfo {author} {\bibfnamefont {T.}~\bibnamefont
  {Jaakkola}},\ }\bibfield  {title} {\bibinfo {title} {Approximate inference in
  additive factorial hmms with application to energy disaggregation},\ }in\
  \href {http://proceedings.mlr.press/v22/zico12.html} {\emph {\bibinfo
  {booktitle} {Proceedings of the Fifteenth International Conference on
  Artificial Intelligence and Statistics}}},\ \bibinfo {series} {Proceedings of
  Machine Learning Research}, Vol.~\bibinfo {volume} {22},\ \bibinfo {editor}
  {edited by\ \bibinfo {editor} {\bibfnamefont {N.~D.}\ \bibnamefont
  {Lawrence}}\ and\ \bibinfo {editor} {\bibfnamefont {M.}~\bibnamefont
  {Girolami}}}\ (\bibinfo  {publisher} {PMLR},\ \bibinfo {address} {La Palma,
  Canary Islands},\ \bibinfo {year} {2012})\ pp.\ \bibinfo {pages}
  {1472--1482}\BibitemShut {NoStop}%
\bibitem [{\citenamefont {Smith}\ \emph {et~al.}(2016)\citenamefont {Smith},
  \citenamefont {Curtis},\ and\ \citenamefont {Zeng}}]{smith2016practical}%
  \BibitemOpen
  \bibfield  {author} {\bibinfo {author} {\bibfnamefont {R.~S.}\ \bibnamefont
  {Smith}}, \bibinfo {author} {\bibfnamefont {M.~J.}\ \bibnamefont {Curtis}},\
  and\ \bibinfo {author} {\bibfnamefont {W.~J.}\ \bibnamefont {Zeng}},\
  }\href@noop {} {\bibinfo {title} {A practical quantum instruction set
  architecture}} (\bibinfo {year} {2016}),\ \Eprint
  {https://arxiv.org/abs/1608.03355} {arXiv:1608.03355 [quant-ph]} \BibitemShut
  {NoStop}%
\bibitem [{\citenamefont {Sivarajah}\ \emph {et~al.}(2020)\citenamefont
  {Sivarajah}, \citenamefont {Dilkes}, \citenamefont {Cowtan}, \citenamefont
  {Simmons}, \citenamefont {Edgington},\ and\ \citenamefont
  {Duncan}}]{sivarajah_tvertketrangle_2020}%
  \BibitemOpen
  \bibfield  {author} {\bibinfo {author} {\bibfnamefont {S.}~\bibnamefont
  {Sivarajah}}, \bibinfo {author} {\bibfnamefont {S.}~\bibnamefont {Dilkes}},
  \bibinfo {author} {\bibfnamefont {A.}~\bibnamefont {Cowtan}}, \bibinfo
  {author} {\bibfnamefont {W.}~\bibnamefont {Simmons}}, \bibinfo {author}
  {\bibfnamefont {A.}~\bibnamefont {Edgington}},\ and\ \bibinfo {author}
  {\bibfnamefont {R.}~\bibnamefont {Duncan}},\ }\bibfield  {title} {\bibinfo
  {title} {t$\vert$ket$\rangle$: a retargetable compiler for {NISQ} devices},\
  }\href {https://doi.org/10.1088/2058-9565/ab8e92} {\bibfield  {journal}
  {\bibinfo  {journal} {Quantum Science and Technology}\ }\textbf {\bibinfo
  {volume} {6}},\ \bibinfo {pages} {014003} (\bibinfo {year}
  {2020})}\BibitemShut {NoStop}%
\bibitem [{\citenamefont {Barber}(2012)}]{barberBayesianReasoningMachine2012}%
  \BibitemOpen
  \bibfield  {author} {\bibinfo {author} {\bibfnamefont {D.}~\bibnamefont
  {Barber}},\ }\href@noop {} {\emph {\bibinfo {title} {Bayesian {{Reasoning}}
  and {{Machine Learning}}}}}\ (\bibinfo  {publisher} {{Cambridge University
  Press}},\ \bibinfo {address} {{Cambridge, UK}},\ \bibinfo {year}
  {2012})\BibitemShut {NoStop}%
\bibitem [{\citenamefont {Arjovsky}\ and\ \citenamefont
  {Bottou}(2017)}]{arjovsky2017principled}%
  \BibitemOpen
  \bibfield  {author} {\bibinfo {author} {\bibfnamefont {M.}~\bibnamefont
  {Arjovsky}}\ and\ \bibinfo {author} {\bibfnamefont {L.}~\bibnamefont
  {Bottou}},\ }\href@noop {} {\bibinfo {title} {Towards principled methods for
  training generative adversarial networks}} (\bibinfo {year} {2017}),\ \Eprint
  {https://arxiv.org/abs/1701.04862} {arXiv:1701.04862 [stat.ML]} \BibitemShut
  {NoStop}%
\bibitem [{\citenamefont {Lucic}\ \emph {et~al.}(2018)\citenamefont {Lucic},
  \citenamefont {Kurach}, \citenamefont {Michalski}, \citenamefont {Gelly},\
  and\ \citenamefont {Bousquet}}]{lucic2018gans}%
  \BibitemOpen
  \bibfield  {author} {\bibinfo {author} {\bibfnamefont {M.}~\bibnamefont
  {Lucic}}, \bibinfo {author} {\bibfnamefont {K.}~\bibnamefont {Kurach}},
  \bibinfo {author} {\bibfnamefont {M.}~\bibnamefont {Michalski}}, \bibinfo
  {author} {\bibfnamefont {S.}~\bibnamefont {Gelly}},\ and\ \bibinfo {author}
  {\bibfnamefont {O.}~\bibnamefont {Bousquet}},\ }\bibfield  {title} {\bibinfo
  {title} {{Are {GANs} Created Equal? {A} Large-Scale Study}},\ }in\ \href@noop
  {} {\emph {\bibinfo {booktitle} {Proceedings of the 32nd International
  Conference on Neural Information Processing Systems}}},\ \bibinfo {series and
  number} {NIPS'18}\ (\bibinfo  {publisher} {Curran Associates Inc.},\ \bibinfo
  {year} {2018})\ pp.\ \bibinfo {pages} {698--707},\ \Eprint
  {https://arxiv.org/abs/1711.10337} {arXiv:1711.10337 [stat.ML]} \BibitemShut
  {NoStop}%
\bibitem [{\citenamefont {Kondor}\ and\ \citenamefont
  {Lafferty}(2002)}]{kondor2002diffusion}%
  \BibitemOpen
  \bibfield  {author} {\bibinfo {author} {\bibfnamefont {R.~I.}\ \bibnamefont
  {Kondor}}\ and\ \bibinfo {author} {\bibfnamefont {J.~D.}\ \bibnamefont
  {Lafferty}},\ }\bibfield  {title} {\bibinfo {title} {Diffusion kernels on
  graphs and other discrete input spaces},\ }in\ \href@noop {} {\emph {\bibinfo
  {booktitle} {Proceedings of the Nineteenth International Conference on
  Machine Learning}}},\ \bibinfo {series and number} {ICML '02}\ (\bibinfo
  {publisher} {Morgan Kaufmann Publishers Inc.},\ \bibinfo {address} {San
  Francisco, CA, USA},\ \bibinfo {year} {2002})\ p.\ \bibinfo {pages}
  {315–322}\BibitemShut {NoStop}%
\bibitem [{\citenamefont {Coecke}\ \emph {et~al.}(2020)\citenamefont {Coecke},
  \citenamefont {de~Felice}, \citenamefont {Meichanetzidis},\ and\
  \citenamefont {Toumi}}]{coecke2020foundations}%
  \BibitemOpen
  \bibfield  {author} {\bibinfo {author} {\bibfnamefont {B.}~\bibnamefont
  {Coecke}}, \bibinfo {author} {\bibfnamefont {G.}~\bibnamefont {de~Felice}},
  \bibinfo {author} {\bibfnamefont {K.}~\bibnamefont {Meichanetzidis}},\ and\
  \bibinfo {author} {\bibfnamefont {A.}~\bibnamefont {Toumi}},\ }\href@noop {}
  {\bibinfo {title} {Foundations for near-term quantum natural language
  processing}} (\bibinfo {year} {2020}),\ \Eprint
  {https://arxiv.org/abs/2012.03755} {arXiv:2012.03755 [quant-ph]} \BibitemShut
  {NoStop}%
\bibitem [{\citenamefont {Meichanetzidis}\ \emph {et~al.}(2020)\citenamefont
  {Meichanetzidis}, \citenamefont {Toumi}, \citenamefont {de~Felice},\ and\
  \citenamefont {Coecke}}]{meichanetzidis2020grammaraware}%
  \BibitemOpen
  \bibfield  {author} {\bibinfo {author} {\bibfnamefont {K.}~\bibnamefont
  {Meichanetzidis}}, \bibinfo {author} {\bibfnamefont {A.}~\bibnamefont
  {Toumi}}, \bibinfo {author} {\bibfnamefont {G.}~\bibnamefont {de~Felice}},\
  and\ \bibinfo {author} {\bibfnamefont {B.}~\bibnamefont {Coecke}},\
  }\href@noop {} {\bibinfo {title} {Grammar-aware question-answering on quantum
  computers}} (\bibinfo {year} {2020}),\ \Eprint
  {https://arxiv.org/abs/2012.03756} {arXiv:2012.03756 [quant-ph]} \BibitemShut
  {NoStop}%
\bibitem [{\citenamefont {Lorenz}\ \emph {et~al.}(2021)\citenamefont {Lorenz},
  \citenamefont {Pearson}, \citenamefont {Meichanetzidis}, \citenamefont
  {Kartsaklis},\ and\ \citenamefont {Coecke}}]{lorenz2021qnlp}%
  \BibitemOpen
  \bibfield  {author} {\bibinfo {author} {\bibfnamefont {R.}~\bibnamefont
  {Lorenz}}, \bibinfo {author} {\bibfnamefont {A.}~\bibnamefont {Pearson}},
  \bibinfo {author} {\bibfnamefont {K.}~\bibnamefont {Meichanetzidis}},
  \bibinfo {author} {\bibfnamefont {D.}~\bibnamefont {Kartsaklis}},\ and\
  \bibinfo {author} {\bibfnamefont {B.}~\bibnamefont {Coecke}},\ }\href@noop {}
  {\bibinfo {title} {Qnlp in practice: Running compositional models of meaning
  on a quantum computer}} (\bibinfo {year} {2021}),\ \Eprint
  {https://arxiv.org/abs/2102.12846} {arXiv:2102.12846 [cs.CL]} \BibitemShut
  {NoStop}%
\bibitem [{\citenamefont {Arute}\ \emph {et~al.}(2019)\citenamefont {Arute},
  \citenamefont {Arya}, \citenamefont {Babbush}, \citenamefont {Bacon},
  \citenamefont {Bardin}, \citenamefont {Barends}, \citenamefont {Biswas},
  \citenamefont {Boixo}, \citenamefont {Brandao}, \citenamefont {Buell},
  \citenamefont {Burkett}, \citenamefont {Chen}, \citenamefont {Chen},
  \citenamefont {Chiaro}, \citenamefont {Collins}, \citenamefont {Courtney},
  \citenamefont {Dunsworth}, \citenamefont {Farhi}, \citenamefont {Foxen},
  \citenamefont {Fowler}, \citenamefont {Gidney}, \citenamefont {Giustina},
  \citenamefont {Graff}, \citenamefont {Guerin}, \citenamefont {Habegger},
  \citenamefont {Harrigan}, \citenamefont {Hartmann}, \citenamefont {Ho},
  \citenamefont {Hoffmann}, \citenamefont {Huang}, \citenamefont {Humble},
  \citenamefont {Isakov}, \citenamefont {Jeffrey}, \citenamefont {Jiang},
  \citenamefont {Kafri}, \citenamefont {Kechedzhi}, \citenamefont {Kelly},
  \citenamefont {Klimov}, \citenamefont {Knysh}, \citenamefont {Korotkov},
  \citenamefont {Kostritsa}, \citenamefont {Landhuis}, \citenamefont
  {Lindmark}, \citenamefont {Lucero}, \citenamefont {Lyakh}, \citenamefont
  {Mandrà}, \citenamefont {McClean}, \citenamefont {McEwen}, \citenamefont
  {Megrant}, \citenamefont {Mi}, \citenamefont {Michielsen}, \citenamefont
  {Mohseni}, \citenamefont {Mutus}, \citenamefont {Naaman}, \citenamefont
  {Neeley}, \citenamefont {Neill}, \citenamefont {Niu}, \citenamefont {Ostby},
  \citenamefont {Petukhov}, \citenamefont {Platt}, \citenamefont {Quintana},
  \citenamefont {Rieffel}, \citenamefont {Roushan}, \citenamefont {Rubin},
  \citenamefont {Sank}, \citenamefont {Satzinger}, \citenamefont {Smelyanskiy},
  \citenamefont {Sung}, \citenamefont {Trevithick}, \citenamefont
  {Vainsencher}, \citenamefont {Villalonga}, \citenamefont {White},
  \citenamefont {Yao}, \citenamefont {Yeh}, \citenamefont {Zalcman},
  \citenamefont {Neven},\ and\ \citenamefont {Martinis}}]{arute_quantum_2019}%
  \BibitemOpen
  \bibfield  {author} {\bibinfo {author} {\bibfnamefont {F.}~\bibnamefont
  {Arute}}, \bibinfo {author} {\bibfnamefont {K.}~\bibnamefont {Arya}},
  \bibinfo {author} {\bibfnamefont {R.}~\bibnamefont {Babbush}}, \bibinfo
  {author} {\bibfnamefont {D.}~\bibnamefont {Bacon}}, \bibinfo {author}
  {\bibfnamefont {J.~C.}\ \bibnamefont {Bardin}}, \bibinfo {author}
  {\bibfnamefont {R.}~\bibnamefont {Barends}}, \bibinfo {author} {\bibfnamefont
  {R.}~\bibnamefont {Biswas}}, \bibinfo {author} {\bibfnamefont
  {S.}~\bibnamefont {Boixo}}, \bibinfo {author} {\bibfnamefont {F.~G. S.~L.}\
  \bibnamefont {Brandao}}, \bibinfo {author} {\bibfnamefont {D.~A.}\
  \bibnamefont {Buell}}, \bibinfo {author} {\bibfnamefont {B.}~\bibnamefont
  {Burkett}}, \bibinfo {author} {\bibfnamefont {Y.}~\bibnamefont {Chen}},
  \bibinfo {author} {\bibfnamefont {Z.}~\bibnamefont {Chen}}, \bibinfo {author}
  {\bibfnamefont {B.}~\bibnamefont {Chiaro}}, \bibinfo {author} {\bibfnamefont
  {R.}~\bibnamefont {Collins}}, \bibinfo {author} {\bibfnamefont
  {W.}~\bibnamefont {Courtney}}, \bibinfo {author} {\bibfnamefont
  {A.}~\bibnamefont {Dunsworth}}, \bibinfo {author} {\bibfnamefont
  {E.}~\bibnamefont {Farhi}}, \bibinfo {author} {\bibfnamefont
  {B.}~\bibnamefont {Foxen}}, \bibinfo {author} {\bibfnamefont
  {A.}~\bibnamefont {Fowler}}, \bibinfo {author} {\bibfnamefont
  {C.}~\bibnamefont {Gidney}}, \bibinfo {author} {\bibfnamefont
  {M.}~\bibnamefont {Giustina}}, \bibinfo {author} {\bibfnamefont
  {R.}~\bibnamefont {Graff}}, \bibinfo {author} {\bibfnamefont
  {K.}~\bibnamefont {Guerin}}, \bibinfo {author} {\bibfnamefont
  {S.}~\bibnamefont {Habegger}}, \bibinfo {author} {\bibfnamefont {M.~P.}\
  \bibnamefont {Harrigan}}, \bibinfo {author} {\bibfnamefont {M.~J.}\
  \bibnamefont {Hartmann}}, \bibinfo {author} {\bibfnamefont {A.}~\bibnamefont
  {Ho}}, \bibinfo {author} {\bibfnamefont {M.}~\bibnamefont {Hoffmann}},
  \bibinfo {author} {\bibfnamefont {T.}~\bibnamefont {Huang}}, \bibinfo
  {author} {\bibfnamefont {T.~S.}\ \bibnamefont {Humble}}, \bibinfo {author}
  {\bibfnamefont {S.~V.}\ \bibnamefont {Isakov}}, \bibinfo {author}
  {\bibfnamefont {E.}~\bibnamefont {Jeffrey}}, \bibinfo {author} {\bibfnamefont
  {Z.}~\bibnamefont {Jiang}}, \bibinfo {author} {\bibfnamefont
  {D.}~\bibnamefont {Kafri}}, \bibinfo {author} {\bibfnamefont
  {K.}~\bibnamefont {Kechedzhi}}, \bibinfo {author} {\bibfnamefont
  {J.}~\bibnamefont {Kelly}}, \bibinfo {author} {\bibfnamefont {P.~V.}\
  \bibnamefont {Klimov}}, \bibinfo {author} {\bibfnamefont {S.}~\bibnamefont
  {Knysh}}, \bibinfo {author} {\bibfnamefont {A.}~\bibnamefont {Korotkov}},
  \bibinfo {author} {\bibfnamefont {F.}~\bibnamefont {Kostritsa}}, \bibinfo
  {author} {\bibfnamefont {D.}~\bibnamefont {Landhuis}}, \bibinfo {author}
  {\bibfnamefont {M.}~\bibnamefont {Lindmark}}, \bibinfo {author}
  {\bibfnamefont {E.}~\bibnamefont {Lucero}}, \bibinfo {author} {\bibfnamefont
  {D.}~\bibnamefont {Lyakh}}, \bibinfo {author} {\bibfnamefont
  {S.}~\bibnamefont {Mandrà}}, \bibinfo {author} {\bibfnamefont {J.~R.}\
  \bibnamefont {McClean}}, \bibinfo {author} {\bibfnamefont {M.}~\bibnamefont
  {McEwen}}, \bibinfo {author} {\bibfnamefont {A.}~\bibnamefont {Megrant}},
  \bibinfo {author} {\bibfnamefont {X.}~\bibnamefont {Mi}}, \bibinfo {author}
  {\bibfnamefont {K.}~\bibnamefont {Michielsen}}, \bibinfo {author}
  {\bibfnamefont {M.}~\bibnamefont {Mohseni}}, \bibinfo {author} {\bibfnamefont
  {J.}~\bibnamefont {Mutus}}, \bibinfo {author} {\bibfnamefont
  {O.}~\bibnamefont {Naaman}}, \bibinfo {author} {\bibfnamefont
  {M.}~\bibnamefont {Neeley}}, \bibinfo {author} {\bibfnamefont
  {C.}~\bibnamefont {Neill}}, \bibinfo {author} {\bibfnamefont {M.~Y.}\
  \bibnamefont {Niu}}, \bibinfo {author} {\bibfnamefont {E.}~\bibnamefont
  {Ostby}}, \bibinfo {author} {\bibfnamefont {A.}~\bibnamefont {Petukhov}},
  \bibinfo {author} {\bibfnamefont {J.~C.}\ \bibnamefont {Platt}}, \bibinfo
  {author} {\bibfnamefont {C.}~\bibnamefont {Quintana}}, \bibinfo {author}
  {\bibfnamefont {E.~G.}\ \bibnamefont {Rieffel}}, \bibinfo {author}
  {\bibfnamefont {P.}~\bibnamefont {Roushan}}, \bibinfo {author} {\bibfnamefont
  {N.~C.}\ \bibnamefont {Rubin}}, \bibinfo {author} {\bibfnamefont
  {D.}~\bibnamefont {Sank}}, \bibinfo {author} {\bibfnamefont {K.~J.}\
  \bibnamefont {Satzinger}}, \bibinfo {author} {\bibfnamefont {V.}~\bibnamefont
  {Smelyanskiy}}, \bibinfo {author} {\bibfnamefont {K.~J.}\ \bibnamefont
  {Sung}}, \bibinfo {author} {\bibfnamefont {M.~D.}\ \bibnamefont
  {Trevithick}}, \bibinfo {author} {\bibfnamefont {A.}~\bibnamefont
  {Vainsencher}}, \bibinfo {author} {\bibfnamefont {B.}~\bibnamefont
  {Villalonga}}, \bibinfo {author} {\bibfnamefont {T.}~\bibnamefont {White}},
  \bibinfo {author} {\bibfnamefont {Z.~J.}\ \bibnamefont {Yao}}, \bibinfo
  {author} {\bibfnamefont {P.}~\bibnamefont {Yeh}}, \bibinfo {author}
  {\bibfnamefont {A.}~\bibnamefont {Zalcman}}, \bibinfo {author} {\bibfnamefont
  {H.}~\bibnamefont {Neven}},\ and\ \bibinfo {author} {\bibfnamefont {J.~M.}\
  \bibnamefont {Martinis}},\ }\bibfield  {title} {\bibinfo {title} {Quantum
  supremacy using a programmable superconducting processor},\ }\href
  {https://doi.org/10.1038/s41586-019-1666-5} {\bibfield  {journal} {\bibinfo
  {journal} {Nature}\ }\textbf {\bibinfo {volume} {574}},\ \bibinfo {pages}
  {505} (\bibinfo {year} {2019})}\BibitemShut {NoStop}%
\bibitem [{\citenamefont {Wu}\ \emph {et~al.}(2021)\citenamefont {Wu},
  \citenamefont {Bao}, \citenamefont {Cao}, \citenamefont {Chen}, \citenamefont
  {Chen}, \citenamefont {Chen}, \citenamefont {Chung}, \citenamefont {Deng},
  \citenamefont {Du}, \citenamefont {Fan}, \citenamefont {Gong}, \citenamefont
  {Guo}, \citenamefont {Guo}, \citenamefont {Guo}, \citenamefont {Han},
  \citenamefont {Hong}, \citenamefont {Huang}, \citenamefont {Huo},
  \citenamefont {Li}, \citenamefont {Li}, \citenamefont {Li}, \citenamefont
  {Li}, \citenamefont {Liang}, \citenamefont {Lin}, \citenamefont {Lin},
  \citenamefont {Qian}, \citenamefont {Qiao}, \citenamefont {Rong},
  \citenamefont {Su}, \citenamefont {Sun}, \citenamefont {Wang}, \citenamefont
  {Wang}, \citenamefont {Wu}, \citenamefont {Xu}, \citenamefont {Yan},
  \citenamefont {Yang}, \citenamefont {Yang}, \citenamefont {Ye}, \citenamefont
  {Yin}, \citenamefont {Ying}, \citenamefont {Yu}, \citenamefont {Zha},
  \citenamefont {Zhang}, \citenamefont {Zhang}, \citenamefont {Zhang},
  \citenamefont {Zhang}, \citenamefont {Zhao}, \citenamefont {Zhao},
  \citenamefont {Zhou}, \citenamefont {Zhu}, \citenamefont {Lu}, \citenamefont
  {Peng}, \citenamefont {Zhu},\ and\ \citenamefont {Pan}}]{wu_strong_2021}%
  \BibitemOpen
  \bibfield  {author} {\bibinfo {author} {\bibfnamefont {Y.}~\bibnamefont
  {Wu}}, \bibinfo {author} {\bibfnamefont {W.-S.}\ \bibnamefont {Bao}},
  \bibinfo {author} {\bibfnamefont {S.}~\bibnamefont {Cao}}, \bibinfo {author}
  {\bibfnamefont {F.}~\bibnamefont {Chen}}, \bibinfo {author} {\bibfnamefont
  {M.-C.}\ \bibnamefont {Chen}}, \bibinfo {author} {\bibfnamefont
  {X.}~\bibnamefont {Chen}}, \bibinfo {author} {\bibfnamefont {T.-H.}\
  \bibnamefont {Chung}}, \bibinfo {author} {\bibfnamefont {H.}~\bibnamefont
  {Deng}}, \bibinfo {author} {\bibfnamefont {Y.}~\bibnamefont {Du}}, \bibinfo
  {author} {\bibfnamefont {D.}~\bibnamefont {Fan}}, \bibinfo {author}
  {\bibfnamefont {M.}~\bibnamefont {Gong}}, \bibinfo {author} {\bibfnamefont
  {C.}~\bibnamefont {Guo}}, \bibinfo {author} {\bibfnamefont {C.}~\bibnamefont
  {Guo}}, \bibinfo {author} {\bibfnamefont {S.}~\bibnamefont {Guo}}, \bibinfo
  {author} {\bibfnamefont {L.}~\bibnamefont {Han}}, \bibinfo {author}
  {\bibfnamefont {L.}~\bibnamefont {Hong}}, \bibinfo {author} {\bibfnamefont
  {H.-L.}\ \bibnamefont {Huang}}, \bibinfo {author} {\bibfnamefont {Y.-H.}\
  \bibnamefont {Huo}}, \bibinfo {author} {\bibfnamefont {L.}~\bibnamefont
  {Li}}, \bibinfo {author} {\bibfnamefont {N.}~\bibnamefont {Li}}, \bibinfo
  {author} {\bibfnamefont {S.}~\bibnamefont {Li}}, \bibinfo {author}
  {\bibfnamefont {Y.}~\bibnamefont {Li}}, \bibinfo {author} {\bibfnamefont
  {F.}~\bibnamefont {Liang}}, \bibinfo {author} {\bibfnamefont
  {C.}~\bibnamefont {Lin}}, \bibinfo {author} {\bibfnamefont {J.}~\bibnamefont
  {Lin}}, \bibinfo {author} {\bibfnamefont {H.}~\bibnamefont {Qian}}, \bibinfo
  {author} {\bibfnamefont {D.}~\bibnamefont {Qiao}}, \bibinfo {author}
  {\bibfnamefont {H.}~\bibnamefont {Rong}}, \bibinfo {author} {\bibfnamefont
  {H.}~\bibnamefont {Su}}, \bibinfo {author} {\bibfnamefont {L.}~\bibnamefont
  {Sun}}, \bibinfo {author} {\bibfnamefont {L.}~\bibnamefont {Wang}}, \bibinfo
  {author} {\bibfnamefont {S.}~\bibnamefont {Wang}}, \bibinfo {author}
  {\bibfnamefont {D.}~\bibnamefont {Wu}}, \bibinfo {author} {\bibfnamefont
  {Y.}~\bibnamefont {Xu}}, \bibinfo {author} {\bibfnamefont {K.}~\bibnamefont
  {Yan}}, \bibinfo {author} {\bibfnamefont {W.}~\bibnamefont {Yang}}, \bibinfo
  {author} {\bibfnamefont {Y.}~\bibnamefont {Yang}}, \bibinfo {author}
  {\bibfnamefont {Y.}~\bibnamefont {Ye}}, \bibinfo {author} {\bibfnamefont
  {J.}~\bibnamefont {Yin}}, \bibinfo {author} {\bibfnamefont {C.}~\bibnamefont
  {Ying}}, \bibinfo {author} {\bibfnamefont {J.}~\bibnamefont {Yu}}, \bibinfo
  {author} {\bibfnamefont {C.}~\bibnamefont {Zha}}, \bibinfo {author}
  {\bibfnamefont {C.}~\bibnamefont {Zhang}}, \bibinfo {author} {\bibfnamefont
  {H.}~\bibnamefont {Zhang}}, \bibinfo {author} {\bibfnamefont
  {K.}~\bibnamefont {Zhang}}, \bibinfo {author} {\bibfnamefont
  {Y.}~\bibnamefont {Zhang}}, \bibinfo {author} {\bibfnamefont
  {H.}~\bibnamefont {Zhao}}, \bibinfo {author} {\bibfnamefont {Y.}~\bibnamefont
  {Zhao}}, \bibinfo {author} {\bibfnamefont {L.}~\bibnamefont {Zhou}}, \bibinfo
  {author} {\bibfnamefont {Q.}~\bibnamefont {Zhu}}, \bibinfo {author}
  {\bibfnamefont {C.-Y.}\ \bibnamefont {Lu}}, \bibinfo {author} {\bibfnamefont
  {C.-Z.}\ \bibnamefont {Peng}}, \bibinfo {author} {\bibfnamefont
  {X.}~\bibnamefont {Zhu}},\ and\ \bibinfo {author} {\bibfnamefont {J.-W.}\
  \bibnamefont {Pan}},\ }\bibfield  {title} {\bibinfo {title} {Strong quantum
  computational advantage using a superconducting quantum processor},\ }\href
  {http://arxiv.org/abs/2106.14734} {\bibfield  {journal} {\bibinfo  {journal}
  {arXiv:2106.14734 [quant-ph]}\ } (\bibinfo {year} {2021})}\BibitemShut
  {NoStop}%
\bibitem [{\citenamefont {Mitarai}\ \emph {et~al.}(2018)\citenamefont
  {Mitarai}, \citenamefont {Negoro}, \citenamefont {Kitagawa},\ and\
  \citenamefont {Fujii}}]{mitaraiQuantumCircuitLearning2018}%
  \BibitemOpen
  \bibfield  {author} {\bibinfo {author} {\bibfnamefont {K.}~\bibnamefont
  {Mitarai}}, \bibinfo {author} {\bibfnamefont {M.}~\bibnamefont {Negoro}},
  \bibinfo {author} {\bibfnamefont {M.}~\bibnamefont {Kitagawa}},\ and\
  \bibinfo {author} {\bibfnamefont {K.}~\bibnamefont {Fujii}},\ }\bibfield
  {title} {\bibinfo {title} {Quantum circuit learning},\ }\href
  {https://doi.org/10.1103/PhysRevA.98.032309} {\bibfield  {journal} {\bibinfo
  {journal} {Phys. Rev. A}\ }\textbf {\bibinfo {volume} {98}},\ \bibinfo
  {pages} {032309} (\bibinfo {year} {2018})}\BibitemShut {NoStop}%
\bibitem [{\citenamefont {Schuld}\ \emph {et~al.}(2019)\citenamefont {Schuld},
  \citenamefont {Bergholm}, \citenamefont {Gogolin}, \citenamefont {Izaac},\
  and\ \citenamefont {Killoran}}]{schuldEvaluatingAnalyticGradients2019}%
  \BibitemOpen
  \bibfield  {author} {\bibinfo {author} {\bibfnamefont {M.}~\bibnamefont
  {Schuld}}, \bibinfo {author} {\bibfnamefont {V.}~\bibnamefont {Bergholm}},
  \bibinfo {author} {\bibfnamefont {C.}~\bibnamefont {Gogolin}}, \bibinfo
  {author} {\bibfnamefont {J.}~\bibnamefont {Izaac}},\ and\ \bibinfo {author}
  {\bibfnamefont {N.}~\bibnamefont {Killoran}},\ }\bibfield  {title} {\bibinfo
  {title} {Evaluating analytic gradients on quantum hardware},\ }\href
  {https://doi.org/10.1103/PhysRevA.99.032331} {\bibfield  {journal} {\bibinfo
  {journal} {Phys. Rev. A}\ }\textbf {\bibinfo {volume} {99}},\ \bibinfo
  {pages} {032331} (\bibinfo {year} {2019})}\BibitemShut {NoStop}%
\bibitem [{\citenamefont {Banchi}\ and\ \citenamefont
  {Crooks}(2021)}]{Banchi2021measuringanalytic}%
  \BibitemOpen
  \bibfield  {author} {\bibinfo {author} {\bibfnamefont {L.}~\bibnamefont
  {Banchi}}\ and\ \bibinfo {author} {\bibfnamefont {G.~E.}\ \bibnamefont
  {Crooks}},\ }\bibfield  {title} {\bibinfo {title} {Measuring {A}nalytic
  {G}radients of {G}eneral {Q}uantum {E}volution with the {S}tochastic
  {P}arameter {S}hift {R}ule},\ }\href
  {https://doi.org/10.22331/q-2021-01-25-386} {\bibfield  {journal} {\bibinfo
  {journal} {{Quantum}}\ }\textbf {\bibinfo {volume} {5}},\ \bibinfo {pages}
  {386} (\bibinfo {year} {2021})}\BibitemShut {NoStop}%
\bibitem [{\citenamefont {Dallaire-Demers}\ and\ \citenamefont
  {Killoran}(2018)}]{Dallaire_2018}%
  \BibitemOpen
  \bibfield  {author} {\bibinfo {author} {\bibfnamefont {P.-L.}\ \bibnamefont
  {Dallaire-Demers}}\ and\ \bibinfo {author} {\bibfnamefont {N.}~\bibnamefont
  {Killoran}},\ }\bibfield  {title} {\bibinfo {title} {Quantum generative
  adversarial networks},\ }\href {https://doi.org/10.1103/PhysRevA.98.012324}
  {\bibfield  {journal} {\bibinfo  {journal} {Phys. Rev. A}\ }\textbf {\bibinfo
  {volume} {98}},\ \bibinfo {pages} {012324} (\bibinfo {year}
  {2018})}\BibitemShut {NoStop}%
\bibitem [{\citenamefont {Benedetti}\ \emph
  {et~al.}(2019{\natexlab{c}})\citenamefont {Benedetti}, \citenamefont {Grant},
  \citenamefont {Wossnig},\ and\ \citenamefont
  {Severini}}]{benedetti2019adversarial}%
  \BibitemOpen
  \bibfield  {author} {\bibinfo {author} {\bibfnamefont {M.}~\bibnamefont
  {Benedetti}}, \bibinfo {author} {\bibfnamefont {E.}~\bibnamefont {Grant}},
  \bibinfo {author} {\bibfnamefont {L.}~\bibnamefont {Wossnig}},\ and\ \bibinfo
  {author} {\bibfnamefont {S.}~\bibnamefont {Severini}},\ }\bibfield  {title}
  {\bibinfo {title} {Adversarial quantum circuit learning for pure state
  approximation},\ }\href {https://doi.org/10.1088/1367-2630/ab14b5} {\bibfield
   {journal} {\bibinfo  {journal} {New Journal of Physics}\ }\textbf {\bibinfo
  {volume} {21}},\ \bibinfo {pages} {043023} (\bibinfo {year}
  {2019}{\natexlab{c}})}\BibitemShut {NoStop}%
\bibitem [{\citenamefont {Chakrabarti}\ \emph {et~al.}(2019)\citenamefont
  {Chakrabarti}, \citenamefont {Yiming}, \citenamefont {Li}, \citenamefont
  {Feizi},\ and\ \citenamefont {Wu}}]{chakrabarti2019quantum}%
  \BibitemOpen
  \bibfield  {author} {\bibinfo {author} {\bibfnamefont {S.}~\bibnamefont
  {Chakrabarti}}, \bibinfo {author} {\bibfnamefont {H.}~\bibnamefont {Yiming}},
  \bibinfo {author} {\bibfnamefont {T.}~\bibnamefont {Li}}, \bibinfo {author}
  {\bibfnamefont {S.}~\bibnamefont {Feizi}},\ and\ \bibinfo {author}
  {\bibfnamefont {X.}~\bibnamefont {Wu}},\ }\bibfield  {title} {\bibinfo
  {title} {Quantum {Wasserstein} generative adversarial networks},\ }in\
  \href@noop {} {\emph {\bibinfo {booktitle} {Advances in Neural Information
  Processing Systems}}},\ Vol.~\bibinfo {volume} {32},\ \bibinfo {editor}
  {edited by\ \bibinfo {editor} {\bibfnamefont {H.}~\bibnamefont {Wallach}},
  \bibinfo {editor} {\bibfnamefont {H.}~\bibnamefont {Larochelle}}, \bibinfo
  {editor} {\bibfnamefont {A.}~\bibnamefont {Beygelzimer}}, \bibinfo {editor}
  {\bibfnamefont {F.}~\bibnamefont {d\textquotesingle {Alch{\'e}-Buc}}},
  \bibinfo {editor} {\bibfnamefont {E.}~\bibnamefont {Fox}},\ and\ \bibinfo
  {editor} {\bibfnamefont {R.}~\bibnamefont {Garnett}}}\ (\bibinfo  {publisher}
  {Curran Associates, Inc.},\ \bibinfo {year} {2019})\ pp.\ \bibinfo {pages}
  {6781--6792},\ \Eprint {https://arxiv.org/abs/1911.00111} {arXiv:1911.00111}
  \BibitemShut {NoStop}%
\bibitem [{\citenamefont {Gorham}\ and\ \citenamefont
  {Mackey}(2018)}]{gorhamMeasuringSampleQuality2018}%
  \BibitemOpen
  \bibfield  {author} {\bibinfo {author} {\bibfnamefont {J.}~\bibnamefont
  {Gorham}}\ and\ \bibinfo {author} {\bibfnamefont {L.}~\bibnamefont
  {Mackey}},\ }\href@noop {} {\bibinfo {title} {Measuring {{Sample Quality}}
  with {{Stein}}'s {{Method}}}} (\bibinfo {year} {2018}),\ \Eprint
  {https://arxiv.org/abs/1506.03039} {arXiv:1506.03039 [cs, math, stat]}
  \BibitemShut {NoStop}%
\end{thebibliography}
\end{document}